\documentclass[11pt]{arxiv-article}

\pdfoutput=1

\bibliography{References}
\bibliography{T-duality}

\usepackage{Definitions}

\usepackage{setspace}
\usepackage{hyperref}

\newcommand*\drawplane[3]{  \draw[thin, densely dotted] (#1,#2,#2) -- (#1,#3,#2) --
  (#1,#3,#3) -- (#1,#2,#3) -- cycle;}

\definecolor[named]{D5col}{RGB}{0,160,176}
\definecolor[named]{D6col}{RGB}{0,160,176}
\definecolor[named]{NScol}{RGB}{106,74,60}
\definecolor[named]{NScol}{RGB}{138,155,15}
\definecolor[named]{O3col}{RGB}{204,51,63}
\definecolor[named]{D3col}{RGB}{0,0,0}
\definecolor[named]{D4col}{RGB}{0,0,0}
\definecolor[named]{Upkcol}{RGB}{235,104,65}
\definecolor[named]{Ukcol}{RGB}{235,104,65}

\newcommand*\minimalHW{
  \tdplotsetmaincoords{30}{30}

  \begin{tikzpicture}[xscale=.8,yscale=.6]
    \begin{scope}[scale=.8,shift={(-6,0)},tdplot_main_coords]
      \coordinate (Xmin) at (-4,0,0);
      \coordinate (Xmax) at (4,0,0);
      \coordinate (Ymin) at (0,-6,0);
      \coordinate (Ymax) at (0,6,0);
      \coordinate (Zmin) at (0,0,-1);
      \coordinate (Zmax) at (0,0,6);

      \draw[-latex,thin] (Xmin) -- (Xmax) node[right] {\(x_6\)};
      \draw[-latex,thin] (Ymin) -- (Ymax) node[right] {\((x_8,x_9)\)};
      \draw[-latex,thin] (Zmin) -- (Zmax) node[above] {\((x_4,x_5)\)};

      \newcommand*\planemin{-4}
      \newcommand*\planemax{4}

      \newcommand*\leftplane{-3}
      \newcommand*\midplane{0}
      \newcommand*\rightplane{3}
  
      \drawplane{\leftplane}{\planemax}{\planemin}
      \drawplane{\midplane}{\planemax}{\planemin}
      \drawplane{\rightplane}{\planemax}{\planemin}

      \draw[D6col,very thick] (\leftplane,\planemax,0) -- (\leftplane,\planemin,0) node[below] {\D6};

      \draw[NScol,very thick] (\rightplane,0,\planemin) -- (\rightplane,0,\planemax)  node[right] {\NS{}};
  
      \node at (0,0,-15) {(a)};
    \end{scope}

    \begin{scope}[scale=.8,shift={(5,0)},tdplot_main_coords]
      \coordinate (Xmin) at (-4,0,0);
      \coordinate (Xmax) at (4,0,0);
      \coordinate (Ymin) at (0,-6,0);
      \coordinate (Ymax) at (0,6,0);
      \coordinate (Zmin) at (0,0,-1);
      \coordinate (Zmax) at (0,0,6);

      \draw[-latex,thin] (Xmin) -- (Xmax) node[right] {\(x_6\)};
      \draw[-latex,thin] (Ymin) -- (Ymax) node[right] {\((x_8,x_9)\)};
      \draw[-latex,thin] (Zmin) -- (Zmax) node[above] {\((x_4,x_5)\)};

      \newcommand*\planemin{-4}
      \newcommand*\planemax{4}

      \newcommand*\leftplane{-3}
      \newcommand*\midplane{0}
      \newcommand*\rightplane{3}
      
      \drawplane{\leftplane}{\planemax}{\planemin}
      \drawplane{\midplane}{\planemax}{\planemin}
      \drawplane{\rightplane}{\planemax}{\planemin}

      \draw[D6col,very thick] (\rightplane,\planemax,0) -- (\rightplane,\planemin,0) node[below] {\D6};

      \draw[D4col,very thick] (\leftplane,0,0) -- (\rightplane,0,0) node[above] {\(D4\)};

      \draw[NScol,very thick] (\leftplane,0,\planemin) -- (\leftplane,0,\planemax)  node[right] {\NS{}};

      \node at (0,0,-15) {(b)};

    \end{scope}

  \end{tikzpicture}
}

\newcommand*\DsixBothSides{
  \tdplotsetmaincoords{30}{30}

  \begin{tikzpicture}[xscale=.8,yscale=.6]
    \begin{scope}[scale=.8,shift={(-6,0)},tdplot_main_coords]
      \coordinate (Xmin) at (-4,0,0);
      \coordinate (Xmax) at (4,0,0);
      \coordinate (Ymin) at (0,-6,0);
      \coordinate (Ymax) at (0,6,0);
      \coordinate (Zmin) at (0,0,-1);
      \coordinate (Zmax) at (0,0,6);

      \draw[-latex,thin] (Xmin) -- (Xmax) node[right] {\(x_6\)};
      \draw[-latex,thin] (Ymin) -- (Ymax) node[right] {\((x_8,x_9)\)};
      \draw[-latex,thin] (Zmin) -- (Zmax) node[above] {\((x_4,x_5)\)};

      \newcommand*\planemin{-4}
      \newcommand*\planemax{4}

      \newcommand*\leftplane{-3}
      \newcommand*\midplane{0}
      \newcommand*\rightplane{3}
  
      \drawplane{\leftplane}{\planemax}{\planemin}
      \drawplane{\midplane}{\planemax}{\planemin}
      \drawplane{\rightplane}{\planemax}{\planemin}

      \draw[D6col,very thick] (\leftplane,\planemax,0) -- (\leftplane,\planemin,0) node[below] {\D6};

      \draw[D4col,very thick] (\leftplane,0,0) -- (\rightplane,0,0) node[above] {\(D4\)};

      \draw[Upkcol,very thick] (\rightplane,\planemax,0) -- (\rightplane,\planemin,0)  node[below] {\NS{}'};

      \draw[NScol,very thick] (\midplane,0,\planemin) -- (\midplane,0,\planemax)  node[right] {\NS{}};
  
      \node at (0,0,-15) {(a)};
    \end{scope}

    \begin{scope}[scale=.8,shift={(5,0)},tdplot_main_coords]
      \coordinate (Xmin) at (-4,0,0);
      \coordinate (Xmax) at (4,0,0);
      \coordinate (Ymin) at (0,-6,0);
      \coordinate (Ymax) at (0,6,0);
      \coordinate (Zmin) at (0,0,-1);
      \coordinate (Zmax) at (0,0,6);

      \draw[-latex,thin] (Xmin) -- (Xmax) node[right] {\(x_6\)};
      \draw[-latex,thin] (Ymin) -- (Ymax) node[right] {\((x_8,x_9)\)};
      \draw[-latex,thin] (Zmin) -- (Zmax) node[above] {\((x_4,x_5)\)};

      \newcommand*\planemin{-4}
      \newcommand*\planemax{4}

      \newcommand*\leftplane{-3}
      \newcommand*\midplane{0}
      \newcommand*\rightplane{3}
      
      \drawplane{\leftplane}{\planemax}{\planemin}
      \drawplane{\midplane}{\planemax}{\planemin}
      \drawplane{\rightplane}{\planemax}{\planemin}

      \draw[D6col,very thick] (\leftplane,\planemax,0) -- (\leftplane,\planemin,0) node[below] {\D6};

      \draw[D4col,very thick] (\leftplane,1,0) -- (\midplane,1,0) node[right] {\(D4\)};
      \draw[D4col,very thick] (\midplane,0,0) -- (\rightplane,0,0) node[below left] {\(D4\)};

      \draw[Upkcol,very thick] (\midplane,\planemax,0) -- (\midplane,\planemin,0)  node[below] {\NS{}'};

      \draw[NScol,very thick] (\rightplane,0,\planemin) -- (\rightplane,0,\planemax)  node[right] {\NS{}};

      \node at (0,0,-15) {(b)};

    \end{scope}

  \end{tikzpicture}
}

\newcommand*\FigureDualityFundamentals{
  \tdplotsetmaincoords{30}{30}

  \begin{tikzpicture}[xscale=.8,yscale=.6]
    \begin{scope}[scale=.8,shift={(-6,0)},tdplot_main_coords]
      \coordinate (Xmin) at (-4,0,0);
      \coordinate (Xmax) at (4,0,0);
      \coordinate (Ymin) at (0,-6,0);
      \coordinate (Ymax) at (0,6,0);
      \coordinate (Zmin) at (0,0,-1);
      \coordinate (Zmax) at (0,0,6);

      \draw[-latex,thin] (Xmin) -- (Xmax) node[right] {\(x_6\)};
      \draw[-latex,thin] (Ymin) -- (Ymax) node[right] {\((x_8,x_9)\)};
      \draw[-latex,thin] (Zmin) -- (Zmax) node[above] {\((x_4,x_5)\)};

      \newcommand*\planemin{-4}
      \newcommand*\planemax{4}

      \newcommand*\leftplane{-3}
      \newcommand*\midplane{0}
      \newcommand*\rightplane{3}
      
      \drawplane{\leftplane}{\planemax}{\planemin}
      \drawplane{\midplane}{\planemax}{\planemin}
      \drawplane{\rightplane}{\planemax}{\planemin}

      \draw[D5col,very thick] (\leftplane,\planemax,0) -- (\leftplane,\planemin,0) node[below] {\(D5\)};

      \draw[O3col,very thick] (\rightplane,0,0) -- (\leftplane,0,0) node[below] {\(O3\)};
      \draw[D3col,very thick] (\leftplane,.1,0) -- (\rightplane,.1,0) node[above] {\(D3\)};

      \draw[Upkcol,very thick] (\rightplane,\planemax,0) -- (\rightplane,\planemin,0)  node[below] {\((1',k)\)};

      \draw[NScol,very thick] (\midplane,0,\planemin) -- (\midplane,0,\planemax)  node[right] {\(NS\)};

    \end{scope}

    \begin{scope}[scale=.8,shift={(5,0)},tdplot_main_coords]
      \coordinate (Xmin) at (-4,0,0);
      \coordinate (Xmax) at (4,0,0);
      \coordinate (Ymin) at (0,-6,0);
      \coordinate (Ymax) at (0,6,0);
      \coordinate (Zmin) at (0,0,-1);
      \coordinate (Zmax) at (0,0,6);
      
      \draw[-latex,thin] (Xmin) -- (Xmax) node[right] {\(x_6\)};
      \draw[-latex,thin] (Ymin) -- (Ymax) node[right] {\(x_7\)};
      \draw[-latex,thin] (Zmin) -- (Zmax) node[above] {\(x_3\)};

      \newcommand*\planemin{-4}
      \newcommand*\planemax{4}

      \newcommand*\leftplane{-3}
      \newcommand*\midplane{0}
      \newcommand*\rightplane{3}
      
      \drawplane{\leftplane}{\planemax}{\planemin}
      \drawplane{\midplane}{\planemax}{\planemin}
      \drawplane{\rightplane}{\planemax}{\planemin}

      \draw[D5col,very thick] (\leftplane,\planemax,0) -- (\leftplane,\planemin,0) node[below] {\(D5\)};

      \draw[O3col,very thick] (\rightplane,0,0) -- (\leftplane,0,0) node[below] {\(O3\)};
      \draw[D3col,very thick] (\leftplane,.1,0) -- (\rightplane,.1,0)  node[above] {\(D3\)};

      \draw[Upkcol,very thick] (\rightplane,\planemax,\planemin) -- (\rightplane,\planemin,\planemax)  node[below] {\((1',k)\)};

      \draw[NScol,very thick] (\midplane,0,\planemin) -- (\midplane,0,\planemax)  node[right] {\(NS\)};
      
    \end{scope}
  \end{tikzpicture} }

\newcommand*\FigureSOSpOfive{
  \tdplotsetmaincoords{20}{70}

  \begin{tikzpicture}[xscale=.8,yscale=.6]

    \begin{scope}[scale=.6,shift={(-7,0)},tdplot_main_coords]

      \coordinate (Xmin) at (-7,0,0);
      \coordinate (Xmax) at (7,0,0);
      \coordinate (Ymin) at (0,-6,0);
      \coordinate (Ymax) at (0,6,0);
      \coordinate (Zmin) at (0,0,-1);
      \coordinate (Zmax) at (0,0,8);
      
      \draw[-latex,thin] (Xmin) -- (Xmax) node[right] {\(x_6\)};
      \draw[-latex,thin] (Ymax) -- (Ymin) node[below] {\((x_4,x_5)\)};
      \draw[-latex,thin] (Zmin) -- (Zmax) node[above] {\((x_8,x_9)\)};

      \newcommand*\planemin{-4}
      \newcommand*\planemax{4}

      \newcommand*\leftleftplane{-5}
      \newcommand*\leftplane{-2.5}
      \newcommand*\midplane{0}
      \newcommand*\rightplane{2.5}
      \newcommand*\rightrightplane{5}
      
      \drawplane{\leftleftplane}{\planemax}{\planemin}
      \drawplane{\leftplane}{\planemax}{\planemin}
      \drawplane{\midplane}{\planemax}{\planemin}
      \drawplane{\rightplane}{\planemax}{\planemin}
      \drawplane{\rightrightplane}{\planemax}{\planemin}

      \draw[Ukcol,very thick] (\leftleftplane,\planemax,\planemax) -- (\leftleftplane,\planemin,\planemin) node[below] {\((n,k)_{\theta}\)};

      \draw[D5col,very thick] (\leftplane,\planemin,\planemin) -- (\leftplane,\planemax,\planemax) node[above] {\(N_f\) \(D5_\theta\)};

      \draw[O3col,very thick] (\midplane,\planemin,0) -- (\midplane,\planemax,0) node[below right] {\(O5\)};
      \draw[D3col,very thick] (\leftleftplane,0,0) -- (\rightrightplane,0,0)  node[above right] {\(D3\)};

      \draw[D5col,very thick] (\rightplane,\planemin,\planemax) -- (\rightplane,\planemax,\planemin) node[right] {\(2N_f\) \(D5_{-\theta}\)};

      \draw[Upkcol,very thick] (\rightrightplane,\planemin,\planemax) --  (\rightrightplane,\planemax,\planemin) node[right] {\((n,-k)_{-\theta}\)};

    \end{scope}

    \begin{scope}[scale=.6,shift={(7,0)},tdplot_main_coords]
      \coordinate (Xmin) at (-7,0,0);
      \coordinate (Xmax) at (7,0,0);
      \coordinate (Ymin) at (0,-6,0);
      \coordinate (Ymax) at (0,6,0);
      \coordinate (Zmin) at (0,0,-1);
      \coordinate (Zmax) at (0,0,8);
      
      \draw[-latex,thin] (Xmin) -- (Xmax) node[right] {\(x_6\)};
      \draw[-latex,thin] (Ymin) -- (Ymax) node[right] {\(x_7\)};
      \draw[-latex,thin] (Zmin) -- (Zmax) node[above] {\(x_3\)};

      \newcommand*\planemin{-4}
      \newcommand*\planemax{4}

      \newcommand*\leftleftplane{-5}
      \newcommand*\leftplane{-2.5}
      \newcommand*\midplane{0}
      \newcommand*\rightplane{2.5}
      \newcommand*\rightrightplane{5}
      
      \drawplane{\leftleftplane}{\planemax}{\planemin}
      \drawplane{\leftplane}{\planemax}{\planemin}
      \drawplane{\midplane}{\planemax}{\planemin}
      \drawplane{\rightplane}{\planemax}{\planemin}
      \drawplane{\rightrightplane}{\planemax}{\planemin}

      \draw[Ukcol,very thick] (\leftleftplane,\planemax,\planemax) -- (\leftleftplane,\planemin,\planemin) node[below] {\((n,k)_{\theta}\)};

      \draw[D5col,very thick] (\leftplane,0,\planemin) -- (\leftplane,0,\planemax) node[above right] {\(2N_f\) \(D5_\theta\)};

      \fill[O3col] (0,0,0) circle [radius=3pt] node[below left] {\(O5\)};
      \draw[D3col,very thick] (\leftleftplane,0,0) -- (\rightrightplane,0,0)  node[above right] {\(D3\)};

      \draw[D5col,very thick] (\rightplane,0,\planemin) -- (\rightplane,0,\planemax) node[right] {\(2N_f\) \(D5_{-\theta}\)};

      \draw[Upkcol,very thick]  (\rightrightplane,\planemax,\planemin) -- (\rightrightplane,\planemin,\planemax) node[below] {\((n,-k)_{-\theta}\)};
      
    \end{scope}

  \end{tikzpicture}
}

\newcommand*\FigureUnitarySymmAntisymm{
  \tdplotsetmaincoords{20}{70}

  \begin{tikzpicture}[xscale=.8,yscale=.6]

    \begin{scope}[scale=.6,shift={(-7,0)},tdplot_main_coords]
      \coordinate (Xmin) at (-7,0,0);
      \coordinate (Xmax) at (7,0,0);
      \coordinate (Ymin) at (0,-6,0);
      \coordinate (Ymax) at (0,6,0);
      \coordinate (Zmin) at (0,0,-1);
      \coordinate (Zmax) at (0,0,8);
      
      \draw[-latex,thin] (Xmin) -- (Xmax) node[right] {\(x_6\)};
      \draw[-latex,thin] (Ymax) -- (Ymin) node[below] {\((x_4,x_5)\)};
      \draw[-latex,thin] (Zmin) -- (Zmax) node[above] {\((x_8,x_9)\)};

      \newcommand*\planemin{-4}
      \newcommand*\planemax{4}

      \newcommand*\leftleftplane{-5}
      \newcommand*\leftplane{-2.5}
      \newcommand*\midplane{0}
      \newcommand*\rightplane{2.5}
      \newcommand*\rightrightplane{5}
      
      \drawplane{\leftleftplane}{\planemax}{\planemin}
      \drawplane{\leftplane}{\planemax}{\planemin}
      \drawplane{\midplane}{\planemax}{\planemin}
      \drawplane{\rightplane}{\planemax}{\planemin}
      \drawplane{\rightrightplane}{\planemax}{\planemin}

      \draw[Ukcol,very thick] (\leftleftplane,\planemax,\planemax) -- (\leftleftplane,\planemin,\planemin) node[below] {\((n,k)_{\theta}\)};

      \draw[D5col,very thick]  (\leftplane,\planemin,\planemin) -- (\leftplane,\planemax,\planemax) node[right] {\(N_f\) \(D5_\theta\)};

      \draw[O3col,very thick] (\midplane,\planemin,-0.2) -- (\midplane,\planemax,-0.2) node[below ] {\(O5\)};
      \draw[NScol,very thick] (\midplane,\planemin,0.2) -- (\midplane,\planemax,0.2) node[above] {\(NS\)};
      \draw[D3col,very thick] (\leftleftplane,0,0) -- (\rightrightplane,0,0)  node[above right] {\(D3\)};

      \draw[D5col,very thick] (\rightplane,\planemin,\planemax) -- (\rightplane,\planemax,\planemin) node[right] {\(2N_f\) \(D5_{-\theta}\)};

      \draw[Upkcol,very thick] (\rightrightplane,\planemin,\planemax) --  (\rightrightplane,\planemax,\planemin) node[right] {\((n,-k)_{-\theta}\)};
    \end{scope}

    \begin{scope}[scale=.6,shift={(7,0)},tdplot_main_coords]
      \coordinate (Xmin) at (-7,0,0);
      \coordinate (Xmax) at (7,0,0);
      \coordinate (Ymin) at (0,-6,0);
      \coordinate (Ymax) at (0,6,0);
      \coordinate (Zmin) at (0,0,-1);
      \coordinate (Zmax) at (0,0,8);
      
      \draw[-latex,thin] (Xmin) -- (Xmax) node[right] {\(x_6\)};
      \draw[-latex,thin] (Ymin) -- (Ymax) node[right] {\(x_7\)};
      \draw[-latex,thin] (Zmin) -- (Zmax) node[above] {\(x_3\)};

      \newcommand*\planemin{-4}
      \newcommand*\planemax{4}

      \newcommand*\leftleftplane{-5}
      \newcommand*\leftplane{-2.5}
      \newcommand*\midplane{0}
      \newcommand*\rightplane{2.5}
      \newcommand*\rightrightplane{5}
      
      \drawplane{\leftleftplane}{\planemax}{\planemin}
      \drawplane{\leftplane}{\planemax}{\planemin}
      \drawplane{\midplane}{\planemax}{\planemin}
      \drawplane{\rightplane}{\planemax}{\planemin}
      \drawplane{\rightrightplane}{\planemax}{\planemin}

      \draw[Ukcol,very thick] (\leftleftplane,\planemin,\planemin) -- (\leftleftplane,\planemax,\planemax) node[right] {\((n,k)_{\theta}\)};

      \draw[D5col,very thick]  (\leftplane,\planemin,0) -- (\leftplane,\planemax,0) node[right] {\(N_f\) \(D5_\theta\)};

      \fill[O3col] (0,0,0) circle [radius=3pt] node[below left] {\(O5\)};
      \draw[NScol,very thick] (\midplane,0,\planemin) -- (\midplane,0,\planemax) node[below right] {\(NS\)};
      \draw[D3col,very thick] (\leftleftplane,0,0) -- (\rightrightplane,0,0)  node[above] {\(D3\)};

      \draw[D5col,very thick] (\rightplane,\planemin,0) -- (\rightplane,\planemax,0) node[right] {\(N_f\) \(D5_{-\theta}\)};

      \draw[Upkcol,very thick] (\rightrightplane,\planemax,\planemin) -- (\rightrightplane,\planemin,\planemax)  node[below] {\((n,-k)_{-\theta}\)};
    \end{scope}
  \end{tikzpicture}
}

\newcommand*\FigureSOSpAdjoint{
  \tdplotsetmaincoords{30}{30}

  \begin{tikzpicture}[xscale=.8,yscale=.6]

    \begin{scope}[scale=.8,shift={(-6,0)},tdplot_main_coords]
      \coordinate (Xmin) at (-4,0,0);
      \coordinate (Xmax) at (4,0,0);
      \coordinate (Ymin) at (0,-6,0);
      \coordinate (Ymax) at (0,6,0);
      \coordinate (Zmin) at (0,0,-1);
      \coordinate (Zmax) at (0,0,6);

      \draw[-latex,thin] (Xmin) -- (Xmax) node[right] {\(x_6\)};
      \draw[-latex,thin] (Ymin) -- (Ymax) node[right] {\((x_8,x_9)\)};
      \draw[-latex,thin] (Zmin) -- (Zmax) node[above] {\((x_4,x_5)\)};

      \newcommand*\planemin{-4}
      \newcommand*\planemax{4}

      \newcommand*\leftplane{-3}
      \newcommand*\midplane{0}
      \newcommand*\rightplane{3}
      
      \drawplane{\leftplane}{\planemax}{\planemin}
      \drawplane{\midplane}{\planemax}{\planemin}
      \drawplane{\rightplane}{\planemax}{\planemin}

      \draw[NScol,very thick] (\leftplane,\planemax,-2) -- (\leftplane,\planemin,-2) node[left] {\(NS\)};
      \draw[NScol,very thick] (\leftplane,\planemax,0) -- (\leftplane,\planemin,0) node[left] {\(NS\)};
      \draw[NScol,very thick] (\leftplane,\planemax,2) -- (\leftplane,\planemin,2) node[left] {\(NS\)};

      \draw[O3col,very thick] (\rightplane,0,0) -- (\leftplane,0,0) node[below] {\(O3\)};
      \draw[D3col,very thick] (\leftplane,.1,-2) -- (\rightplane,.1,-2) node[above] {\(r_i\) \(D3\)};
      \draw[D3col,very thick] (\leftplane,.1,0) -- (\rightplane,.1,0) node[above] {\(2 r_0\) \(D3\)};
      \draw[D3col,very thick] (\leftplane,.1,2) -- (\rightplane,.1,2) node[above] {\(r_i\) \(D3\)};

      \draw[D5col,very thick] (\midplane,0,\planemax) -- (\midplane,0,\planemin) node[below] {\(2N_f\) \(D5\)};
      \draw[Upkcol,very thick] (\rightplane,0,\planemax) -- (\rightplane,0,\planemin)  node[below] {\((1',2k)\)};

    \end{scope}

    \begin{scope}[scale=.8,shift={(5,0)},tdplot_main_coords]
      \coordinate (Xmin) at (-4,0,0);
      \coordinate (Xmax) at (4,0,0);
      \coordinate (Ymin) at (0,-6,0);
      \coordinate (Ymax) at (0,6,0);
      \coordinate (Zmin) at (0,0,-1);
      \coordinate (Zmax) at (0,0,6);

      \draw[-latex,thin] (Xmin) -- (Xmax) node[right] {\(x_6\)};
      \draw[-latex,thin] (Ymin) -- (Ymax) node[right] {\(x_7\)};
      \draw[-latex,thin] (Zmin) -- (Zmax) node[above] {\(x_3\)};

      \newcommand*\planemin{-4}
      \newcommand*\planemax{4}

      \newcommand*\leftplane{-3}
      \newcommand*\midplane{0}
      \newcommand*\rightplane{3}
      
      \drawplane{\leftplane}{\planemax}{\planemin}
      \drawplane{\midplane}{\planemax}{\planemin}
      \drawplane{\rightplane}{\planemax}{\planemin}

      \draw[NScol,very thick] (\leftplane,0,\planemin) -- (\leftplane,0,\planemax) node[above] {\((2n+1)\) \(NS\)};

      \draw[O3col,very thick] (\rightplane,0,0) -- (\leftplane,0,0) node[below] {\(O3\)};
      \draw[D3col,very thick] (\leftplane,.1,0) -- (\rightplane,.1,0) node[above] {\(2 N\) \(D3\)};

      \draw[D5col,very thick] (\midplane,\planemax,0) -- (\midplane,\planemin,0) node[below] {\(2N_f\) \(D5\)};
      \draw[Upkcol,very thick] (\rightplane,\planemax,\planemin) -- (\rightplane,\planemin,\planemax)  node[below] {\((1',2k)\)};

    \end{scope}

  \end{tikzpicture}
} 

\newcommand*\FigureCS{

\begin{tikzpicture}[]

  \begin{scope}[shift={(-3.5,0)}]
  \coordinate (Xmin) at (-3,0);
  \coordinate (Xmax) at (3,0);
  \coordinate (Ymin) at (0,-2);
  \coordinate (Ymax) at (0,2);
    \draw[-latex] (Xmin) -- (Xmax) node [right] {\(x_7\)};
    \draw[-latex] (Ymin) -- (Ymax) node [above] {\(x_3\)};

    \draw[D5col, very thick] (-2.5,0) -- (2.5,0) node [above] {\D5};
    \draw[NScol, very thick] (0,-1.5) -- (0,1.5) node [right] {\NS{}};

    \node at (0,-2.5) {(a)};
  \end{scope}

  \begin{scope}[shift={(3.5,0)}]
  \coordinate (Xmin) at (-3,0);
  \coordinate (Xmax) at (3,0);
  \coordinate (Ymin) at (0,-2);
  \coordinate (Ymax) at (0,2);
    \draw[-latex] (Xmin) -- (Xmax) node [right] {\(x_7\)};
    \draw[-latex] (Ymin) -- (Ymax) node [above] {\(x_3\)};

    \draw[D5col, very thick] (0,-.5) -- (-2.5,-.5) node [above] {\D5};
    \draw[D5col, very thick] (0,.5) -- (2.5,.5) node [above] {\D5};

    \draw[NScol, very thick] (0,-1.5) -- (0,1.5) node [right] {\NS{}};
    \node at (0,-2.5) {(b)};

  \end{scope}

  \begin{scope}[shift={(0,-5)}]
  \coordinate (Xmin) at (-3,0);
  \coordinate (Xmax) at (3,0);
  \coordinate (Ymin) at (0,-2);
  \coordinate (Ymax) at (0,2);
    \draw[-latex] (Xmin) -- (Xmax) node [right] {\(x_7\)};
    \draw[-latex] (Ymin) -- (Ymax) node [above] {\(x_3\)};
    \draw[D5col, very thick] (-.5,-.5) -- (-2.5,-.5) node [above] {\D5};
    \draw[D5col, very thick] (.5,.5) -- (2.5,.5) node [above] {\D5};

    \draw[Ukcol, very thick] (-.5,-.5) -- (.5,.5) node [below right] {\((1,1)\)};
    \draw[NScol, very thick] (-.5,-.5) -- (-.5,-1.5) node [right] {\NS{}};
    \draw[NScol, very thick] (.5,.5) -- (.5,1.5) node [right] {\NS{}};

    \node at (0,-2.5) {(c)};

  \end{scope}
  
\end{tikzpicture}
}

\newcommand*\FigureSOSpOsix{
  \tdplotsetmaincoords{20}{70}

  \begin{tikzpicture}[xscale=.8,yscale=.6]

    \begin{scope}[scale=.6,shift={(-7,0)},tdplot_main_coords]

      \coordinate (Xmin) at (-7,0,0);
      \coordinate (Xmax) at (7,0,0);
      \coordinate (Ymin) at (0,-6,0);
      \coordinate (Ymax) at (0,6,0);
      \coordinate (Zmin) at (0,0,-1);
      \coordinate (Zmax) at (0,0,8);
      
      \draw[-latex,thin] (Xmin) -- (Xmax) node[right] {\(x_6\)};
      \draw[-latex,thin] (Ymax) -- (Ymin) node[below] {\((x_4,x_5)\)};
      \draw[-latex,thin] (Zmin) -- (Zmax) node[above] {\((x_8,x_9)\)};

      \newcommand*\planemin{-4}
      \newcommand*\planemax{4}

      \newcommand*\leftleftplane{-5}
      \newcommand*\leftplane{-2.5}
      \newcommand*\midplane{0}
      \newcommand*\rightplane{2.5}
      \newcommand*\rightrightplane{5}
      
      \drawplane{\leftleftplane}{\planemax}{\planemin}
      \drawplane{\leftplane}{\planemax}{\planemin}
      \drawplane{\midplane}{\planemax}{\planemin}
      \drawplane{\rightplane}{\planemax}{\planemin}
      \drawplane{\rightrightplane}{\planemax}{\planemin}

      \draw[NScol,very thick] (\leftleftplane,\planemax,\planemax) -- (\leftleftplane,\planemin,\planemin) node[below] {\(\NS{}_{\theta}\)};

      \draw[D5col,very thick] (\leftplane,\planemin,\planemin) -- (\leftplane,\planemax,\planemax) node[above] {\(N_f\) \(D6_\theta\)};

      \draw[O3col,very thick] (\midplane,\planemin,0) -- (\midplane,\planemax,0) node[below right] {\(O6\)};
      \draw[D3col,very thick] (\leftleftplane,0,0) -- (\rightrightplane,0,0)  node[above right] {\(D4\)};

      \draw[D5col,very thick] (\rightplane,\planemin,\planemax) -- (\rightplane,\planemax,\planemin) node[right] {\(2N_f\) \(D6_{-\theta}\)};

      \draw[NScol,very thick] (\rightrightplane,\planemin,\planemax) --  (\rightrightplane,\planemax,\planemin) node[right] {\(\NS{}_{-\theta}\)};

    \end{scope}

    \begin{scope}[scale=.6,shift={(7,0)},tdplot_main_coords]

      \coordinate (Xmin) at (-7,0,0);
      \coordinate (Xmax) at (7,0,0);
      \coordinate (Ymin) at (0,-6,0);
      \coordinate (Ymax) at (0,6,0);
      \coordinate (Zmin) at (0,0,-1);
      \coordinate (Zmax) at (0,0,8);
      
      \draw[-latex,thin] (Xmin) -- (Xmax) node[right] {\(x_6\)};
      \draw[-latex,thin] (Ymax) -- (Ymin) node[below] {\((x_4,x_5)\)};
      \draw[-latex,thin] (Zmin) -- (Zmax) node[above] {\((x_8,x_9)\)};

      \newcommand*\planemin{-4}
      \newcommand*\planemax{4}

      \newcommand*\leftleftplane{-5}
      \newcommand*\leftplane{-2.5}
      \newcommand*\midplane{0}
      \newcommand*\rightplane{2.5}
      \newcommand*\rightrightplane{5}
      
      \drawplane{\leftleftplane}{\planemax}{\planemin}
      \drawplane{\leftplane}{\planemax}{\planemin}
      \drawplane{\midplane}{\planemax}{\planemin}
      \drawplane{\rightplane}{\planemax}{\planemin}
      \drawplane{\rightrightplane}{\planemax}{\planemin}

      \draw[NScol,very thick] (\leftplane,\planemax,\planemax) -- (\leftplane,\planemin,\planemin) node[below] {\(\NS{}_{\theta}\)};

      \draw[D5col,very thick] (\leftleftplane,\planemin,\planemin) -- (\leftleftplane,\planemax,\planemax) node[above] {\(N_f\) \(D6_\theta\)};

      \draw[O3col,very thick] (\midplane,\planemin,0) -- (\midplane,\planemax,0) node[below right] {\(O6\)};
      \draw[D3col,very thick] (\leftleftplane,0,0) -- (\rightrightplane,0,0)  node[above right] {\(D4\)};

      \draw[D5col,very thick] (\rightrightplane,\planemin,\planemax) -- (\rightrightplane,\planemax,\planemin) node[right] {\(2N_f\) \(D6_{-\theta}\)};

      \draw[NScol,very thick] (\rightplane,\planemin,\planemax) --  (\rightplane,\planemax,\planemin) node[right] {\(\NS{}_{-\theta}\)};

    \end{scope}

  \end{tikzpicture}
}

\newcommand*\FigureGKele{
  \tdplotsetmaincoords{30}{30}

  \begin{tikzpicture}[xscale=.8,yscale=.6]
    \begin{scope}[scale=.8,shift={(-6,0)},tdplot_main_coords]
      \coordinate (Xmin) at (-4,0,0);
      \coordinate (Xmax) at (4,0,0);
      \coordinate (Ymin) at (0,-6,0);
      \coordinate (Ymax) at (0,6,0);
      \coordinate (Zmin) at (0,0,-1);
      \coordinate (Zmax) at (0,0,6);

      \draw[-latex,thin] (Xmin) -- (Xmax) node[right] {\(x_6\)};
      \draw[-latex,thin] (Ymin) -- (Ymax) node[right] {\((x_8,x_9)\)};
      \draw[-latex,thin] (Zmin) -- (Zmax) node[above] {\((x_4,x_5)\)};

      \newcommand*\planemin{-4}
      \newcommand*\planemax{4}

      \newcommand*\leftplane{-3}
      \newcommand*\midplane{0}
      \newcommand*\rightplane{3}
      
      \drawplane{\leftplane}{\planemax}{\planemin}
      \drawplane{\midplane}{\planemax}{\planemin}
      \drawplane{\rightplane}{\planemax}{\planemin}

      \draw[D5col,very thick] (\leftplane,\planemax,0) -- (\leftplane,\planemin,0) node[below] {\(N_f\) \D5};

      \draw[D3col,very thick] (\leftplane,.1,0) -- (\rightplane,.1,0);
      \node[D3col] at (-2,1,0) {\(N_f\) \D3};
      \node[D3col] at (2,1,0) {\(N_c\) \D3};

      \draw[NScol,very thick] (\rightplane,\planemax,0) -- (\rightplane,\planemin,0)  node[below] {\(\NS{}'\)};

      \draw[Ukcol,very thick] (\midplane,0,\planemin) -- (\midplane,0,\planemax)  node[right] {\((1,k)\)};

    \end{scope}

    \begin{scope}[scale=.8,shift={(5,0)},tdplot_main_coords]
      \coordinate (Xmin) at (-4,0,0);
      \coordinate (Xmax) at (4,0,0);
      \coordinate (Ymin) at (0,-6,0);
      \coordinate (Ymax) at (0,6,0);
      \coordinate (Zmin) at (0,0,-1);
      \coordinate (Zmax) at (0,0,6);
      
      \draw[-latex,thin] (Xmin) -- (Xmax) node[right] {\(x_6\)};
      \draw[-latex,thin] (Ymin) -- (Ymax) node[right] {\(x_7\)};
      \draw[-latex,thin] (Zmin) -- (Zmax) node[above] {\(x_3\)};

      \newcommand*\planemin{-4}
      \newcommand*\planemax{4}

      \newcommand*\leftplane{-3}
      \newcommand*\midplane{0}
      \newcommand*\rightplane{3}
      
      \drawplane{\leftplane}{\planemax}{\planemin}
      \drawplane{\midplane}{\planemax}{\planemin}
      \drawplane{\rightplane}{\planemax}{\planemin}

      \draw[D5col,very thick] (\leftplane,\planemax,0) -- (\leftplane,\planemin,0) node[below] {\(N_f\) \D5};

      \draw[D3col,very thick] (\leftplane,0,0) -- (\rightplane,0,0);
      \node[D3col] at (-2,1,0) {\(N_f\) \D3};
      \node[D3col] at (2,1,0) {\(N_c\) \D3};

      \draw[Ukcol,very thick] (\midplane,\planemax,\planemax) -- (\midplane,\planemin,\planemin)  node[below] {\((1,k)\)};

      \draw[NScol,very thick] (\rightplane,\planemin,0) -- (\rightplane,\planemax,0)  node[right] {\(\NS{}'\)};
      
    \end{scope}
    \node at (0,-5) {(a)};
  \end{tikzpicture}
}

\newcommand*\FigureGKmag{
  \tdplotsetmaincoords{30}{30}

  \begin{tikzpicture}[xscale=.8,yscale=.6]
    \begin{scope}[scale=.8,shift={(-6,0)},tdplot_main_coords]
      \coordinate (Xmin) at (-4,0,0);
      \coordinate (Xmax) at (4,0,0);
      \coordinate (Ymin) at (0,-6,0);
      \coordinate (Ymax) at (0,6,0);
      \coordinate (Zmin) at (0,0,-1);
      \coordinate (Zmax) at (0,0,6);

      \draw[-latex,thin] (Xmin) -- (Xmax) node[right] {\(x_6\)};
      \draw[-latex,thin] (Ymin) -- (Ymax) node[right] {\((x_8,x_9)\)};
      \draw[-latex,thin] (Zmin) -- (Zmax) node[above] {\((x_4,x_5)\)};

      \newcommand*\planemin{-4}
      \newcommand*\planemax{4}

      \newcommand*\leftplane{-3}
      \newcommand*\midplane{0}
      \newcommand*\rightplane{3}
      
      \drawplane{\leftplane}{\planemax}{\planemin}
      \drawplane{\midplane}{\planemax}{\planemin}
      \drawplane{\rightplane}{\planemax}{\planemin}

      \draw[D5col,very thick] (\leftplane,\planemax,0) -- (\leftplane,\planemin,0) node[below] {\(N_f\) \D5};

      \draw[D3col,very thick] (\leftplane,.1,0) -- (\rightplane,.1,0);
      \node[D3col] at (-2,1,0) {\(N_f\) \D3};
      \node[D3col] at (2,-1,0) {\(N_f - N_c + \abs{k}\) \D3};

      \draw[NScol,very thick] (\midplane,\planemax,0) -- (\midplane,\planemin,0)  node[below] {\(\NS{}'\)};

      \draw[Ukcol,very thick] (\rightplane,0,\planemin) -- (\rightplane,0,\planemax)  node[right] {\((1,k)\)};

    \end{scope}

    \begin{scope}[scale=.8,shift={(5,0)},tdplot_main_coords]
      \coordinate (Xmin) at (-4,0,0);
      \coordinate (Xmax) at (4,0,0);
      \coordinate (Ymin) at (0,-6,0);
      \coordinate (Ymax) at (0,6,0);
      \coordinate (Zmin) at (0,0,-1);
      \coordinate (Zmax) at (0,0,6);
      
      \draw[-latex,thin] (Xmin) -- (Xmax) node[right] {\(x_6\)};
      \draw[-latex,thin] (Ymin) -- (Ymax) node[right] {\(x_7\)};
      \draw[-latex,thin] (Zmin) -- (Zmax) node[above] {\(x_3\)};

      \newcommand*\planemin{-4}
      \newcommand*\planemax{4}

      \newcommand*\leftplane{-3}
      \newcommand*\midplane{0}
      \newcommand*\rightplane{3}
      
      \drawplane{\leftplane}{\planemax}{\planemin}
      \drawplane{\midplane}{\planemax}{\planemin}
      \drawplane{\rightplane}{\planemax}{\planemin}

      \draw[D5col,very thick] (\leftplane,\planemax,0) -- (\leftplane,\planemin,0) node[below] {\(N_f\) \D5};

      \draw[D3col,very thick] (\leftplane,0,0) -- (\rightplane,0,0);
      \node[D3col] at (-2,1,0) {\(N_f\) \D3};
      \node[D3col] at (2,-1,0) {\(N_f - N_c + \abs{k}\) \D3};

      \draw[Ukcol,very thick] (\rightplane,\planemax,\planemax) -- (\rightplane,\planemin,\planemin)  node[below] {\((1,k)\)};

      \draw[NScol,very thick] (\midplane,\planemin,0) -- (\midplane,\planemax,0)  node[right] {\(\NS{}'\)};
      
    \end{scope}
    \node at (0,-5) {(b)};

  \end{tikzpicture}
}

\newcommand*\FigureNiarchosele{
  \tdplotsetmaincoords{30}{30}

  \begin{tikzpicture}[xscale=.8,yscale=.6]

    \begin{scope}[scale=.8,shift={(-6,0)},tdplot_main_coords]
      \coordinate (Xmin) at (-4,0,0);
      \coordinate (Xmax) at (4,0,0);
      \coordinate (Ymin) at (0,-6,0);
      \coordinate (Ymax) at (0,6,0);
      \coordinate (Zmin) at (0,0,-1);
      \coordinate (Zmax) at (0,0,6);

      \draw[-latex,thin] (Xmin) -- (Xmax) node[right] {\(x_6\)};
      \draw[-latex,thin] (Ymin) -- (Ymax) node[right] {\((x_8,x_9)\)};
      \draw[-latex,thin] (Zmin) -- (Zmax) node[above] {\((x_4,x_5)\)};

      \newcommand*\planemin{-4}
      \newcommand*\planemax{4}

      \newcommand*\leftplane{-3}
      \newcommand*\midplane{0}
      \newcommand*\rightplane{3}
      
      \drawplane{\leftplane}{\planemax}{\planemin}
      \drawplane{\midplane}{\planemax}{\planemin}
      \drawplane{\rightplane}{\planemax}{\planemin}

      \draw[D5col,very thick] (\leftplane,\planemax,-2) -- (\leftplane,\planemin,-2);
      \draw[D5col,very thick] (\leftplane,\planemax,0) -- (\leftplane,\planemin,0);
      \draw[D5col,very thick] (\leftplane,\planemax,2) -- (\leftplane,\planemin,2);
      \draw [D5col,decorate,decoration={brace,amplitude=5pt}] (-3.3,\planemin,-2) -- (-3.3,\planemin,2) node[D5col,midway,xshift=-0.6cm]  {\(n\) \D5};
      
      \draw[NScol,very thick] (\rightplane,\planemax,-2) -- (\rightplane,\planemin,-2);
      \draw[NScol,very thick] (\rightplane,\planemax,0) -- (\rightplane,\planemin,0);
      \draw[NScol,very thick] (\rightplane,\planemax,2) -- (\rightplane,\planemin,2);
      \draw [NScol,decorate,decoration={brace,amplitude=5pt}] (2.7,\planemin,-2) -- (2.7,\planemin,2) node[NScol,midway,xshift=-0.6cm]  {\(n\) \NS{}'};

      \draw[D3col,very thick] (\leftplane,0,-2) -- (\rightplane,0,-2);
      \draw[D3col,very thick] (\leftplane,0,0) -- (\rightplane,0,0);
      \draw[D3col,very thick] (\leftplane,0,2) -- (\rightplane,0,2);

      \draw[Ukcol,very thick] (\midplane,0,\planemax) -- (\midplane,0,\planemin)  node[below] {\((1,k)\)};

    \end{scope}

    \begin{scope}[scale=.8,shift={(5,0)},tdplot_main_coords]
      \coordinate (Xmin) at (-4,0,0);
      \coordinate (Xmax) at (4,0,0);
      \coordinate (Ymin) at (0,-6,0);
      \coordinate (Ymax) at (0,6,0);
      \coordinate (Zmin) at (0,0,-1);
      \coordinate (Zmax) at (0,0,6);

      \draw[-latex,thin] (Xmin) -- (Xmax) node[right] {\(x_6\)};
      \draw[-latex,thin] (Ymin) -- (Ymax) node[right] {\(x_7\)};
      \draw[-latex,thin] (Zmin) -- (Zmax) node[above] {\(x_3\)};

      \newcommand*\planemin{-4}
      \newcommand*\planemax{4}

      \newcommand*\leftplane{-3}
      \newcommand*\midplane{0}
      \newcommand*\rightplane{3}
      
      \drawplane{\leftplane}{\planemax}{\planemin}
      \drawplane{\midplane}{\planemax}{\planemin}
      \drawplane{\rightplane}{\planemax}{\planemin}

      \draw[D5col,very thick] (\leftplane,\planemin,0) -- (\leftplane,\planemax,0) node[above] {\(n N_f\) \(\D5\)};

      \draw[D3col,very thick] (\leftplane,0,0) -- (\rightplane,0,0);
      \node at (-2,1,0) {\(n N_f\) \D3};
      \node at (1.5,1,0) {\(n N_c\) \D3};

      \draw[NScol,very thick] (\rightplane,0,\planemax) -- (\rightplane,0,\planemin) node[below] {\(n\) \NS{}'};
      \draw[Ukcol,very thick] (\midplane,\planemax,\planemax) -- (\midplane,\planemin,\planemin)  node[below] {\((1,k)\)};

    \end{scope}
  \node at (0,-5) {(a)};

  \end{tikzpicture}
} 

\newcommand*\FigureNiarchosmag{
  \tdplotsetmaincoords{30}{30}

  \begin{tikzpicture}[xscale=.8,yscale=.6]

    \begin{scope}[scale=.8,shift={(-6,0)},tdplot_main_coords]
      \coordinate (Xmin) at (-4,0,0);
      \coordinate (Xmax) at (4,0,0);
      \coordinate (Ymin) at (0,-6,0);
      \coordinate (Ymax) at (0,6,0);
      \coordinate (Zmin) at (0,0,-1);
      \coordinate (Zmax) at (0,0,6);

      \draw[-latex,thin] (Xmin) -- (Xmax) node[right] {\(x_6\)};
      \draw[-latex,thin] (Ymin) -- (Ymax) node[right] {\((x_8,x_9)\)};
      \draw[-latex,thin] (Zmin) -- (Zmax) node[above] {\((x_4,x_5)\)};

      \newcommand*\planemin{-4}
      \newcommand*\planemax{4}

      \newcommand*\leftplane{-3}
      \newcommand*\midplane{0}
      \newcommand*\rightplane{3}
      
      \drawplane{\leftplane}{\planemax}{\planemin}
      \drawplane{\midplane}{\planemax}{\planemin}
      \drawplane{\rightplane}{\planemax}{\planemin}

      \draw[D5col,very thick] (\leftplane,\planemax,-2) -- (\leftplane,\planemin,-2);
      \draw[D5col,very thick] (\leftplane,\planemax,0) -- (\leftplane,\planemin,0);
      \draw[D5col,very thick] (\leftplane,\planemax,2) -- (\leftplane,\planemin,2);
      \draw [D5col,decorate,decoration={brace,amplitude=5pt}] (-3.3,\planemin,-2) -- (-3.3,\planemin,2) node[D5col,midway,xshift=-0.6cm]  {\(n\) \D5};
      
      \draw[NScol,very thick] (\midplane,\planemax,-2) -- (\midplane,\planemin,-2);
      \draw[NScol,very thick] (\midplane,\planemax,0) -- (\midplane,\planemin,0);
      \draw[NScol,very thick] (\midplane,\planemax,2) -- (\midplane,\planemin,2);
      \draw [NScol,decorate,decoration={brace,amplitude=5pt}] (-0.3,\planemin,-2) -- (-0.3,\planemin,2) node[NScol,midway,xshift=-0.6cm]  {\(n\) \NS{}'};

      \draw[D3col,very thick] (\leftplane,0,-2) -- (\rightplane,0,-2);
      \draw[D3col,very thick] (\leftplane,0,0) -- (\rightplane,0,0);
      \draw[D3col,very thick] (\leftplane,0,2) -- (\rightplane,0,2);

      \draw[Ukcol,very thick] (\rightplane,0,\planemax) -- (\rightplane,0,\planemin)  node[below] {\((1,k)\)};

    \end{scope}

    \begin{scope}[scale=.8,shift={(5,0)},tdplot_main_coords]
      \coordinate (Xmin) at (-4,0,0);
      \coordinate (Xmax) at (4,0,0);
      \coordinate (Ymin) at (0,-6,0);
      \coordinate (Ymax) at (0,6,0);
      \coordinate (Zmin) at (0,0,-1);
      \coordinate (Zmax) at (0,0,6);

      \draw[-latex,thin] (Xmin) -- (Xmax) node[right] {\(x_6\)};
      \draw[-latex,thin] (Ymin) -- (Ymax) node[right] {\(x_7\)};
      \draw[-latex,thin] (Zmin) -- (Zmax) node[above] {\(x_3\)};

      \newcommand*\planemin{-4}
      \newcommand*\planemax{4}

      \newcommand*\leftplane{-3}
      \newcommand*\midplane{0}
      \newcommand*\rightplane{3}
      
      \drawplane{\leftplane}{\planemax}{\planemin}
      \drawplane{\midplane}{\planemax}{\planemin}
      \drawplane{\rightplane}{\planemax}{\planemin}

      \draw[D5col,very thick] (\leftplane,\planemin,0) -- (\leftplane,\planemax,0) node[above] {\(n N_f\) \(\D5\)};

      \draw[D3col,very thick] (\leftplane,0,0) -- (\rightplane,0,0);
      \node at (-2,1,0) {\(n N_f\) \D3};
      \node at (2.5,1.5,0) {\(n(N_f+|k|)-N_c\) \D3};

      \draw[NScol,very thick] (\midplane,0,\planemax) -- (\midplane,0,\planemin) node[below] {\(n\) \NS{}'};
      \draw[Ukcol,very thick] (\rightplane,\planemax,\planemax) -- (\rightplane,\planemin,\planemin)  node[below] {\((1,k)\)};

    \end{scope}
  \node at (0,-5) {(b)};

  \end{tikzpicture}
} 

\newcommand*\FigureMultipleNS{
\begin{tikzpicture}
  \begin{scope}[shift={(-2,-3.5)}]
    \begin{footnotesize}
      \draw[->] (0,3) -- (1,3);
      \draw[->] (.2,2.8) -- (.2,3.8);
      \draw (1,3.2) node[]{\(x_6\)};
      \draw (.2,4) node[]{\(x_{4,5}\)};
    \end{footnotesize}
  \end{scope}
  
  \begin{scope}[thick]
    \begin{footnotesize}
      \draw (0,3) -- (0,8);
      \draw [dashed] (0,0) -- (0,3);
      \draw (-0.5,0) node[]{\NS5'};
      
      \draw (0,7) -- (2.5,7);
      \draw (1.5,6.5) node[]{\( r_2\) \D4};
      \draw[fill=black] (2.5,7) node{} circle (5pt);
      \draw (2.5,7.5) node[]{\NS5};
      \draw[fill=white] (0,7) circle (5pt);
      \draw (0,7) node[cross=4pt]{};
      \draw (-0.75,7) node[]{\( N_f\) \D6};
      
      \draw (0,5) -- (2.5,5);
      \draw (1.5,4.5) node[]{\( r_1\) \D4};
      \draw[fill=black] (2.5,5) node{} circle (5pt);
      \draw (2.5,5.5) node[]{\NS5};
      \draw[fill=white] (0,5) circle (5pt);
      \draw (0,5) node[cross=4pt]{};
      \draw (-0.75,5) node[]{\( N_f\) \D6};
      
      \draw[] (-.5,3) -- (3,3);
      \draw (1.5,2.5) node[]{\(2 r_0\) \D4};
      \draw[fill=black] (2.5,3) node{} circle (5pt);
      \draw (2.5,3.5) node[]{\NS5};
      \draw[fill=white] (0,3) circle (5pt);
      \draw (0,3) node[cross=4pt]{};
      \draw (-0.75,3) node[]{\(2 N_f\) \D6};

      \draw[dashed] (0,1) -- (2.5,1);
      \draw[fill=black] (2.5,1) node{} circle (5pt);
      \draw (2.5,1.5) node[]{\NS5};
      \draw[fill=white] (0,1) circle (5pt);
      \draw (0,1) node[cross=4pt]{};
      \draw (-0.75,1) node[]{\( N_f\) \D6};

      \node at (1,-1) {(a)};
    \end{footnotesize}
  \end{scope}

  \begin{scope}[shift={(7,0)},thick]
    \begin{footnotesize}

      \draw[NScol] (2.5,3) -- (2.5,8);
      \draw [NScol,dashed] (2.5,0) -- (2.5,3);
      
      \draw[D4col] (0,7) -- (2.5,7);
      \draw[D4col] (1.2,6.5) node[]{\((N_f - r_2)\) \D4};
      \draw[NScol,fill=NScol] (0,7) node{} circle (5pt);
      \draw[D6col] (1.9,7.5) node[]{\(N_f\) \D6};
      \draw[D6col,fill=white] (2.5,7) circle (5pt);
      \draw[D6col] (2.5,7) node[D6col,cross=4pt]{};
      \draw[NScol] (0,7.5) node[]{\NS5};
      
      \draw[D4col] (0,5) -- (2.5,5);
      \draw[D4col] (1.2,4.5) node[]{\((N_f - r_1)\) \D4};
      \draw[fill=NScol] (0,5) node{} circle (5pt);
      \draw[D6col] (1.9,5.5) node[]{\(N_f\) \D6};
      \draw[D6col,fill=white] (2.5,5) circle (5pt);
      \draw[D6col] (2.5,5) node[D6col,cross=4pt]{};
      \draw[NScol] (0,5.5) node[]{\NS5};

      \draw[D4col] (-.5,3) -- (3,3);

      \draw[D4col] (1.2,2.5) node[]{\((N_f - r_0)\) \D4};
      \draw[NScol,fill=NScol] (0,3) node{} circle (5pt);
      \draw[D6col] (1.9,3.5) node[]{\(N_f\) \D6};
      \draw[D6col,fill=white] (2.5,3) circle (5pt);
      \draw (2.5,3) node[fill=D6col,cross=4pt]{};
      \draw[NScol] (0,3.5) node[]{\NS5};
      
      \draw[NScol] [dashed] (0,1) -- (2.5,1);
      \draw[NScol,fill=NScol] (0,1) node{} circle (5pt);
      \draw[D6col] (1.9,1.5) node[]{\(N_f\) \D6};
      \draw[D6col,fill=white] (2.5,1) circle (5pt);
      \draw [D6col] (2.5,1) node[D6col,cross=4pt]{};
      \draw[NScol] (0,1.5) node[]{\NS5};

      \node at (1,-1) {(b)};
            
    \end{footnotesize}
  \end{scope}
  
\end{tikzpicture}

}

\titleformat{\chapter}
  {\normalfont\huge\bfseries}{\thechapter}{20pt}{\Huge\raggedright}
\titlespacing*{\chapter} {0pt}{3.5ex plus 1ex minus .2ex}{2.3ex plus .2ex}

\newcommand{\OfficialTitle}{
  String theory and the 4D/3D reduction of Seiberg duality. A~Review
}

\title{%
  {\color{Thoughtless}\Huge\textbf{\dosserif\OfficialTitle}}
}

\hypersetup{pdfauthor={Antonio Amariti and Domenico Orlando and Susanne Reffert},pdftitle={\OfficialTitle}}

\author{%
  \begin{minipage}{.8\linewidth}
    \vspace{1cm}
    \begin{center} \dosserif
      {\small 
      \textbf{Antonio~Amariti}, 
         \textbf{Domenico~Orlando} and 
        \textbf{Susanne~Reffert}}
    \end{center}
    \vspace{1cm}
    \authorBlock{}{Albert Einstein Center for Fundamental Physics\\
      Institute for Theoretical Physics\\
      University of Bern,\\
      Sidlerstrasse 5, \textsc{ch}-3012 Bern, Switzerland}
  \end{minipage}
}

\date{} 

\begin{document}

\numberwithin{equation}{section}

\begin{titlepage}

  \newgeometry{top=23.1mm,bottom=46.1mm,left=34.6mm,right=34.6mm}

  \setstretch{1.25}

  \maketitle
  \setstretch{1.15}
  \thispagestyle{empty}

  \vfill\dosserif

 \abstract{\normalfont \noindent
We review the reduction of four-dimensional \(\mathcal{N}=1\) Seiberg duality
to three dimensions focusing on the \D~brane engineering approach.
We start with an overview of four-dimensional Seiberg duality for theories with
various types of gauge groups and matter content both from a field-theoretic and a brane engineering point of view.
Then we describe two families of \(\mathcal{N}=2\) three-dimensional dualities, 
namely Giveon--Kutasov-like and Aharony-like dualities.
The last part of our discussion 
is  devoted to the 4D/3D reduction of the dualities studied above. We discuss both the analysis
at finite radius,  crucial for preserving the duality in the dimensional 
reduction, and the zero-size limit that must be supported by a
real mass flow and a Higgsing, which can differ case by case. 
We show that this mechanism is reproduced in the brane description by T-duality,
supplying a unified picture for all the different cases.
As a bonus we show that this analysis provides a brane description for Aharony-like dualities.
}

\vfill

\end{titlepage}

\restoregeometry

\tableofcontents

\def\O#1{\ensuremath{\mathrm{O}#1}}

\chapter{Introduction}
\label{sec:intro}

Dualities have played an extremely important role in theoretical
physics in the last decades. In this review, we bring together
four-dimensional Seiberg-like dualities, three-dimensional dualities
and the reduction from four to three dimensions, with a
string-theoretic outlook on the problem, namely brane constructions.
These brane constructions give rise to a clear picture that unifies
all the reductions of four-dimensional dualities to dualities in three dimensions.%

\medskip A \emph{duality} is an exact equivalence of two seemingly different
physical systems. Especially so-called strong/weak dualities are an
important tool for accessing non-perturbative regimes of quantum field
theories. In general, we can approach a physical problem via perturbation theory, where we have a Hamiltonian
\begin{equation}
H=H_0+g H_1,
\end{equation}
where $H_0$ is solvable. Observables are given by a perturbation series in the parameter $g$, which is valid for small values of $g$. For large $g$, the perturbative approach loses its applicability, and often we are at a complete loss of how to study the system in the strongly coupled regime. Sometimes, however, the system allows two different descriptions,
\begin{equation}
  \begin{aligned}
    H &= H_0+g H_1,\\
    &= H_0'+g'H_1',
  \end{aligned}
\end{equation}
leading to two different perturbative expansions.\footnote{For an elementary introduction to the concept of duality, see~\cite{Polchinski:2014mva}.} We speak of a strong/weak duality, when the coupling constants $g$, $g'$ are related via
\begin{equation}
g' = 1/g.
\end{equation}
When $g$ becomes large, the perturbation series in $g'$ becomes an accurate description of the system, allowing important insights into the theory at a regime which was inaccessible in the first description.

Maybe the first prototype of such a duality in four dimensions is electric/mag\-netic duality, \emph{i.e.} the invariance of source-free Maxwell theory under the mapping
\begin{align}
  \vec E &\to \vec B, & \vec B &\to -\vec E,
\end{align}
which in the quantum theory includes the transformation of the coupling 
\begin{equation}
e \to e'=\frac{2\pi}{e}.
\end{equation}
This duality becomes more interesting when sources are present, \emph{i.e.} for non-Abelian gauge theories which include monopole solutions. While the electric charges become  more strongly coupled and less point-like for larger $e$, the mono\-poles appear to become lighter and smaller, suggesting that they could be quantized as fundamental fields for very large $e$. These observations led to the Montonen--Olive conjecture~\cite{Montonen:1977sn} stating that non-Abelian gauge theories with monopoles are invariant under a strong/weak electric/magnetic duality. It turned however out that without adding supersymmetry, even gathering only circumstantial evidence was impossible. Witten and Olive~\cite{Witten:1978mh} showed that for $\mathcal{N}=2$ \ac{sym} the calculation of the monopole masses is exact. In $\mathcal{N}=4$ \ac{sym}, also the spins of the electric charges and the monopoles can be shown to match~\cite{Osborn:1979tq}, and the mapping of the coupling constant gets extended to the full S--duality group $SL(2,\mathbb{Z})$. 

\medskip
In 1994, Seiberg~\cite{Seiberg:1994pq} proposed a duality for $\mathcal{N}=1$ \ac{sqcd}, which unlike the versions with higher supersymmetry which are exact for all values of the coupling constant, is a \ac{ir} duality. In this review, we will concentrate on Seiberg-duality and its many generalizations and extensions. These four-dimensional dualities have sparked interest also in their three-dimensional analogs, starting with the work of Aharony~\cite{Aharony:1997gp} in 1997. The question of how the three-dimensional dualities are related to their four-dimensional cousins is obvious. As was discussed at length in~\cite{Aharony:2013dha}, a naive dimensional reduction does not yield the correct results.

\medskip
A large amount of literature and good reviews exist about dualities of four-dimensional supersymmetric gauge theories (see \emph{e.g.}~\cite{Intriligator:1995au,Giveon:1998sr,Argyres:2001eva,Terning:2003th,Strassler:2005qs}). Also, the literature on three-dimensional dualities is extensive 
\cite{Aharony:1997bx,Karch:1997ux,Aharony:1997gp,Jafferis:2008em,Aharony:2008gk,Giveon:2008zn,Niarchos:2008jb,Jafferis:2011ns,Kapustin:2011gh,Hwang:2011ht,Aharony:2011ci,Kapustin:2011vz, Kim:2013cma,Park:2013wta,Aharony:2014uya}, even though no real reviews exist on the subject. 
The aim of this review is to bring together the field theoretic material of the above topics and to add as a new ingredient the string-theoretic point of view. We will discuss the brane constructions not only for the relevant gauge theories, but also their reduction.
This approach has allowed us to obtain a brane picture for Aharony duality, which had been missing in the literature.
The reason for this lack is intrinsic in the brane engineering of a field theory.
Via this picture, it is quite easy to capture the classical dynamics,
for example the global symmetries and the classical moduli space.
Extracting quantum information is however not obvious. To
understand which of the global symmetries become anomalous at the quantum level for example requires
more information, such as the uplift to M-theory.
The key role in capturing the quantum aspects of the reduction of four-dimensional dualities and 
the brane picture of Aharony duality is played by T-duality in string theory.
T--duality provides the missing quantum information, necessary to reconstruct the three-dimensional physics
in this picture.

\medskip

The plan of the review is as follows. 
\begin{itemize}
\item In Section~\ref{sec:4d} we review some aspects of
  four-dimensional $\mathcal{N}=1$ gauge theories: the algebra, the
  matter content and the interactions (Sec.~\ref{sec:4dn=1}). We also
  discuss some aspects of the analysis of the moduli space, mass
  deformations and anomalies. The discussion is focused on the case of
  \ac{sqcd} (Sec.~\ref{sec:4dsqcd}). After a brief introduction, we
  introduce some basic aspects of Seiberg duality for \ac{sqcd} and
  some generalizations, restricting our analysis to theories with a
  single gauge group. We consider theories with unitary or real gauge
  groups, fundamental and tensor matter (Sec.~\ref{sec:4dseiberg}). We
  review the main aspects of these dualities, \emph{i.e.} the dual
  gauge groups, the matter content of the dual theory and the
  superpotential. Finally, we discuss the engineering of these systems
  in \tIIA string theory by introducing the realization in terms of
  \NS{} branes, \D{} branes and orientifold \O{}-planes. We discuss
  the relation between Seiberg duality and the \ac{hw} transition,
  showing how to reproduce the field theory results in the brane
  language (Sec.~\ref{sec:4dbranes}).
\item In Section~\ref{sec:3d}, we discuss some known fact about three-dimensional $\mathcal{N}=2$ gauge theories with a single gauge
  group. We start by reviewing the main aspects of these theories,
  \emph{i.e.} their field content, their interactions and the moduli
  space. In particular, we stress the role of the monopole operators
  and their relation to the Coulomb branch. We also discuss the
  \ac{cs} interactions, and their relation with the real masses that
  exist in three dimensions (Sec.~\ref{sec:3dsusy} and
  Sec.~\ref{sec:3dsqcd}). After this review, we start discussing two
  classes of dualities: the one found by Aharony
  in~\cite{Aharony:1997gp} and the one found by Giveon and Kutasov
  in~\cite{Giveon:2008zn}. Our focus is on the similarities and the
  differences between these three-dimensional dualities and
  four-dimensional Seiberg duality
  (Sec.~\ref{sec:3ddualities}).
  We also discuss the generalization of these dualities to real gauge
  groups and tensor matter and we conclude the section by reviewing
  the brane engineering of the three-dimensional dualities
  (Sec.~\ref{sec:3dbranes}). We discuss in detail the dualities of
  \ac{gk} type and we introduce some brane engineerings of more
  involved configurations that (to our knowledge) have not yet
  appeared in the literature. We conclude commenting on the absence of
  a brane engineering for Aharony duality and for its generalizations.
\item In Section~\ref{sec:redux}, we discuss the dimensional reduction
  of four-dimensional Seiberg duality to three dimensions along the
  lines of~\cite{Aharony:2013dha}. We review the main aspects of this
  reduction, discussing the problems of a naive reduction on $S^1$ and how
  the standard prescription has to be modified by considering
  effective dualities at finite radius. We also show that these are related
  to the more conventional dualities of Aharony and of Giveon--Kutasov
  via a real mass flow. The cases with unitary and real gauge groups
  require a very different treatment of the decoupling limit on the
  field theory side and apparently an \emph{ad hoc} procedure has to
  be adopted (Sec.~\ref{sec:redfields}). In the second part of this
  section, we show that these reductions can be understood via brane
  engineering (Sec.~\ref{sec:redbranes}) %
  and a 
  unified picture emerges. All the different cases are obtained by considering a double-scaling limit on the real
  masses and on the radius of the compactification circle.
    By defining the circle along the spatial direction $x_3$ and considering the
  three-dimensional gauge theory living at $x_3=0$,
  a very important role is played by the \emph{mirror} point on the circle,
  $x_3^\circ \equiv R \pi$, where $R$ is the radius of $S^1$.
  The double-scaling  limit can be performed in the same way in each case
  because the physics at that point can be described in a unified way
  in terms of branes.
\item In Section~\ref{sec:loc}, we discuss checks of the dualities via localization. We review the derivation of the three-dimensional partition function on the squashed sphere via the circle reduction of the four-dimensional superconformal index  (Sec.~\ref{sec:3dindex}); we apply this technique to the reduction of dualities (Sec.~\ref{sec:relationloc}) and  we show as an application the reduction of \ac{kss} duality.
\end{itemize}

We hope to close a gap in the literature with this review by collecting not only 
the various generalizations of Seiberg duality in four dimensions, but by finally providing 
an overview over the Seiberg-like dualities in three dimensions and their relation to the four-dimensional case.
Last but not least, we believe that the string theory point of view succeeds in unifying the understanding
of these various dualities.

\chapter{The four-dimensional case}\label{sec:4d}

In this section, we discuss the four-dimensional case. We first recall the general properties, multiplets and invariant Lagrangians for $\mathcal{N}=1$ gauge theories. Then we specialize to \ac{sqcd} with gauge group $SU(N_c)$, our main example. Then, we discuss the dualities, starting from Seiberg duality for \ac{sqcd} and discussing its extensions. Finally, we move to the string theory point of view and recreate the dualities via D~brane constructions.

\section{$\mathcal{N}=1$ supersymmetric gauge theories in four dimensions}\label{sec:4dn=1}

In the following, we will summarize the basics of supersymmetric gauge theories in four dimensions, following in presentation and notation Ch.~9 of~\cite{Dine:2007book}. Then we will review the specific case of \ac{sqcd}, which is the most basic example of Seiberg duality.

\subsection{Generalities}

The $\mathcal{N}=1$ supersymmetry algebra is given by
\begin{equation}
\acomm{ Q_\alpha}{ \overline Q_{\dot\beta}}=2\, \sigma_{\alpha\dot\beta}^\mu P_\mu,
\end{equation}
where the supersymmetry generators $Q,\,\overline Q$ are two-component spinors with spinor indices $\alpha,\,\dot{\beta}$, $\sigma^\mu=(1,\vec{\sigma})$ are the Pauli matrices, and $P_\mu$ is the four-momentum. 

The smallest irreducible representations which can describe massless fields include the chiral multiplet which contains as propagating degrees of freedom a complex scalar $\phi$ and a chiral fermion $\psi_\alpha$, and the vector multiplet containing a chiral fermion $\lambda_\alpha$ and a vector field $A_\mu$ which are both in the adjoint representation of the gauge group $G$.

We will be using the \emph{superspace formulation}, which extends spacetime with coordinates $x^\mu$ by two anti-commuting Grassmann variables $\theta_\alpha,\ \bar\theta_{\dot\beta}$. The supersymmetry generators $Q,\,\overline Q$ act on functions of the superspace variables $f(x^\mu,\theta, \bar\theta)$ as differential operators:
\begin{align}\label{eq:Q}
  Q_\alpha &=\frac{\del}{\del\theta_\alpha}-i\sigma_{\alpha\dot\alpha}^\mu \theta^{*\dot{\alpha}}\del_\mu, &  \overline Q_{\dot\alpha} &= -\frac{\del}{\del\bar\theta_{\dot\alpha}} + i\theta^{*{\alpha}} \sigma_{\alpha\dot\alpha}^\mu \del_\mu.
\end{align}
The covariant derivatives are defined by
\begin{align}\label{eq:D}
  D_\alpha &=\frac{\del}{\del\theta_\alpha} + i\sigma_{\alpha\dot\alpha}^\mu \theta^{*\dot{\alpha}}\del_\mu, &  \overline D_{\dot\alpha} &= -\frac{\del}{\del\bar\theta_{\dot\alpha}} - i\theta^{*{\alpha}} \sigma_{\alpha\dot\alpha}^\mu \del_\mu.
\end{align}
They anti-commute with the supersymmetry generators and are thus useful for constructing invariant expressions. 
The condition
\begin{equation}
\overline D_{\dot\alpha}\Phi = 0
\end{equation}
on a superfield $\Phi$ is invariant under supersymmetry transformations. Fields satisfying this condition are called \emph{chiral superfields}.
A chiral superfield $\Phi$ has the expansion in $\theta_\alpha,\ \bar\theta_{\dot\beta}$
\begin{equation}\label{eq:chiral4}
\Phi = \phi(x) + i\theta \sigma^\mu \bar\theta \del_\mu \phi + \tfrac{1}{4}\theta^2\bar\theta^2\del^2\phi+\sqrt{2} \theta\psi -\tfrac{i}{\sqrt 2}\theta\theta\del_\mu \sigma^\mu\bar\theta+\theta^2 F,
\end{equation}
where as mentioned before,  $\phi$ is a complex scalar, $\psi_\alpha$ a chiral fermion, and $F$ is a scalar field, which as we will see in the following, does not propagate and is thus called \emph{auxiliary}. The spinor indices are all fully contracted and thus suppressed.

As mentioned, \emph{vector fields} form another irreducible representation of the algebra, they satisfy
\begin{equation}
V= V^\dagger.
\end{equation}
Also this condition is preserved under supersymmetry transformations. The expansion of a vector superfield $V$ in a power series in $\theta$ is given by
\begin{equation}\label{eq:vec}
V = i\chi - i\chi^\dagger - \theta\sigma^\mu\theta^*A_\mu+i\theta^2\bar\theta\bar\lambda- i\bar\theta^2\theta\lambda+\tfrac{1}{2}\theta^2 \bar\theta^2 D,
\end{equation}
where $D$ is a real scalar field, which like $F$ is auxiliary as it does not propagate.
The vector field $A_\mu$ hints at an underlying gauge symmetry: let us take $\Phi$ to transform as
\begin{equation}
\Phi \to e^{-i\Lambda} \Phi
\end{equation}
under gauge transformations, where $\Lambda$ is a matrix-valued chiral superfield.
In order for the combination
\begin{equation}
\Phi^\dagger e^V \Phi,\quad V \ \text{a matrix valued superfield}
\end{equation}
to be invariant,
$V$ must transform as
\begin{equation}
e^V \to e^{-i\Lambda^*} e^V e^{i\Lambda}
\end{equation}
under gauge transformations.
We can use the gauge freedom to set the $\chi$ component of $V$ to zero, this is known as the Wess--Zumino gauge.
The gauge-covariant field strength is given by the expression
\begin{equation}\label{eq:gfs}
W_\alpha = - \tfrac{1}{4} \overline D^2 e^{-V} D_\alpha e^V.
\end{equation}
It transforms under gauge transformations like a chiral field in the adjoint representation, 
\begin{equation}
W_\alpha \to e^{-i\Lambda^*} W_\alpha e^{i\Lambda}.
\end{equation}

\paragraph{Supersymmetric Lagrangians.} As we have now collected all the ingredients, we are ready to construct manifestly supersymmetric and gauge invariant Lagrangians, using the fact that $\theta$ has dimension $-1/2$, while chiral superfields have dimension $1$ and vector superfields dimension $0$.

The most general renormalizable Lagrangian consists of the following pieces:
\begin{equation}
  \mathcal{L}_{\text{kin}} = \int \dd[4]{\theta} \sum_i \Phi_i^\dagger e^V \Phi_i,
\end{equation}
where the sum runs over all the matter multiplets and $V$ is in a representation that is appropriate for the field $\Phi_i$. 
\begin{equation}
\mathcal{L}_{\text{W}} = \int \dd[2]{\theta} W(\Phi_i) + \text{c.c},
\end{equation}
where $W(\Phi_i)$ is a holomorphic function of the $\Phi_i$, the so-called \emph{superpotential}. For the theory to be renormalizable\footnote{In the following we will deal with non-renormalizable superpotentials as well. They will be related to dangerously irrelevant 
interactions that will lead the flow to strongly coupled fixed points in the~\ac{ir}.},
the superpotential must take the form
\begin{equation}\label{eq:renW}
W = \tfrac{1}{2} m_{ij} \Phi_i \Phi_j + \tfrac{1}{3} \lambda_{ijk} \Phi_i\Phi_j\Phi_k.
\end{equation}
The gauge part of the Lagrangian is given by
\begin{equation}
\mathcal{L}_{\text{gauge}} = \frac{1}{g^{(i)2}} \int \dd[2]{\theta} W_\alpha^{(i)}W^{\alpha(i)}.
\end{equation}
The full Lagrangian density is given by
\begin{equation}
\mathcal{L} = \mathcal{L}_{\text{kin}} + \mathcal{L}_{\text{W}} + \mathcal{L}_{\text{gauge}}.
\end{equation}
In case the gauge group $G$ has $U(1)$ factors, we can include a supersymmetric and gauge invariant \emph{Fayet--Iliopoulos term} for each of them:
\begin{equation}\label{eq:FI}
\xi \int \dd[4]{\theta}  V,
\end{equation}
where $\xi$ is a real Fayet--Iliopoulos parameter. %

\medskip
Thanks to the superspace formalism, the above Lagrangian is very compact, but in order to make the physical meaning of these terms more transparent, we express them in terms of the component fields:
\begin{equation}
\mathcal{L}_{\text{kin}} = | \del_\mu \phi_i |^2 + i \psi_i \del_\mu \sigma^\mu \psi^*_i + F^*_iF_i.
\end{equation}
We see now that the $F_i$ appear without derivatives, which is why they are called auxiliary fields.
\begin{equation}
  \begin{aligned}
    \mathcal{L}_{\text{W}} &= \frac{\del W}{\del \Phi_i} F_i +\frac{\del^2 W}{\del \Phi_i \del \Phi_j} \psi_i \psi_j \\
    &= F_i \left( m_{ij} \Phi_j + \lambda_{ijk}\Phi_j\Phi_k \right) +
    \left( m_{ij}+ \lambda_{ijk}\Phi_k \right)\psi_i \psi_j +
    \text{c.c},
  \end{aligned}
\end{equation}
where the second equality holds for the choice of superpotential Eq.~\eqref{eq:renW}.
The full Lagrangian in Wess--Zumino gauge in terms of component fields is given by
\begin{equation}
  \begin{aligned}
    \mathcal{L} ={}&- \frac{1}{4 g_a^2} {F^a_{\mu\nu}}^2 - i \lambda^a \sigma^\mu D_\mu \lambda^{a*} + | D_\mu \phi_i |^2 -  i \psi_i^a \sigma^\mu D_\mu \psi^*_i   \\
    &+  \frac{1}{2 g_a^2} (D^a)^2+D^a\sum_i \phi_i^* T^a \phi_i +F^*_iF_i-F_i \frac{\del W}{\del \phi_i}+ \text{c.c}\\
    &+\sum_{i,j} \frac{1}{2} \frac{\del^2 W}{\del \phi_i \del
      \phi_j}\psi_i \psi_j + i\sqrt 2 \sum_{i,a} \lambda^a \psi_i
    T^a\phi_i^* + \text{c.c},
  \end{aligned}
\end{equation}
where $a$ is the gauge index and $T^a$ are the generators of the Lie algebra associated to $G$.

We can solve for the auxiliary fields $F_i$ and $D$ via their equations of motion and find
\begin{equation}
F_i=\frac{\del W}{\del \phi_i^*}, \quad D^a = \sum_i (g^a\phi^*_iT^a\phi_i).
\end{equation}
Substituted back into the Lagrangian, we find the scalar potential
\begin{align}\label{eq:scalpot}
V= & |F_i|^2 + \frac{1}{2g^{a2}}(D^a)^2.
\end{align}
Allowing also non-renormalizable terms, the most general, globally  $\mathcal{N}=1$ supersymmetric theory with at most two derivatives has the form
 \begin{equation}
\mathcal{L} = \int \dd[4]{\theta} K(\Phi_i, \Phi^\dagger_i) + \int \dd[2]{\theta} W(\Phi_i) + \text{c.c.} + \int \dd[2]{\theta} f_a(\Phi_i) ( W_\alpha^{(a)})^2 + \text{c.c.},
\end{equation}
where the \emph{Kähler potential} $K(\Phi_i, \Phi^\dagger_i)$ gives rise to the kinetic terms, and the superpotential and the \emph{gauge kinetic function} $ f_a(\Phi_i)$ are functions of the chiral fields only.
In $\mathcal{N}=1$ supersymmetric theories, non-renormalization theorems state that the superpotential is not corrected in perturbation theory beyond its tree-level value and that $f_a$ is at most renormalized at one loop.

\paragraph{R--symmetry.} 
Supersymmetric theories have global symmetries which rotate the
supercharges, so-called \emph{$R$--symmetries}. For $\mathcal{N}=1$
supersymmetric theories, the R--symmetry group is isomorphic to a
global $U(1)_R$, or a discrete subgroup thereof. The defining feature
of a continuous R--symmetry is that $\theta$, $\bar\theta$ transform as
\begin{align}
  \theta &\to e^{i\alpha}\theta, & \bar\theta &\to e^{-i\alpha}\bar\theta,
\end{align}
with $\alpha$ the transformation parameter, meaning that they have R--charges $1, -1$.
The supersymmetry generators transform thus as 
\begin{align}
  Q &\to e^{-i\alpha} Q, & \overline Q &\to e^{i\alpha}\overline Q,
\end{align}
\emph{i.e.} they have R--charges $-1,+1$ and do not commute with the R--symmetry generator:
\begin{align}
  \comm{R}{Q} &= -Q, & \comm{R}{\overline Q} &= \overline Q.
\end{align}
The different component fields within a superfield carry different R--charges. If the theory is invariant under R--symmetry, each superfield $S$ can be assigned an R--charge $r_S$ defined by
\begin{equation}
S(x^\mu, \theta, \bar\theta) \to e^{ir_S\alpha} S(x^\mu, e^{-i\alpha}\theta, e^{i\alpha}\bar\theta).
\end{equation}
The R--charge of a product of superfields is the sum of the individual charges of the fields. 
In Table~\ref{tab:R}, the R--charges of the various quantities are collected.

\begin{table}[]
  \centering
  \begin{tabular}{cccccccccccccc}
    \toprule
    & $\theta_\alpha$ & $\bar\theta_{\dot\alpha}$  & $Q_\alpha$ &  $\dd[2]{\theta}$ & $D_\alpha$ & $W_\alpha$ & $A^\mu$ & $\lambda_\alpha$ & $D$ & $W$ & $\phi$ & $\psi_\alpha$ & $F$\\ \midrule
$r$ & $1$ & $-1$ &$-1$ & $-2$ & $-1$ & $1$ & $0$ & $1$ & $0$ & $2$ & $r_\Phi$ & $r_\Phi-1$ & $r_\Phi-2$ \\ \bottomrule
  \end{tabular}
  \caption{$U(1)_R$ R--charges of the different quantities appearing in $\mathcal{N}=1$ supersymmetric gauge theories.}
  \label{tab:R}
\end{table}

All Lagrangian terms involving gauge superfields are automatically R--sym\-met\-ric, including the couplings to the chiral fields. The superpotential, however, must carry R--charge $+2$ in order to conserve R--symmetry which is not automatic and often not the case.

\paragraph{Moduli space.}
Supersymmetric theories in general do not have isolated vacua but exhibit a continuous family of connected vacua. The \emph{moduli space} of inequivalent vacua is the set of all zero-energy field configurations modulo gauge transformations. The vacuum degeneracy of the classical moduli space $\mathcal{M}_{cl}$ is however not protected by any symmetries and can be lifted in the quantum theory by a dynamically generated effective superpotential.\footnote{Such a superpotential is generated non-perturbatively and does not violate the non-renormalization theorems.}
For an $\mathcal{N}=1$ supersymmetric gauge theory with zero tree-level superpotential, the classical squark potential is given by
\begin{equation}\label{eq:squarkpot}
V_{\text{squark}} = i\sqrt 2 \sum_{i,a} \lambda^a \psi_i T^a\phi_i^* + \text{c.c} + \frac{1}{2}\sum_a \left( \sum_i \phi_i^*T^a\phi_i\right)^2,
\end{equation}
where the last term comes from the $D$--term in the scalar potential given in Eq.~\eqref{eq:scalpot}. 
The classical moduli space $\mathcal{M}_{cl}$ is the space of squark expectation values $\langle \phi_i\rangle$ modulo gauge equivalence, along which the potential Eq.~\eqref{eq:squarkpot} vanishes. $\mathcal{M}_{cl}$ can be given a gauge-invariant description in terms of expectation values of gauge-invariant combinations of the fields (such as \emph{e.g.} mesons), subject to classical constraints. A more detailed description of the moduli space can be found \emph{e.g.} in~\cite{Intriligator:1995au}.

\subsection{SQCD with $G=SU(N_c)$}\label{sec:4dsqcd}

In this section, we will study \acl{sqcd} with gauge group $U(N_c)$, where $N_c$ stands for the number of colors.\footnote{For more details, see~\cite{Seiberg:1994pq,Intriligator:1995au}.} The gauge field is encoded in the $\mathcal{N}=1$ vector multiplet 
\begin{equation}
V=(\lambda^a_\alpha, A^a_\mu, D^a)
\end{equation} 
with gauge index $a$ as discussed in the last section. The matter content consists of $N_f$ flavors of quarks, encoded in the chiral multiplets 
\begin{equation}
Q^i_a=(\phi_{Q_a}^i, \psi^i_{\alpha,Q_a}, F^i_{Q_a}),\quad  i=1,\dots,N_f,
\end{equation}
 and $N_f$ flavors of anti-quarks, encoded in the chiral multiplets 
\begin{equation}
\widetilde Q_{\tilde i}^a=(\tilde\phi^{\widetilde Q_a}_{\tilde i}, \tilde\psi_{\alpha,\tilde i}^{\widetilde Q_a}, \tilde F_{\tilde i}^{\widetilde Q_a}),\quad \tilde i=1,\dots,N_f.
\end{equation}
This means that in total, there are $2N_cN_f$ chiral degrees of freedom.
The quarks transform in the fundamental representation $\mathbf{N}_c$ of the gauge group, while the anti-quarks transform in the anti-fundamental 
$\overline{\mathbf{N}}_c$.
There is no perturbative superpotential, $W=0$.
We will see in the following that the qualitative features of \ac{sqcd} change very drastically for different relative values of $N_f$ and $N_c$.

\paragraph{Global symmetries.}
Classically, this theory has the following global symmetries:
\begin{itemize}
\item the flavor symmetry $SU(N_f)_L \times SU(N_f)_R$
\item the baryonic symmetry $U(1)_B$
\item the axial symmetry $U(1)_A$
\item the R--symmetry $U(1)_R$.
\end{itemize}
At the quantum level, only the $U(1)_A$ is anomalous.
In Table~\ref{tab:global}, we have collected  the transformation properties of the quark multiplets under the global symmetries.\footnote{At this point, the assignment of $r_{Q_a}$ is unmotivated.}
\begin{table}[]
  \centering
  \begin{tabular}{cccccc}
    \toprule
                                & $SU(N_f)_L$    & $SU(N_f)_R$               & $U(1)_B$ & $U(1)_A$ & $U(1)_R$              \\ \midrule
    $Q^i_a$                     & $\mathbf{N}_f$ & $\mathbf{1}$              & $1$      & $1$      & $\frac{N_f-N_c}{N_f}$ \\
    $\widetilde Q_{\tilde i}^a$ & $\mathbf{1}$   & $\overline{\mathbf{N}}_f$ & $-1$     & $1$      & $\frac{N_f-N_c}{N_f}$ \\ \bottomrule
  \end{tabular}
  \caption{Transformation properties of the quarks under the global symmetries of \ac{sqcd}.}
  \label{tab:global}
\end{table}

\paragraph{Moduli space.} For $N_f<N_c$, the classical moduli space is parameterized by gauge-invariant mesons
\begin{equation}
M^i_{\tilde j} = Q^i_a\widetilde Q_{\tilde j}^a,
\end{equation}
which transform in the bifundamental representation $\mathbf{N}_f \times \overline{\mathbf{N}}_f$ of $SU(N_f)_L \times SU(N_f)_R$. The gauge group $SU(N_c)$ is partially broken to $SU(N_c-N_f)$. It turns out, however, that $\mathcal{M}_{cl}$  is lifted completely in this range of $N_f$ by a dynamically generated superpotential, so the quantum theory has no supersymmetric ground state.

The classical moduli space for $N_f \geq N_c$ is parameterized in a gauge-invariant way by
\begin{align}
M^i_{\tilde j} &= Q^i_a\widetilde Q_{\tilde j}^a,\\
B^{i_1\dots i_{N_c}} &= Q^{i_1}_{a_1}\dots Q^{i_{N_c}}_{a_{N_c}}\epsilon^{a_1\dots a_{N_c}},\\
\widetilde B_{i_1\dots i_{N_c}} &=\widetilde Q_{\tilde i_1}^{a_1}\dots \widetilde Q_{\tilde i_{N_c}}^{a_{N_c}}\epsilon_{a_1\dots a_{N_c}},
\end{align}
where $B$ is a baryon and $\widetilde B$ an anti-baryon. These basic generators are however not independent, but subject to classical constraints. At a generic point of $\mathcal{M}_{cl}$, the gauge group is completely broken ("Higgsed").
In the range $N_f \geq N_c$, the vacuum degeneracy cannot be lifted and $\mathcal{M}_{q}$ exists.  While $N_f=N_c$, $N_f=N_c+1$ are special cases (see \emph{e.g.} the discussion in~\cite{Intriligator:1995au}), for $N_f \geq N_c+2$, the quantum and classical moduli spaces are identical, $\mathcal{M}_{q}=\mathcal{M}_{cl}$.

\paragraph{Phases of SQCD.} \ac{sqcd} behaves very differently for different ranges of $N_f$. We can see this by studying the exact $\beta$--function of \ac{sqcd}~\cite{Novikov:1983uc}:\footnote{This applies at the origin of moduli space.}
\begin{align}
\beta(g) &= -\frac{g^3}{16\pi^2}\frac{3N_c-N_f(1-\gamma_0)}{1-\frac{g^2}{8\pi^2}N_c},\label{eq:beta}\\
\gamma_0(g) &= -\frac{g^2}{8\pi^2}\frac{N_c^2-1}{N_c}+\mathcal{O}(g^4),
\end{align}
where $\gamma_0$ is the \emph{anomalous dimension} of the mass. We see clearly that $N_f=3N_c$ is a special point. In fact, for $N_f<3N_c$, the beta-function is negative. If was shown that for $N_c, N_f\gg 1$ and $3N_c-N_f\ll N_c$, a conformal fixed point exists. This fixed point 
\begin{equation}
g=g_*\approx \sqrt{\frac{(3N_c-N_f)}{3c}}\frac{1}{N_c}
\end{equation}
 is a \emph{stable infrared fixed point} ($c$ is a positive number of order one). We see that $g_*$ increases for diminishing $N_f$. 
 Via arguments involving the operator dimensions, it can be shown that the description of the theory must change completely for $N_f < 3/2 N_c$. It has been argued that such a conformal fixed point exists in the whole range $3/2 N_c < N_f < 3 N_c$.
  In this range, the symmetry is enhanced to the superconformal algebra and we speak of the \emph{conformal window} of \ac{sqcd}. In the conformal window, the infrared theory is a non-trivial interacting \ac{scft}. The elementary quarks and gluons are not confined but appear as interacting massless particles. The potential between external electric sources at distance $R$ behaves like $V(R)  \sim 1/R$, so we are in a \emph{non-Abelian Coulomb phase}.
 
 \medskip
 In the range $N_f\geq 3N_c$, the theory is not asymptotically free: the coupling decreases at large distances due to screening effects. The potential between external electric sources behaves like $V(R) \sim \frac{1}{R \log(R\Lambda)}$, so we are in a \emph{non-Abelian free electric phase} ($g^2 \sim \frac{1}{\log(R\Lambda)}$).
 
  \medskip
 In the range $N_c+2\leq N_f \leq 3/2 N_c$, the theory becomes infinitely coupled and we need a different description. It is in fact in a \emph{non-Abelian free magnetic phase} and should be described by a different set of degrees of freedom.
 
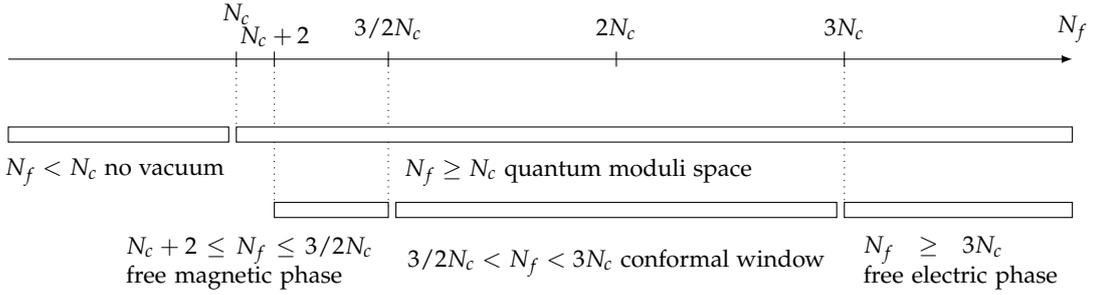
\begin{figure}
  \centering
  \begin{footnotesize}
    \begin{tikzpicture}
      \node at (3,.6) {\(N_c\)};
      \node at (3.5,.3) {\(N_c + 2\)};
      \node at (5,.4) {\(3/2 N_c\)};
      \node at (8,.4) {\(2 N_c\)};
      \node at (11,.4) {\(3 N_c\)};
      \node at (14,.4) {\(N_f\)};
      
      \draw[-latex] (0,0) -- (14,0);
      \draw (3,.1) -- (3,-.1);
      \draw (3.5,.1) -- (3.5,-.1);
      \draw (5,.1) -- (5,-.1);
      \draw (8,.1) -- (8,-.1);
      \draw (11,.1) -- (11,-.1);

      \draw[dotted] (3,0) -- (3,-1);
      \draw[dotted] (3.5,0) -- (3.5,-2);
      \draw[dotted] (5,0) -- (5,-2);
      \draw[dotted] (11,0) -- (11,-2);

      \draw (0,-1.1) rectangle (2.9,-.9);
      \node at (1.4,-1.5) {\(N_f < N_c\) no vacuum};
      \draw[fill=white] (3,-1.1) rectangle (14,-.9);
      \node at (7.5,-1.5) {\(N_f \ge N_c\) quantum moduli space};

      \draw[fill=white] (3.5, -2.1) rectangle (5,-1.9);
      \node [text width=3.5cm] at (3.3,-2.7) {\(N_c + 2 \le N_f \le 3/2 N_c\) free magnetic phase};
      \draw[fill=white] (5.1, -2.1) rectangle (10.9,-1.9);
      \node [] at (8,-2.7) {\(3/2 N_c < N_f < 3 N_c\) conformal window};
      \draw[fill=white] (11, -2.1) rectangle (14,-1.9);
      \node [text width=3.5 cm] at (13,-2.7) {\( N_f \ge 3 N_c\) \hfill \hspace{3cm} free electric phase};

    \end{tikzpicture}
  \end{footnotesize}
  \caption{Properties of the moduli space and phases of \ac{sqcd} for different values of $N_f$.}
  \label{fig:sqcd-chart}
\end{figure}

 \medskip
In Figure~\ref{fig:sqcd-chart}, the different properties of \ac{sqcd} for different values of $N_f$ are summarized.

\subsection{SQCD with real gauge groups}

So far we have discussed theories with unitary gauge groups.
We will also study theories with real gauge groups, namely symplectic 
$Sp(2N_c)$ and orthogonal $SO(N_c)$ gauge theories.

Let us briefly summarize some properties of these theories.
We refer to the unitary symplectic group as $Sp(2N_c)$, \emph{i.e.} the subgroup of $SU(2N_c)$ leaving 
an antisymmetric tensor $J$ invariant.
This group has dimension $N_c(2N_c+1)$ and there are $2N_f$ quarks in the fundamental 
representation of $Sp(2N_c)$. This representation has dimension $2N_c$.
There is a meson operator of dimension $N_f(2N_f-1)$ and, differently from the unitary case,
there are no baryons.
We refer the reader to~\cite{Intriligator:1995ne} for the details of the dynamics of this model.

If the gauge group is orthogonal, it corresponds to a model with an $so(N_c)$ algebra.
Depending on the global properties, there are different choices for the gauge group.
These groups can be constructed starting from an $SO(N_c)$ gauge group and gauging a discrete $\mathbb{Z}_2$ symmetry.
There are different choices for this gauging, leading to either an $O(N_c)_-$, an $O(N_c)_+$, or a $Spin(N_c)$ gauge group.
These cases are slightly different, \emph{e.g.}
matter fields in the spinor representation are allowed
in the $Spin(N_c)$ case.
We will ignore this issue in the rest of this review, referring the reader to~\cite{Aharony:2013kma}
for details.

We consider $SO(N_c)$ \ac{sqcd}, and in this case the gauge group has dimension $N_c(N_c-1)/2$.
We consider $N_f$ flavors in the vector representation of the gauge group. There is a mesonic operator of dimension
$\pqty{N_f+1}/\pqty{2N_f}$, and there are baryons as well.

\section{Seiberg duality in four dimensions and some generalizations}\label{sec:4dseiberg}

In this section we review some salient features of four-dimensional Seiberg duality relevant for our discussion.
First we introduce the notion of Seiberg duality for $SU(N_c)$ \ac{sqcd} with $N_f$ flavors
then we turn the discussion to the generalization of this construction in presence of
tensor matter and for real gauge groups.

\subsection{SQCD with $G=SU(N_c)$}

In his original paper~\cite{Seiberg:1994pq}, Seiberg considered \ac{sqcd} as discussed in Section~\ref{sec:4dsqcd}
with superpotential
\begin{equation}
\label{eleSei}
W = 0 
\end{equation}
We have seen that this theory has different \ac{ir} behaviors, depending on the ratio $x = N_f/N_c$.
We will focus on the case of $3/2 \leq x \leq 3$, where the theory is superconformal.
One of the main consequences of superconformality is that the  $R$-current 
belongs to a supermultiplet, which contains the stress
energy tensor as higher component. 
In this window, the theory can be studied via conformal perturbation theory. In general, there is a weakly coupled regime for 
$x>2$ and a strongly coupled regime for $x<2$.
In the special case of $x=2$, corresponding to $N_f = 2 N_c$, one can equivalently describe the physics, \emph{i.e.} the partition functions
and the correlations in terms of a \emph{different} set of fundamental degrees of freedom, the fundamentals and antifundamentals $q$ and $\tilde q$. In this description, it is necessary to introduce
a gauge singlet, that we will refer to as $M$. This singlet is in the bifundamental representation of the flavor symmetry group and it couples
to the fields $q$ and $\tilde q$ through a superpotential coupling, 
\begin{equation}
\label{magSei}
W = h M q \tilde q.
\end{equation}
The transformation properties of these fields under the global
symmetries of \ac{sqcd} are summarized in
Table~\ref{tab:global-magnetic}.

\begin{table}[]
  \centering
  \begin{tabular}{CCCCCC}
    \toprule
      & SU(N_f)_L    & SU(N_f)_R               & U(1)_B & U(1)_A & U(1)_R              \\ \midrule
    M & \mathbf{N}_f & \overline{\mathbf{N}}_f & 0 & 2 & 2\frac{N_f - N_c}{N_f} \\
    q & \overline{\mathbf{N}}_f & \mathbf{1} & \frac{N_c}{N_f - N_c} & -1 & \frac{N_c}{N_f} \\
    \tilde q &  \mathbf{1} & \mathbf{N}_f & -\frac{N_c}{N_f - N_c} & -1 & \frac{N_c}{N_f} \\
    \bottomrule
  \end{tabular}
  \caption{Transformation properties of the fields in the magnetic
    description under the global symmetries of \ac{sqcd}.}
  \label{tab:global-magnetic}
\end{table}

One can study the properties of the fixed point of these theories in both  cases: in the first, \emph{``electric''}, case there is one coupling involved, the gauge coupling $\lambda$,
while in the second, \emph{``magnetic''}, case there are two couplings, the gauge coupling $\tilde \lambda$ and the superpotential coupling $h$.
The two descriptions are physically equivalent.
This equivalence can be reformulated in terms of a duality, by
observing that the singlet $M$ has the same quantum numbers of the
gauge invariant operator $Q \widetilde Q$. %
One can claim a mapping 
\begin{equation}
\label{eq:meson}
M = \frac{1}{\mu}Q \widetilde Q.
\end{equation}
between the elementary singlet of the magnetic theory and the composite operator of the electric theory.
In equation (\ref{eq:meson}), $\mu$ represents the \ac{rg} scale at which we consider the theory.
This mapping can be extended to the full conformal window. In general, for $x \neq 2$, the magnetic theory has gauge group $SU(N_f-N_c)$.
For $x<2$, the magnetic theory is more weakly coupled than the electric theory, while this regime is inverted in the $x>2$ window, see Figure~\ref{fig:Seiberg-Chart}.
This behavior is reminiscent of \emph{electric/magnetic duality} (see \emph{e.g.} the case of $\mathcal{N}=4$ \ac{sym}, where S-duality maps the holomorphic gauge coupling 
as $\tau' = -1/\tau$).

The holomorphic scale $\Lambda$ is defined in terms of the holomorphic gauge coupling $\tau = \frac{4 \pi i }{g^2} + \frac{\theta}{2 \pi}$
as
\begin{equation}
\label{eq:lambdaolo}
\Lambda = \mu\, e^{ \pi i b/\tau},
\end{equation}
where $b$ is the numerator of the 1-loop $\beta$-function for the gauge coupling, 
\begin{equation}
b=3N_c-N_f
\end{equation}
for \ac{sqcd}, see also Eq.~\eqref{eq:beta}.

\definecolor[named]{magneticcolor}{RGB}{78,205,196}
\definecolor[named]{electriccolor}{RGB}{196,77,88}

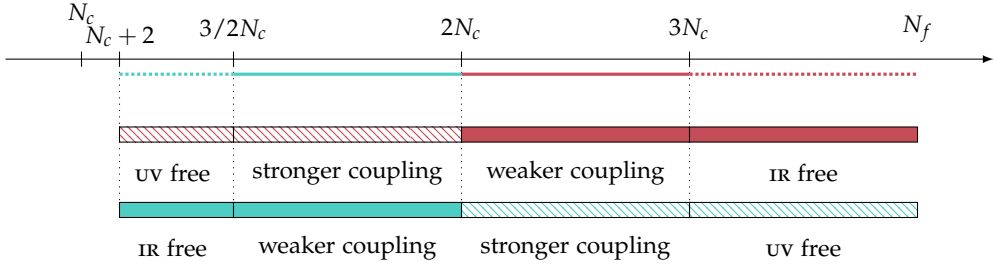
\begin{figure}
  \centering
  \begin{footnotesize}
    \begin{tikzpicture}
      \node at (3,.6) {\(N_c\)};
      \node at (3.5,.3) {\(N_c + 2\)};
      \node at (5,.4) {\(3/2 N_c\)};
      \node at (8,.4) {\(2 N_c\)};
      \node at (11,.4) {\(3 N_c\)};
      \node at (14,.4) {\(N_f\)};
      
      \draw[-latex] (2,0) -- (15,0);
      \draw (3,.1) -- (3,-.1);
      \draw (3.5,.1) -- (3.5,-.1);
      \draw (5,.1) -- (5,-.1);
      \draw (8,.1) -- (8,-.1);
      \draw (11,.1) -- (11,-.1);

      \draw[dotted] (3.5,0) -- (3.5,-2);
      \draw[dotted] (5,0) -- (5,-2);
      \draw[dotted] (8,0) -- (8,-2);
      \draw[dotted] (11,0) -- (11,-2);

      \draw[densely dotted, very thick, magneticcolor] (3.5,-.2) -- (5,-.2);
      \draw[very thick, magneticcolor] (5,-.2) -- (8,-.2);
      \draw[very thick, electriccolor] (8,-.2) -- (11,-.2);
      \draw[densely dotted, very thick, electriccolor] (11,-.2) -- (14,-.2);
      
      \draw[pattern=north west lines, pattern color=electriccolor] (3.5,-1.1) rectangle (5,-.9);
      \node at (4.2,-1.5) {\acs{uv} free};
      \draw[pattern=north west lines, pattern color=electriccolor] (5,-1.1) rectangle (8,-.9);
      \node at (6.5,-1.5) {stronger coupling};
      \draw[fill=electriccolor] (8,-1.1) rectangle (11,-.9);
      \node at (9.5,-1.5) {weaker coupling};
      \draw[fill=electriccolor] (11,-1.1) rectangle (14,-.9);
      \node at (12.5,-1.5) {\acs{ir} free};

      \draw[fill=magneticcolor] (3.5,-2.1) rectangle (5,-1.9);
      \node at (4.2,-2.5) {\acs{ir} free};
      \draw[fill=magneticcolor] (5,-2.1) rectangle (8,-1.9);
      \node at (6.5,-2.5) {weaker coupling};
      \draw[pattern=north west lines, pattern color=magneticcolor] (8,-2.1) rectangle (11,-1.9);
      \node at (9.5,-2.5) {stronger coupling};
      \draw[pattern=north west lines, pattern color=magneticcolor] (11,-2.1) rectangle (14,-1.9);
      \node at (12.5,-2.5) {\acs{uv} free};

    \end{tikzpicture}
  \end{footnotesize}
  \caption{Electric [red] and magnetic [blue] descriptions for \ac{sqcd}.}
  \label{fig:Seiberg-Chart}
\end{figure}
In general, we can associate the holomorphic scale $\Lambda$ of the electric theory  
to the scale of the magnetic theory via the relation
\begin{equation}
\label{eq:scalematching}
\Lambda^b \tilde\Lambda^{\tilde b} = (-1)^{N_f} \mu^{2 N_f}.
\end{equation}

The duality has been extended beyond the conformal window: for $1< x <3/2$, the electric theory is \ac{uv} free and strongly coupled in the \ac{ir}, while the
magnetic theory is \ac{ir} free. This regime is inverted in the $x>3$ window.
\subsection{Checks}

In the last two decades, many checks of Seiberg duality have been performed.
The duality relates a weakly coupled description to a strongly coupled one, which requires 
the development of non-perturbative techniques in order to perform some of the checks.

One of the first checks examined the structure of the moduli spaces. In some cases, the moduli space 
of one of the descriptions is classical while in the dual theory, there are quantum 
effects that have to be taken into account.
Another class of tests analyzes the behavior of the duality under mass deformations and Higgsing.
We refer the reader to~\cite{Intriligator:1995au} for an exhaustive analysis.

Another powerful check of Seiberg duality consists in matching the 't Hooft anomalies 
for the global symmetries between the dual descriptions.
It has been performed with success
on all generalizations that we will discuss below.
The basic idea underlying the anomaly matching is that during an \ac{rg} flow from a \ac{uv} fixed point to the \ac{ir}, the value of the chiral anomaly
has to be independent of the energy scale.
This condition can be equivalently applied to a duality: despite the different degrees of freedom used to 
describe the model, the chiral anomalies have to match. In other words, the dual theories have to respond 
in the same manner to an external background gauge perturbation.

The anomalies in the case of \ac{sqcd} are summarized in Table~\ref{tab:sqcd-anomalies}.
\begin{table}
  \centering
  \begin{tabular}{LCLC}
    \toprule
    SU(N_f)^3:& N_c & U(1)_B SU(N_f)^2:& \frac{N_c}{2}  \\
    U(1)_R SU(N_f)^2:& -\frac{N_c^2}{2 N_f}&U(1)_B^3:&  0 \\
    U(1)_B:& 0 &U(1)_B U(1)_R^2:& 0\\
    U(1)_R:& -N_c^2-1&U(1)_R^3:& -\frac{2 N_c^4}{N_f^2}+N_c^2-1 \\
    \bottomrule
  \end{tabular}
  \caption{Anomalies in \ac{sqcd}.}
  \label{tab:sqcd-anomalies}
\end{table}
They have been shown to match in the electric and in the magnetic phases. Also this test has
been passed with success by all the other dualities presented in this review.
Note that the 't Hooft anomaly matching is a necessary but not sufficient condition for 
a duality (see~\cite{Brodie:1998vv} for further elaboration on this point).

Other possible checks consist in matching other exact quantities such as the Witten index~\cite{Witten:1982df} 
and the Hilbert series~\cite{Pouliot:1998yv}.

In the recent past, a new set of checks has been furnished by localization: many
tests have been performed by matching the superconformal index of the dual phases (see for example~\cite{Dolan:2008qi}).
A detailed analysis of the identities that relate the dualities discussed in this review is given in~\cite{Spiridonov:2009za,Spiridonov:2011hf}. 
Note that in some of the cases, the identities
have been proven, while in some other cases only partial results are known.

\subsection{Extensions}
\label{sec:4d-extensions}

Seiberg duality has been extended in many directions over the years. 
It has been shown for example that a similar pattern exists for \emph{orthogonal} and \emph{symplectic} gauge groups.
Another extension of the duality includes the presence of tensor matter in the adjoint, symmetric and antisymmetric representations.
Another interesting extension is related to product groups. In the following, we will discuss the former two situations, and briefly 
comment on the latter.

\subsubsection{Tensor matter}
\begin{description}[style=nextline]
\item[{$SU(N_c)$ with adjoint}]
This was the first generalization of Seiberg duality with tensor matter, introduced in~\cite{Kutasov:1995ss}.
The electric theory corresponds to the \ac{sqcd} considered above with an extra field $X$ in the adjoint representation of the $SU(N_c)$ gauge group turned on.
There is also a superpotential interaction 
\begin{equation}
\label{KSSe}
W = \Tr X^{n+1},
\end{equation} where $n$ is a positive integer.
For $n=1$, the adjoint can be integrated out and the original \ac{sqcd} is recovered.
The case $n>1$ is more interesting.
In this case the singlets, generalizing the meson $M$ of \ac{sqcd}, are 
\begin{equation}
M_j = Q X^j \widetilde Q
\end{equation}
and the chiral ring relations (\emph{i.e.} the F-term of $X$) classically constrain $j<n$.

The dual theory in this case has an $SU(n N_f - N_c)$ gauge group, a dual adjoint $Y$ and
dual fundamentals and antifundamentals interacting with the singlets $M_j$. The superpotential for
this theory is  
\begin{equation}
\label{KSSm}
W = \Tr Y^{n+1} + \sum_{j=0}^{n-1} Y^j q M_{n-j-1} \tilde q. 
\end{equation}

\item[{$SU(N_c)$ with symmetric flavor}]
This duality was introduced in~\cite{Intriligator:1995ax}.
In this case one can consider the field $S$ in the symmetric representation of the gauge group,
and its conjugate $\tilde S$,
with an interaction 
\begin{equation}
\label{eleS}
W = \Tr (S \tilde S)^{n+1}.
\end{equation}
This induces a truncation on the chiral ring, such that the mesons that we have to consider are
\begin{equation}
M = Q (\widetilde S S)^{j+1} \tilde Q,\quad P = Q (\widetilde S S )^{j} \widetilde s Q, \quad \tilde P =   \widetilde Q S (\widetilde S S )^{j}  \widetilde Q,\quad j=0,\dots,n-1,
\end{equation}
where $M$ is a bifundamental
operator while $P$ and $\tilde P$  are symmetric in the flavor indices.

The dual theory has $SU((2n+1) N_f+4n -N_c)$ gauge group, there are dual fundamentals
and antifundamentals $q$ and $\tilde q$, dual symmetric and conjugate $s$ and $\tilde s$,
and the singlet fields are the mesons $M$, $P$ and $\tilde P$.
The dual superpotential is
\begin{equation}
\label{magS}
W = (s \tilde s)^{n+1}
+
\sum_{j=0}^{n} M_{n-j} q (s \tilde s)^{j} \tilde q
+\sum_{j=0}^{n-1}
\left(P_{n-j-1} q (s \tilde s)^j \tilde s q +
\widetilde P_{n-j-1} \widetilde q s (s \tilde s)^j \tilde a \widetilde q \right).
\end{equation}

\item[{$SU(N_c)$ with antisymmetric flavor}]
This duality was introduced in~\cite{Intriligator:1995ax}.
It has an $SU(N_c)$ gauge groups, $N_f$ pairs of fundamentals and antifundamentals $Q$, $\widetilde Q$, and two
tensor matter fields $A$, $\tilde A$, in the antisymmetric and in the conjugate antisymmetric representations. Note that the presence of the conjugate antisymmetric is required 
in order for the theory to be anomaly free.
There is also a superpotential term
\begin{equation}
\label{eleA}
W = \Tr(A \tilde A)^{n+1}.
\end{equation}
This induces a truncation on the chiral ring, such that the mesons that we have to consider are
\begin{equation}
M = Q (\tilde A A)^{j+1} \widetilde Q,\quad P = Q (\tilde A A)^{j} \widetilde A Q,\quad \tilde P =   \widetilde Q A (\tilde A A)^{j}  \widetilde Q,\quad j=0,\dots,n-1,
\end{equation} where $M$ is a bifundamental
operator while $P$ and $\tilde P$  are antisymmetric in the flavor indices.

The dual theory has $SU((2n+1) N_f-4n -N_c)$ gauge group, there are dual fundamentals
and antifundamentals, $q$ and $\tilde q$, dual antisymmetric and conjugate, $a$ and $\tilde a$
and the singlet fields are the mesons $M$, $P$ and $\tilde P$.
The dual superpotential is
\begin{equation}
\label{magA}
W = \pqty{a \tilde a}^{n+1}
+
\sum_{j=0}^{n} M_{n-j} q \pqty{a \tilde a}^{j} \tilde q
+\sum_{j=0}^{n-1}
\pqty{P_{n-j-1} q \pqty{a \tilde a}^j \tilde a q +
\widetilde P_{n-j-1} \widetilde q a \pqty{a \tilde a}^j \tilde a \widetilde q} .
\end{equation}
\end{description}
\subsubsection{Orthogonal case}

Another extension of Seiberg duality consists of theories with real gauge group.
We will be interested in these realizations of the duality as they 
can be easily realized in terms of \D~branes and orientifold planes, 
and play a crucial role in understanding the role of the orientifold in the reduction.

We will start by discussing the case of \ac{sqcd} with orthogonal gauge group and $N_f$
vectors (corresponding to the fundamental and the antifundamental representations
in the unitary case). Then we turn to the case of tensor matter. The two most interesting cases
are those with symmetric and with antisymmetric (\emph{i.e.} adjoint) tensors.
\begin{description}[style=nextline]
\item[{$SO(N_c)$ with $N_f$ vectors}]
    In this case, the electric theory has gauge group $SO(N_c)$ and
    $N_f$ vectors $q$. Observe that here, we do not distinguish the
    $N_c$ even and odd cases.
The superpotential is 
\begin{equation} 
\label{eleSO}
W = 0
\end{equation}
and does not impose constraints on the chiral ring;
there is a meson operator of the form 
\begin{equation}
M = Q Q.
\end{equation}

The dual theory was worked out in~\cite{Intriligator:1995id} and it corresponds to an  $SO(N_f-N_c+4)$ gauge stheory with $N_f$ vectors $q$
and the meson $M$. The dual superpotential is 
\begin{equation} 
\label{magSO}
W = M q q .
\end{equation}
\item[{$SO(N_c)$ with $N_f$ vectors and an adjoint}]
In this case, the electric theory has gauge symmetry $SO(N_c)$, with $N_f$ vectors
$Q$ and an adjoint $A$ (in this case the adjoint corresponds to an antisymmetric tensor).
The superpotential is
\begin{equation}
\label{eleSOAdj}
W = \Tr A^{2(n+1)},
\end{equation}
and the mesons in the chiral ring are identified as 
\begin{equation}
M_j = Q A^j Q, \quad j=0,\dots,2n,
\end{equation}
symmetric in the flavor indices for even $j$ and antisymmetric for odd $j$.

The dual theory, obtained in~\cite{Leigh:1995qp}, has $SO((2n+1) N_f - N_c+4)$  gauge group, $N_f$ vectors $q$, an
adjoint $a$ and the mesons $M_j$.
The superpotential of this dual theory is  
\begin{equation}
\label{magSOAdj}
W =a^{2(n+1)} +  \sum_{j=0}^{2n}  M_{2n-j} q a^j q .
\end{equation}
\item[{$SO(N_c)$ with  $N_f$ vectors and a traceless symmetric tensor}]
In this case the electric theory has $SO(N_c)$ gauge symmetry, with $N_f$ vectors
$Q$ and a traceless  symmetric tensor $S$
\begin{equation}
\label{eleSOS}
W = \Tr S^{n+1},
\end{equation}
and the mesons in the chiral ring are identified as 
\begin{equation}M_j = Q S^j \tilde Q,\quad j=0,\dots,n-1.
\end{equation}
The dual theory, obtained in~\cite{Intriligator:1995ax}, has $SO(n (N_f+4) - N_c)$  gauge group, $N_f$ vectors $q$, a traceless symmetric tensor
$s$ and the mesons $M_j$.
The superpotential of this dual theory is  
\begin{equation}
\label{magSOS}
W =s^{n+1} +  \sum_{j=0}^{n-1}  M_{k-j} q s^j q .
\end{equation}

\end{description}
\subsubsection{Symplectic case}

Here we discuss the case of symplectic gauge groups, by fixing the convention
$Sp(2) = SU(2)$. We first discuss the case of gauge group $Sp(2N_c)$ in presence for $2 N_f$ 
fundamentals. Then we generalize the construction by considering also a symmetric (adjoint)
or antisymmetric tensor.
\begin{description}[style=nextline]
\item[{$Sp(2N_c)$ with $2N_f$ fundamentals}]
In this case the electric theory has gauge group $Sp(2N_c)$ and there are $2N_f$ fundamental
quarks $Q$. The superpotential is 
\begin{equation}
\label{eleSp}
W=0
\end{equation}
 and there is a meson 
\begin{equation}
M = Q Q.
\end{equation}
The dual theory, found in~\cite{Intriligator:1995ne}, has $Sp(2(N_f-N_c-2))$ gauge group with $2N_f$ fundamentals $q$ and the meson $M$.
The superpotential of this dual theory is
\begin{equation}
\label{magSp}
W = M q q .
\end{equation}
\item[{$Sp(2N_c)$ with $2N_f$ fundamentals and an adjoint}]

In this case the electric theory has gauge group $Sp(2N_c)$ and there are $2N_f$ fundamental
quarks $Q$ and an adjoint $S$ (in this case we use this terminology because the adjoint 
is a symmetric tensor). The superpotential is 
\begin{equation}
\label{eleSpAdj}
W = \Tr S^{2(n+1)},
\end{equation}
and there are mesons 
\begin{equation}
M_j = Q S^j Q, \quad j=0,\dots 2n,
\end{equation}
symmetric in the flavor indices for odd $j$ and antisymmetric for even $j$.
The dual theory, obtained in~\cite{Leigh:1995qp},  has $Sp(2((2n+1) N_f-N_c-2))$ gauge group with $2N_f$ fundamentals $q$,
a dual adjoint $s$ and the mesons $M_j$.
The superpotential of this dual theory is
\begin{equation}
\label{magSpAdj}
W =s^{2(n+1)} +  \sum_{j=0}^{2n}  M_{2n-j} q s^j q .
\end{equation}
\item[{$Sp(2N_c)$ with $2N_f$ fundamentals and an antisymmetric tensor}]
In this case the electric theory has gauge group $Sp(2N_c)$ and there are $2N_f$ fundamental
quarks $Q$ and a traceless antisymmetric tensor $A$. The superpotential is 
\begin{equation}
\label{eleSpA}
W = \Tr A^{n+1}.
\end{equation}
and there are mesons 
\begin{equation}
M_j = Q A^j Q,\quad j=0,\dots n-1.
\end{equation}
The dual theory, obtained in~\cite{Intriligator:1995ax},
has $Sp(2(n(N_f-2)-N_c))$ gauge group with $2N_f$ fundamentals $q$,
a dual antisymmetric $a$ and the mesons $M_j$.
The superpotential of this dual theory is
\begin{equation}
\label{magSpA}
W =a^{n+1} +  \sum_{j=0}^{n-1}  M_{n-j} q a^j q .
\end{equation}
\end{description}

\begin{sidewaystable}
  \centering
  \begin{tabular}{LLLLLl}
    \toprule
    G        & \text{Charged Fields}                             & \tilde G                & W_e              & W_m              & \text{Ref.}                \\
    \midrule
    SU(N_c)  & N_f(F \oplus \tilde F)                            & SU(N_f-N_c)             & (\ref{eleSei})   & (\ref{magSei})   & \cite{}                    \\
    SU(N_c)  & N_f(F \oplus \tilde F) \oplus X                   & SU(n(N_f)-N_c)          & (\ref{KSSe})     & (\ref{KSSm})     & \cite{Kutasov:1995ss}      \\
    SU(N_c)  & N_f(F \oplus \tilde F) \oplus (S \oplus \tilde S) & SU((2n+1) N_f+4n -N_c)  & (\ref{eleS})     & (\ref{magS})     & \cite{Intriligator:1995ax} \\
    SU(N_c)  & N_f(F \oplus \tilde F) \oplus (A \oplus \tilde A) & SU((2n+1) N_f-4n -N_c)  & (\ref{eleA})     & (\ref{magA})     & \cite{Intriligator:1995ax} \\
    Sp(2N_c) & 2 N_f F                                           & Sp(2(N_f-N_c-2))        & (\ref{eleSp})    & (\ref{magSp})    & \cite{Intriligator:1995ne} \\
    Sp(2N_c) & 2 N_f F \oplus X                                  & Sp(2((2n+1)N_f-N_c -2)) & (\ref{eleSpAdj}) & (\ref{magSpAdj}) & \cite{Leigh:1995qp}        \\
    Sp(2N_c) & 2N_f F \oplus A                                   & Sp(2(n (N_f-2)-N_c))    & (\ref{eleSpA})   & (\ref{magSpA})   & \cite{Intriligator:1995ax} \\
    SO(N_c)  & N_f V                                             & SO(N_f-N_c+4)           & (\ref{eleSO})    & (\ref{magSO})    & \cite{Intriligator:1995id} \\
    SO(N_c)  & N_f V \oplus X                                    & SO((2n+1)N_f-N_c +4)    & (\ref{eleSOAdj}) & (\ref{magSOAdj}) & \cite{Leigh:1995qp}        \\
    SO(N_c)  & N_f V \oplus S                                    & SO(n (N_f+4) - N_c)     & (\ref{eleSOS})   & (\ref{magSOS})   & \cite{Intriligator:1995ax} \\
    \bottomrule
  \end{tabular}  
  \caption{Summary of the four dimensional dimensional dualities reviewed in this section. Here $F$ is the  fundamental representation,
$\tilde F$ the antifundamental, $S$ the symmetric, and $\tilde S$ its conjugate, $A$ the antisymmetric,
$\tilde A$ its conjugate, $X$ the adjoint and $V$ the vector.
There are also singlets in the spectrum of the dual theories that we do not report in the table, corresponding to the generalized mesons.
}
  \label{tab:four-d-dualities}
\end{sidewaystable}

\section{The~brane picture in four dimensions}\label{sec:4dbranes}

In this section we discuss the realization of four-dimensional Seiberg duality and 
its generalization in terms of \tIIA D~brane configurations.
We start by reviewing some general aspects of the~brane 
realization of four-dimensional $\mathcal{N}=1$ theories in Sec.~\ref{sec:generalities}.
Then we show how this picture is modified when flavors are added, realizing the
electric \ac{sqcd} in terms of D~branes and NS~branes (Sec.~\ref{sec:sqcd}).
The dual theory is constructed via a \ac{hw} transition that we review in Sec.~\ref{sec:hw}.
One generalization of this duality is obtained by adding an adjoint field with
a power-law superpotential, leading to the \ac{kss} duality (Sec.~\ref{sec:kss}).
Another generalization involves theories with real gauge group and with
tensor matter, realized by introducing orientifold planes in the~brane setup. 
We review some of these dualities explicitly in Sec.~\ref{sec:orientifolds} and sketch some further generalizations
in Sec.~\ref{sec:other-realizations}.

\subsection{Generalities}
\label{sec:generalities}

Here we introduce some basic aspects of~brane dynamics necessary to study the 
realization of Seiberg duality and its extensions in four-dimensional $\mathcal{N}=1$ theories.
As this subject has been extensively described in the literature (see \emph{e.g.}~\cite{Giveon:1998sr}), we will give a general overview 
as opposed to a description in full detail.
We will try nevertheless to be self-consistent, discussing the details necessary to the 
analysis of the reduction of the four-dimensional dualities to three dimensions.

The D~brane realization of four-dimensional $\mathcal{N}=1$ theories
requires a ten-dimensional \tIIA description.
The ten-dimensional theory is non-chiral: there are two supercharges $Q_L$ and 
$Q_R$, generated respectively by the left and right-moving worldsheet degrees of freedom.
The supercharges satisfy the relation 
\begin{equation}
  \Gamma^{0123456789} Q_{L/R} = \pm Q_{L/R} ,
\end{equation}
where $\Gamma^{0123456789} = \Gamma^0 \Gamma^1 \cdots \Gamma^9$.

The theory has two sectors, the \ac{ns} and \ac{r} sectors, in which we can have different potentials.
There are two main~branes charged under these potentials that will play a crucial role
in the description, namely Neveu--Schwarz (\NS{}) and Dirichlet-$p$ (\D{p})~branes.
Let us discuss their main properties separately.
\begin{itemize}
\item The \NS{}~branes are five-dimensional membranes, magnetically coupled to the two-dimensional antisymmetric 
\(B\) field of the \ac{ns} sector. They have tension $(2 \pi)^{-5} g_s^{-2} (\alpha')^{-3}$ and do not admit a simple \ac{cft} description
because the dilaton blows up close to their core.
They preserve some of the original \ac{susy}: for an \NS{}~brane
extended along $x_{12345}$, the preserved supercharge is $\epsilon_L Q_L + \epsilon_R Q_R$,
where the spinors satisfy $\epsilon_{L/R} = \Gamma^{012345} \epsilon_{L/R}$.
Unless explicitly stated, we will denote with \NS{} a~brane extended along
\(x_{12345}\) and with \NS{}' a~brane extended in  $x_{12389}$.
\item The other necessary ingredient are the \D{p}~branes. They are defined with respect to their charge 
  under the $p+1$ forms of the R sector, where $p=0,2,4,6,8$.  Their tension is $(2\pi)^{-p}(\alpha')^{-(p+1)/2}$ and
they admit a \ac{cft} description at weak string coupling.
We will mostly focus on \D4~branes along $x_{01236}$ and on \D6~branes
along $x_{0123789}$; we will also introduce  \D6'~branes along
$x_{0123457}$ when necessary. The~brane directions are summarized in Table~\ref{tab:4d-brane-engineering}.
\begin{table}
  \centering
  \begin{tabular}{lCCCCCCCCCC}
    \toprule
           & 0      & 1      & 2      & 3      & 4      & 5      & 6      & 7      & 8      & 9      \\
    \midrule
    \NS{}  & \times & \times & \times & \times & \times & \times &        &        &                 \\
    \NS{}' & \times & \times & \times & \times &        &        &        &        & \times & \times \\
    \D4    & \times & \times & \times & \times &        &        & \times &        &                 \\
    \D6    & \times & \times & \times & \times &        &        &        & \times & \times & \times \\
    \D6'   & \times & \times & \times & \times & \times & \times &        & \times &        &        \\ \bottomrule
  \end{tabular}
  \caption{Brane configuration for four-dimensional~brane engineering.}
  \label{tab:4d-brane-engineering}
\end{table}

The $\epsilon_L Q_L + \epsilon_R Q_R$ supercharges preserved by a \D4~brane  requires
$\epsilon_L = \Gamma^{01236} \epsilon_R$ while for a D6~brane we have
$\epsilon_L = \Gamma^{0123789} \epsilon_R$.
The worldvolume theory on a \D{p}~brane  corresponds to a field theory in \(p+1\) dimensions
with \(16\) supercharges. In the spectrum there is a massless
\(U(1)\) gauge field and $9-p$ massless scalars, corresponding to the position of the \(Dp\)
brane in the transverse geometry. This description is valid as long as the~brane is 
infinite, but we will deal with \D4~branes extended along a segments in $x_6$. This
is possible because at the extremal points we will put an \NS{} or a
\D6~brane. In this case, we will interpret the theory as effectively four dimensional. 
\end{itemize}
Combining the two types of~branes, in the simplest case of a \D4
suspended between two \NS{}~branes, the resulting theory is $\mathcal{N}=2$
\ac{sym}. In fact, in this case the system of~branes preserves \(8\) supercharges.
The distance between the \NS{}~branes (the length of the $x_6$ segment) is
inversely proportional to the gauge coupling:
\begin{equation}
  g_{\textsc{sym}}^2 = \frac{g_s (2 \pi)^2 \sqrt{\alpha'}}{\ell_6} .  
\end{equation}
There are in this case two types of massless fields, a $U(1)$ gauge field (and its supersymmetric completion)
and a complex scalar, related to the freedom of moving the \D4 segment along the \NS{}~branes (\emph{i.e.} along the
directions $x_{45}$), together with its supersymmetric fermionic partner.

We can also consider more than one \D4~brane (a stack of \(N\) \D4~branes). In this case
the \tIIA description is dressed with Chan--Paton indices 
to obtain a $U(N)$ theory; now we have a $U(N)$ gauge field and the 
scalar field becomes an $N \times N$ matrix, transforming in the adjoint representation of the
gauge group. This is not the end of the story. In the four-dimensional picture, the $U(1) \subset U(N)$ decouples through a Green--Schwarz
mechanism because one component of the massive vector
boson becomes massive in the \ac{ir}. In geometrical terms, the decoupling of the $U(1)$ corresponds to the freedom in the choice
of the position of the center of mass of the stack of \D4~branes.
In this way we are left with an $SU(N)$ gauge theory in the \ac{ir}.

We can also break half of the supersymmetry by rotating one of the
\NS{}~branes into an \NS{}'~brane (extended in \(x_{012389}\)).
Geometrically we lose the freedom to slide the \D4~brane along the \NS{}~branes, and on the field theory side this
corresponds to integrating out the adjoint field. This configuration corresponds to $\mathcal{N}=1$ \ac{sym} with 
an $SU(N)$ gauge group.

\subsection{Adding flavor:  SQCD}
\label{sec:sqcd}

In this section we discuss the flavoring of the model discussed above.
Flavors are introduced by adding a stack of $N_f$ \D6~branes
to the setup. This will lead us directly to the discussion of
physically equivalent configurations and to the idea of \acl{hw}
transitions, \emph{i.e.} the~brane description of Seiberg duality.

The \D6s are considered as heavy objects and they have four dimensions
common with the \D4~branes (they are extended in \(x_{0123789}\)). 
The dynamical degrees of freedom in the four-dimensional picture are $N_f$ fundamental fields of the $SU(N_c)$
gauge group, describing the 4--6 strings (stretching from the \D4~branes to the \D6~branes) plus the adjoint fields from
the 4--4 strings. There are two inequivalent cases: in one case the
\NS{} is placed between the \D6 and the \NS{}' along the direction
\(x_6\); in the other, the positions of the two \NS{}~branes are
interchanged (see Figure~\ref{fig:D6-two-sides}) %

\begin{figure}
  \centering
  \begin{footnotesize}
    \DsixBothSides
  \end{footnotesize}
  \caption{The two possible configurations of \D4--\D5--\NS{}--\NS{}'.}
  \label{fig:D6-two-sides}
\end{figure}

In the first case, we cannot move the \D4~branes between the \NS{}'
and the \D6~branes in any direction, as they do not share directions orthogonal to the \D4. 
In the second case, we can actually consider two separate stacks of \D4s, one stretched between the stack of \D6s and the \NS',
the other stretched between the \NS' and the \NS.  The first stack of \D4s can move in the $x_{89}$ direction. 
The relative position in \(x_{89}\) of the two stacks of \D4s
translates into a mass term for the quarks $Q$ and $\widetilde Q$. This
mass term is associated to the presence of a coupling of the
fundamentals with a singlet $M$, and moving the~branes apart corresponds to
assigning a \ac{vev} to this singlet. In other words, in the first
D~brane configuration, we have vanishing superpotential $W=0$, while in
the second case, we have 
\begin{equation}
W= M Q \widetilde Q.
\end{equation}

Let us discuss the global symmetries of the model. There is a $U(1)$ symmetry associated to the decoupling
of the center of mass of the stack of \D4s, which is the overall $U(1)$ in $U(N_c)$
discussed above. There is also a flavor symmetry $U(N_f)^2$ associated
to the stack of $N_f$ \D4~branes ending on the \D6
branes\footnote{\label{foot:flavor-in-branes-4d}The
brane picture only realizes a \(U(N_f)\) symmetry. The full flavor
symmetry is obtained by superimposing the \D6 to the \NS{} as
discussed in~\cite{Giveon:2008zn}. The situation will be different in
three dimensions, see Sec.~\ref{subsec:fivebranes}}. The two overall
$U(1)$s contained in the flavor groups however are not independent, since only the relative
displacement of the two stacks of \D4~branes is physically relevant.
They form the baryonic $U(1)$ of \ac{sqcd} in terms of the gauge theory description.

More interestingly, there are two $U(1)$
symmetries associated to rotations in the $x_{45}$ and $x_{89}$
planes. 
This $U(1)_{45} \times U(1)_{89}$ is part of the Lorentz group in \(9+1\) dimensions and rotates the supercharges.
It is hence an \(R\)--symmetry.
However, the axial $U(1)$ subgroup leaves the supercharges invariant while it rotates $Q$ and $\widetilde Q$ in the same way,
appearing as axial symmetry $U(1)_A$ in the field theory. 
While both $U(1)_R$ and $U(1)_A$ are classically present, the latter
is anomalous. In the brane construction, this can be seen if one uplifts the system
to M-theory, where the \NS{}~brane bends away from the \D4
branes. This is a classical effect in the M-theory picture and a
quantum effect in the \tIIA setup and signals the breakdown of one of
the $U(1)$ symmetries discussed above. We obtain a geometrical
realization of the axial anomaly in four dimensions. We will come
back to this issue later, when discussing the reduction of the duality
to three dimensions, showing how this breaking is realized in the
T--dual case, when $x_3$ is compact.

In the following section we will show how the two configurations in
Figure~\ref{fig:D6-two-sides} are related via the so-called \acl{hw}
transition, which is the~brane realization of Seiberg duality.

\subsection{Hanany--Witten transition as Seiberg duality}
\label{sec:hw}

\begin{figure}
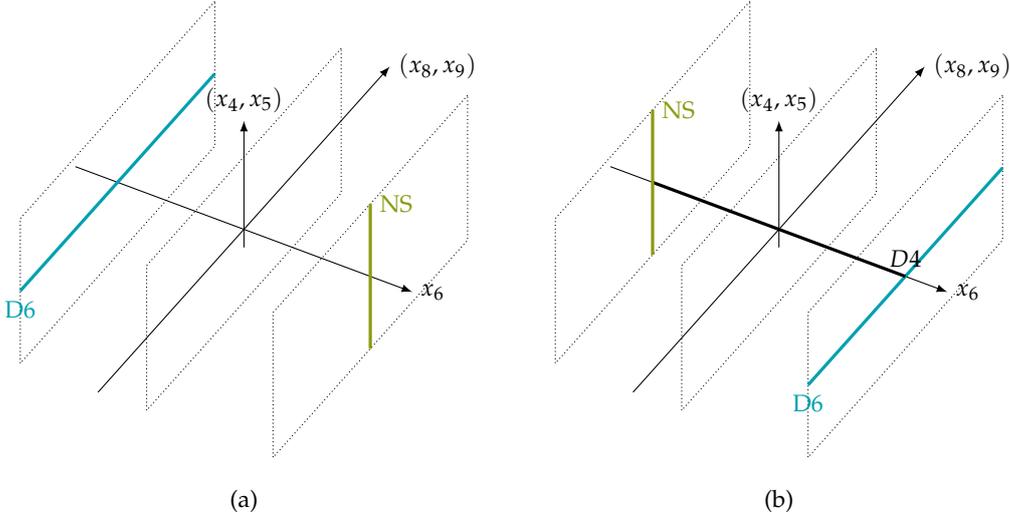

  \begin{footnotesize}
    \centering \minimalHW
  \end{footnotesize}
  \caption{\ac{hw} transition. When the \D6~brane crosses the \NS{}
   ~brane a new \D4~brane is created. The two configurations are
    physically equivalent.}
  \label{fig:HW-transition}
\end{figure}

In this section we review the realization of Seiberg duality as an equivalence between two
brane realizations.
As we will see, the two flavored models discussed above
(the two sides in Figure~\ref{fig:D6-two-sides}) can be transformed into each other
via a~brane transition. This transformation corresponds to 
the duality on the field theory side.
The~brane transition has been obtained in~\cite{Hanany:1996ie} by Hanany and Witten, and we will refer to it as the 
\ac{hw} transition.
The main idea behind the transition is the following\footnote{Here we adapt the discussion to the \tIIA setup, even if the original
discussion was done in the \tIIB case.}. Consider a \D6~brane and an NS~brane (or equivalently a \D6' and an \NS{}'~brane).
If we move the \D6 along the $x_6$ direction to the other side of the \NS{}~brane, 
we obtain an equivalent configuration, 
in which a \D4~brane has been created between the \D6 and the \NS{}~brane.
The converse is also true: if there was a \D4 in the initial configuration between the \D6 and the \NS{}~brane,
after moving the \D6 to the left of the \NS{}~brane, this~brane is
destroyed (see Figure~\ref{fig:HW-transition}).

For more general configurations, one has to consider the flux 
sourced by each~brane and study its effect on the action of the other~branes.
This defines the so called \emph{linking number},
representing the total magnetic charge of the gauge fields coupled to
the worldvolume of the~branes~\cite{Hanany:1996ie}.
This number
has to be preserved in the transition in order to make the~brane 
action well defined. This will be important later, when we will add other characters to the picture, \emph{i.e.} other
extended objects carrying D~brane charge.
In this discussion, there is always an underlying assumption that can be formulated in terms of a constraint,
the so-called \emph{s-rule}. This principle states that the
configurations that preserve supersymmetry have necessarily
\(0\) or \(1\) \D4~branes stretched between  a \D6 and an \NS{}~brane~\cite{Hanany:1996ie}.

\bigskip

We are now ready to discuss the Seiberg duality in terms of a \ac{hw} transition.
Let us start considering the setup in Figure~\ref{fig:D6-two-sides}(a).
There are $N_c$ \D4~branes between an \NS{} and an \NS{}'~brane and,
on the left side, $N_f$ \D4s between the $N_f$ \D6s and the \NS{}'~brane.
Let us now recombine $N_c$ \D4~branes, stretching them between the \NS{} and the $N_f$ \D6~branes.
We can now move the \NS{}~branes along $x_6$ and cross the \NS{}'~brane (Figure~\ref{fig:D6-two-sides}(b)). After this, we can recombine the \D4~branes 
between the \NS{}' and the \D6~brane. In this way, we obtain a configuration with $N_f-N_c$ \D4s between the \NS{} and the \NS{}'
brane and $N_f$ \D4s between the \NS{}' and the \D6~branes. This configuration corresponds to the one discussed in the section above
with superpotential $W= M Q \widetilde Q$, \emph{i.e.} the Seiberg-dual configuration expected from the field theory discussion.

\subsection{Multiple NS and KSS duality}
\label{sec:kss}

In this section, we discuss in some detail a generalization of Seiberg duality in presence of 
a power law superpotential for an adjoint matter field, \emph{i.e.} the \acl{kss} duality.

First we need to understand how to engineer a power-law superpotential 
for the adjoint field. This is done by considering a stack of $n$ \NS{}~branes
(or equivalently \NS{}'~branes).
A stack of \NS{}~branes corresponds to a singular limit, and the correct physical interpretation 
is obtained by considering the desingularization of this geometry.

In order to understand the origin of the superpotential, we need to take a step back.
As discussed above, the case of $\mathcal{N}=2$ \ac{sym} is associated to a pair of \NS{}
branes and $N_c$ \D4s stretched between them.
The mass term of the adjoint breaking supersymmetry to $\mathcal{N}=1$ was 
obtained by rotating the \NS{} into an \NS{}'~brane.
In other words, the case $n=1$ corresponds to $W = X^{n+1} = X^2$, where $X$ is an adjoint field
of $SU(N_c)$. For general $n$, the claim is then that the
superpotential is $W = X^{n+1}$.

There is a classical way to verify this result. It consists in studying a polynomial deformation 
\begin{equation}
\label{polyX}
\Delta W = \sum_{i=1}^{k-1} \lambda_i \Tr X^i
\end{equation}
in the $X$, and assigning a \ac{vev} to  $X$, breaking 
$SU(N_c)$ to $SU(r_1) \times \dots \times SU(r_n)$ with $r_1+\dots r_n =N_c$.
This corresponds to studying a set of decoupled \ac{sqcd} sectors, each with $N_f$ 
flavors.

On the~brane side, this is precisely the desingularization of the geometry, and it is obtained by separating
the \NS{}~branes along the $x_{89}$ direction. In this way we obtain $n$ different \ac{sqcd} sectors, as expected.
This corresponds on the field theory side to considering the superpotential deformation (\ref{polyX}).
and breaking the theory into a set of \ac{sqcd} sectors.
There is a subtlety when we consider the flavor~branes, though. Having $N_f$ pairs of fundamentals and antifundamentals
in each sector we must consider in total $n N_f$ flavor \D4 and \D6~branes in this setup even if 
it does not enhance the global symmetry to $SU(n N_f)^2$.
There are in total  
 $n N_f^2$ mesonic components, corresponding to 
 \begin{equation}
M_j = Q X^j \widetilde Q, \quad j=0,\dots,n.
\end{equation}
Even if it is unclear how to realize this possibility when the $n$ \NS{}~branes are coincident, when the geometry is desingularized,
the counting becomes more obvious, because we have a single meson $M = Q^{\alpha_{l}} Q_{\alpha_{l}}$,
where the subscript $l$ refers to the fact that we are summing over the gauge index in each \ac{sqcd} sector,
$\alpha_l=1,\dots,r_l$, for $l=1,\dots,n$.

\begin{figure}
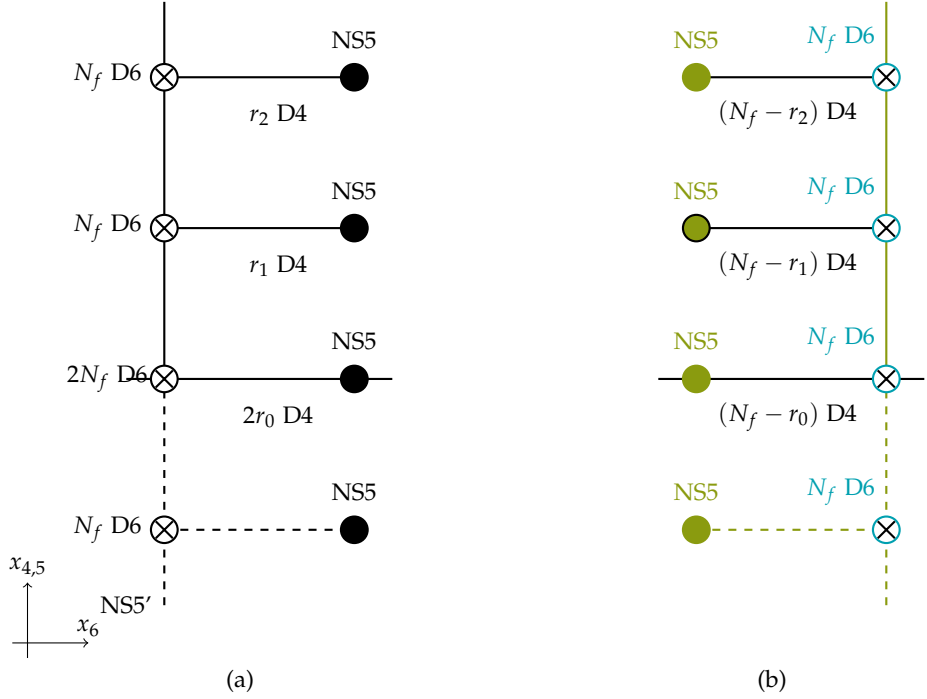

  \centering
  \begin{footnotesize}
    \FigureMultipleNS
  \end{footnotesize}
  \caption{\ac{hw} transition for the \ac{kss} duality.}
  \label{fig:KSS}
\end{figure}

We can now proceed as above, by performing the \ac{hw} transition in
each sector and recombining the~branes at the end.
The final configuration has $n N_f - Nc$ \D4~branes stretched between the \NS{}' and the \(n\) \NS{}~branes, and
$N_f$ \D4~branes stretched between the \NS{}~brane and the $N_f$ \D6~branes.
This configuration corresponds to the expected \ac{kss} dual theory,
and realizes the field theory duality in the~brane setup (see Figure~\ref{fig:KSS}).

\subsection{Orientifold planes}
\label{sec:orientifolds}

In this section, we introduce another extended object that plays a crucial role in realizing
four-dimensional $\mathcal{N}=1$ field theories from the~brane perspective: the orientifold plane.
Orientifolds allow us to construct real orthogonal and symplectic gauge groups
and matter fields in the symmetric and antisymmetric representation of the gauge group.

An orientifold 
can be defined from its action in perturbative string theory.
It is the combined action of
\begin{itemize} 
\item a parity inversion $\sigma$ of the coordinates transverse to the
plane 
\item a world-sheet parity $\Omega$
\item $(-1)^{F_L}$, $F_L$ being the left-moving fermion number.
\end{itemize}
We can have in general \(p\)-dimensional orientifold planes, denoted
as \O{p} planes, and we can use the definition both in \tIIA (even \(p\)) 
and in \tIIB (odd \(p\)). In the following, we will restrict our analysis to the case of $p=4,6$ in \tIIA.

It is important to distinguish the action of the orientifold on the \ac{ns} and on the \ac{r} sectors.
They are two different $\mathbb{Z}_2$ parities:
\begin{itemize}
\item in the \ac{ns} sector, the $\mathbb{Z}_2$ action corresponds to a perturbative action on the string theory side and we  denote it by a \(\pm\) sign;
\item in the \ac{r} sector, the $\mathbb{Z}_2$ action is non-perturbative, and we denote it by a tilde (\(\sim\)) sign.
\end{itemize}
There are in total four possibilities, $Op^{\pm}$ and
$\widetilde{Op}^{\pm}$
\cite{Hanany:1999sj,Hanany:2000fq}.
By specifying the $\mathbb{Z}_2$ charges, we completely characterize the action 
of the orientifold on the gauge theory
from the string theory perspective (geometrically we specify the so-called ``discrete torsion'').

It is important to note that the orientifolds are charged under the \ac{r} sector, \emph{i.e.} they 
carry D~brane charge. This fact will be crucial when studying the
duality in terms of  \ac{hw}
transition, as will will see in the following.

\subsubsection{O4 planes}

In this section we add an \O4 plane on top of the stack of \D4~branes
that realizes the \ac{sqcd} model as discussed above.

There are some subtleties arising in this case:
\begin{itemize}
\item The charge of the \O{p} plane in the \ac{r} sector enters the
  calculation of the linking number in the \ac{hw} transition. These
  charges are $c_R[\O4^{\pm}] = \pm 1/2$.  This forces
  the number of \D4~branes to be even.
\item In presence of an $\widetilde{O4}^{-}$, the
  orientifold charge has an extra $1/2$ factor. This forces the
  number of \D4 gauge~branes to be odd, $2N_c+1$. %
\end{itemize}
Putting these things together, we find that for \(\O4^- \) we have
gauge group \(SO(2N_c)\), for \(\O4^+\) we have \(Sp(2N_c)\), for
\(\widetilde{\O4}^-\) we have \(SO(2N_c+1)\) and for
\(\widetilde{\O4}^+\) we have again \(Sp(2N_c)\) but with a different
non-perturbative sector.

One more subtlety comes from the \ac{ns} sector: the \ac{ns} charge of the \O{p} plane has to be flipped when
  crossing an \NS{}~brane, so that an $\O4^+$ plane on the left of an \NS{}
 ~brane is equivalent to an $O4^-$ plane on the right.

We can now proceed with the duality, by performing the \ac{hw} transition.
When a \D6~brane crosses an \NS{}~brane,  the number of \D4 that is created is shifted by
 $\pm 4$ depending on the $\pm$ orientifold charge.
This reproduces the field theory results $Sp(2N_c) \rightarrow Sp(2(N_f-N_c-2)$
and $SO(N_c) \rightarrow SO(N_f-N_c+4)$ as in Sec.~\ref{sec:4d-extensions}.

\subsubsection{O6 planes}

Adding \O6~branes to the picture proves very useful in the construction of models with tensor matter but requires some
modifications to the setup discussed above.

Consider for example a configuration with $2 N_c$ \D4~branes and two
\NS{}~branes, that we refer to as \(\NS{}_{\pm \theta}\), rotated by an angle $\theta$
in the $x_{45}$ and in the $x_{89}$ plane.
Similarly, we can also add $\D6_{\pm \theta}$~branes.
\begin{figure}
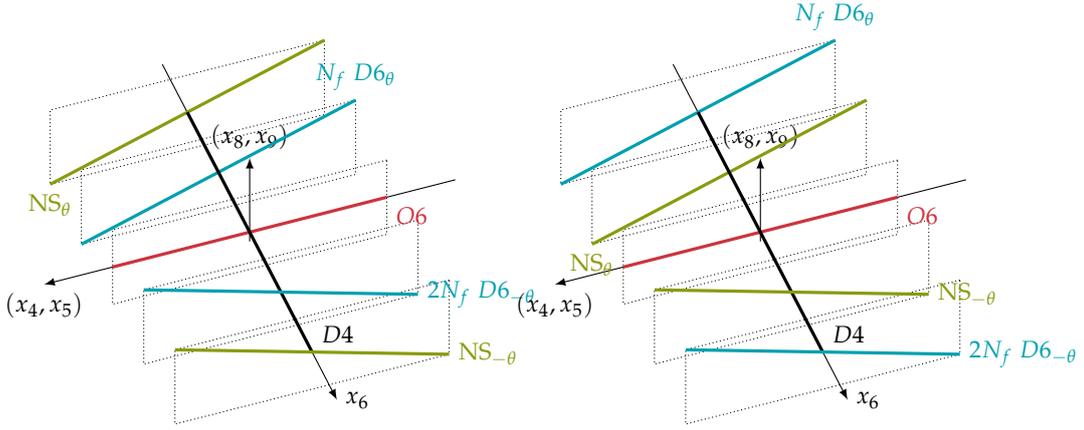

  \centering
  \begin{footnotesize}
    \FigureSOSpOsix
  \end{footnotesize}
  \caption{\ac{hw} transition in the presence of an \O6 plane.}
  \label{fig:HW-O6}
\end{figure}

One possibility corresponds to considering the~branes as shown in Figure~\ref{fig:HW-O6}.
In this case, we can place the O6 plane symmetrically as in the Figure.
In this case the original theory corresponds to $SU(2N_c)$ \ac{sqcd} with $2F$ flavors.
Depending on the charge of the \O6 plane (here $c_R[O6^{\pm}] =\pm 2$)
we obtain a real or a symplectic gauge group.
Again we can perform the \ac{hw} transition and obtain the expected result
for the rank of the dual gauge group.

\subsection{Realization of other dualities}
\label{sec:other-realizations}

We have now collected all the ingredients needed to represent the models described on the field 
theory side in terms of their~brane realizations.

Real gauge groups and tensor matter require the presence of \NS{}, \D4,
\D6 and \O{p} planes.
The dual model is obtained by performing an \ac{hw} transition, exchanging the
\NS{}~branes.
We will not discuss in detail all of these examples, and refer the reader to the original
literature~\cite{
Brodie:1997sz,Csaki:1998mx,Ahn:1998rj,Landsteiner:1998pb} for their explicit embeddings.

\begin{figure}
  \centering
  \begin{tikzpicture}
  \begin{scope}[shift={(-2,-3.5)}]
    \begin{footnotesize}
      \draw[->] (0,3) -- (1,3);
      \draw[->] (.2,2.8) -- (.2,3.8);
      \draw (1,3.2) node[]{\(x_6\)};
      \draw (.2,4) node[]{\(x_{4,5}\)};
    \end{footnotesize}
  \end{scope}
  
  \begin{scope}[thick]
    \begin{footnotesize}
      \draw [NScol] (0,3) -- (0,8);
      \draw [NScol, dashed] (0,0) -- (0,3);
      \draw [NScol] (-0.5,0) node[]{\NS5'};
      
      \draw (0,7) -- (2.5,7);
      \draw (1.5,6.5) node[]{\( r_2\) \D4};
      \draw[fill=black] (2.5,7) node{} circle (5pt);
      \draw (2.5,7.5) node[]{\NS5};
      \draw[fill=white] (0,7) circle (5pt);
      \draw (0,7) node[cross=4pt]{};
      \draw (-0.75,7) node[]{\( F\) \D6};
      
      \draw (0,5) -- (2.5,5);
      \draw (1.5,4.5) node[]{\( r_1\) \D4};
      \draw[fill=black] (2.5,5) node{} circle (5pt);
      \draw (2.5,5.5) node[]{\NS5};
      \draw[fill=white] (0,5) circle (5pt);
      \draw (0,5) node[cross=4pt]{};
      \draw (-0.75,5) node[]{\( F\) \D6};
      
      \draw[O3col] (-.5,3) -- (3,3);
      \draw (3,3) node[right=1ex] {\(\O4^-\)};
      \draw (1.5,2.5) node[]{\(2 r_0\) \D4};
      \draw[fill=black] (2.5,3) node{} circle (5pt);
      \draw (2.5,3.5) node[]{\NS5};
      \draw[fill=white] (0,3) circle (5pt);
      \draw (0,3) node[cross=4pt]{};
      \draw (-0.75,3) node[]{\(2 F\) \D6};

      \draw[dashed] (0,1) -- (2.5,1);
      \draw[fill=black] (2.5,1) node{} circle (5pt);
      \draw (2.5,1.5) node[]{\NS5};
      \draw[fill=white] (0,1) circle (5pt);
      \draw (0,1) node[cross=4pt]{};
      \draw (-0.75,1) node[]{\( F\) \D6};

      \node at (1,-1) {(a)};
    \end{footnotesize}
  \end{scope}

  \begin{scope}[shift={(7,0)},thick]
    \begin{footnotesize}

      \draw[NScol] (2.5,3) -- (2.5,8);
      \draw [NScol, dashed] (2.5,0) -- (2.5,3);
      \draw [NScol] (3,0) node[]{\NS5'};
      
      \draw (0,7) -- (2.5,7);
      \draw (1.2,6.5) node[]{\(F - r_2\) \D4};
      \draw[fill=black] (0,7) node{} circle (5pt);
      \draw (1.9,7.5) node[]{\(F\) \D6};
      \draw[fill=white] (2.5,7) circle (5pt);
      \draw (2.5,7) node[cross=4pt]{};
      \draw (0,7.5) node[]{\NS5};
      
      \draw (0,5) -- (2.5,5);
      \draw (1.2,4.5) node[]{\(F - r_1\) \D4};
      \draw[fill=black] (0,5) node{} circle (5pt);
      \draw (1.9,5.5) node[]{\(F\) \D6};
      \draw[fill=white] (2.5,5) circle (5pt);
      \draw (2.5,5) node[cross=4pt]{};
      \draw (0,5.5) node[]{\NS5};

      \draw[O3col] (-.5,3) -- (3,3);
      \draw (3,3) node[right=1ex] {\(\O4^-\)};

      \draw (1.2,2.5) node[]{\(2(F - r_0)\) \D4};
      \draw[fill=black] (0,3) node{} circle (5pt);
      \draw (1.9,3.5) node[]{\(2F\) \D6};
      \draw[fill=white] (2.5,3) circle (5pt);
      \draw (2.5,3) node[cross=4pt]{};
      \draw (0,3.5) node[]{\NS5};
      
      \draw [dashed] (0,1) -- (2.5,1);
      \draw[fill=black] (0,1) node{} circle (5pt);
      \draw (1.9,1.5) node[]{\(F\) \D6};
      \draw[fill=white] (2.5,1) circle (5pt);
      \draw [] (2.5,1) node[cross=4pt]{};
      \draw (0,1.5) node[]{\NS5};

      \node at (1,-1) {(b)};
            
    \end{footnotesize}
  \end{scope}
  
\end{tikzpicture}
  \caption{Electric and magnetic sides of the duality for \(Sp(2N_c)\) gauge theories with adjoint matter.}
  \label{fig:Sp-adjoint}
\end{figure}
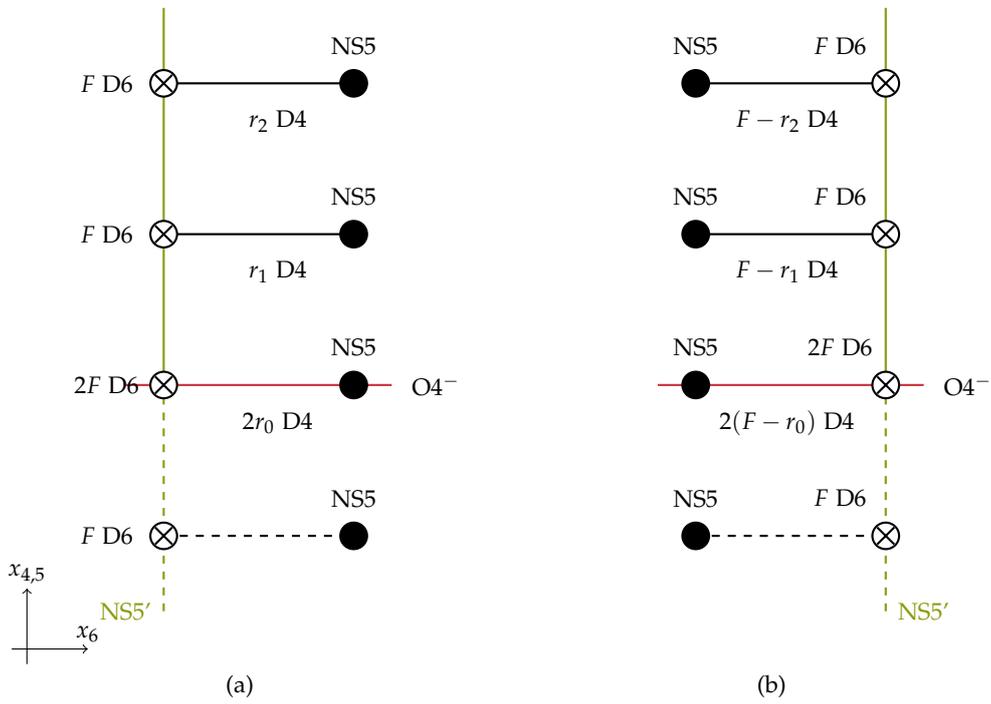

Here we will instead concentrate on a single case, which we will use as a toy example also when discussing the reduction
of the four-dimensional dualities to three dimensions.
It consists of an $Sp(2N_c)$ gauge group with an adjoint and $2N_f$ fundamentals.
This is the theory discussed in Section~\ref{sec:4d-extensions}.
This example can be engineered  with  $2N_c$ \D4~branes and an $O4^-$ plane stretched
between $2n + 1$ \NS{}~branes and one \NS{}'~brane. 
The fundamentals are obtained by considering $2N_f$ \D6~branes on the
\NS{}~branes.
The gauge group is broken by a polynomial superpotential for the adjoint into 
the product 
\begin{equation}
Sp(2 r_0) \times \prod_{i=1}^{n} U(r_i),\quad \text{with } \sum_{i=0}^n r_i = N_c.
\end{equation}
This polynomial deformation is obtained
in the~brane picture by  separating the \NS{}~branes along the $x_{45}$ plane (Figure~\ref{fig:Sp-adjoint}(a)).

The magnetic theory is obtained by a \ac{hw} transition in each sector
and eventually reconnecting the \NS5s, see Figure~\ref{fig:Sp-adjoint}.
Before reconnecting the~branes, the gauge group is broken to 
\begin{equation}
Sp(2 (N_f-r_0-2)) \times \prod_{i=1}^{n} U(N_f-r_i).
\end{equation}
In the final step the dual gauge group becomes
$Sp(2((2n+1)N_f-N_c-2))$ as expected (Figure~\ref{fig:Sp-adjoint}(b)).

\chapter{The three-dimensional case}\label{sec:3d}

In the following, we will repeat the four-dimensional discussion for the three-dimensional case, \emph{i.e.} we will first discuss the generalities of supersymmetric gauge theories with four supercharges in three dimensions and then specialize to the case of three-dimensional \ac{sqcd} with gauge group $U(N_c)$, following largely~\cite{Aharony:1997bx, deBoer:1997kr}. After this, we discuss the various three-dimensional dualities and go on to describe them in the~brane picture.

\section{$\mathcal{N}=2$ supersymmetric gauge theories in three dimensions }

In this section, we will discuss the basics of gauge theories with
four supercharges in three dimensions. This is the same number of
supercharges as for $\mathcal{N}=1$ in four dimensions, but in three dimensions, it corresponds to $\mathcal{N}=2$. These theories can be obtained by dimensional reduction from four dimensions. While theories with $\mathcal{N}=1$ have no holomorphy properties in three dimensions, for $\mathcal{N}=2$, there are holomorphic objects and non-renormalization theorems.

As the gauge coupling is dimensionful in three dimensions, none of the theories are conformal. Their \ac{ir} fixed points are in general strongly interacting theories, so when we study \ac{ir} dualities, we do not have effective \ac{ir} descriptions.
The analysis of the conformal window in $\mathcal{N}=2$ three-dimensional theories is more complicated
than for the four-dimensional case. This is mostly due to the fact that in three dimensions, there are strong coupling effects
in the \ac{ir} and many of the techniques applicable in the four-dimensional case are not available.
It has been nevertheless possible to gain new insights into the analysis of the scaling dimensions of the 
fields from localization techniques via the F-maximization principle~\cite{Jafferis:2010un,Jafferis:2011zi,Closset:2012vg,Pufu:2016zxm}.
One of the main issues in this analysis is the presence of accidental symmetries in the \ac{ir}~\cite{Morita:2011cs,Agarwal:2012wd}.
We refer the reader to~\cite{Safdi:2012re,Lee:2016zud} for an analysis of the conformal window and 
of the accidental symmetries for Aharony duality and its generalizations.

\subsection{Generalities}\label{sec:3dsusy}

A representation of the Clifford algebra in three dimensions with $\eta=(-,+,+)$ is given by
\begin{equation}
\gamma_{\alpha\beta}^i =(i \sigma_2, \sigma_3, \sigma_1),\quad i=0,1,2,
\end{equation}
with $\sigma_i$ the Pauli matrices. We see that here, the $\gamma^i$ are all real. The fundamental fermion representation in $2+1$ dimensions is a two-component Majorana fermion $\psi_i^\alpha$, $\alpha=1,2$.
When performing the dimensional reduction, the scalars remain scalars, $\phi(x^\mu) \to \phi(x^i)$, where $\mu=0,\dots ,3$ and $i=0,1,2$. The vector field reduces as
\begin{equation}
A_\mu(x^\mu) \to A_i(x^i), \ \sigma(x^i)=A_3(x^i),
\end{equation}
and a four-dimensional Weyl spinor $\psi$ with four real components turns into two independent Majorana spinors with two real components each, which can also be combined into a complex spinor.

The dimensional reduction of the four-dimensional  $\mathcal{N}=1$ supersymmetry algebra is given by
\begin{align}
\{ Q_\alpha, \overline Q_\beta \} &= 2\gamma^i_{\alpha\beta}\, P_i + 2i\,\epsilon_{\alpha\beta}Z,\\
\{ Q_\alpha,  Q_\beta \} &= \{ \overline Q_\alpha, \overline Q_\beta \} =0.
\end{align}
The central term $Z$ is given by the $P_3$ component of the four-dimensional momentum.
As in the four-dimensional case, the R--symmetry is isomorphic to a global $U(1)_R$. The operators $Q_\alpha,\, \overline Q_\alpha$ and $D_\alpha,\, \overline D_\alpha$ are defined from the $\theta,\, \bar\theta$ analogously to Eq.~\eqref{eq:Q} and \eqref{eq:D}.
\medskip
In three dimensions, there are chiral superfields $Q$ satisfying $\overline D_\alpha Q=0$, anti-chiral fields $\overline Q$ satisfying $D_\alpha \overline Q=0$ and vector superfields $V$ satisfying $V=V^\dagger$. Both chiral and vector multiplets contain two real bosonic degrees of freedom and two Majorana fermionic degrees of freedom. In addition, there can be \emph{linear superfields} $\Sigma$ which satisfy
\begin{equation}
\epsilon_{\alpha\beta} D_\alpha D_\beta \Sigma = \epsilon_{\alpha\beta} \overline D_\alpha \overline D_\beta \Sigma = 0.
\end{equation}
The lowest component of $\Sigma$ is a real scalar.

\medskip
The four-dimensional chiral superfield Eq.~\eqref{eq:chiral4} reduces straightforwardly as discussed. In the reduction of the four-dimensional vector field Eq.~\eqref{eq:vec}, the real scalar $\sigma$ appears:
\begin{equation}\label{eq:vec3}
V = - i\theta\bar\theta\sigma - \theta\gamma^i\bar\theta A_i+i\theta^2\bar\theta\bar\lambda+i\bar\theta^2\theta\lambda+\tfrac{1}{2}\theta^2 \bar\theta^2 D.
\end{equation}
Unlike in the four-dimensional theory, $\sigma$ can acquire a \ac{vev} that breaks the gauge group to its maximal Abelian torus, leading to a Coulomb branch of the moduli space.

\medskip
In three dimensions, in a free Abelian theory with gauge coupling $e_3$, the photon $A_i$ can be dualized into a scalar $\gamma$ via
\begin{equation}\label{eq:dualphoton}
\del_i \gamma =\frac{\pi}{e_3^2} \epsilon_{ijk}F^{jk},
\end{equation}
where $F^{jk}$ is the field strength of $A_i$. The quantization of the magnetic charge requires that $\gamma$, which is known as the \emph{dual photon}, is periodic, $\gamma \sim \gamma + 2\pi$. The dual field strength $J=\epsilon_{ijk}F^{jk}$ is a one-form. From Maxwell's, equation, we see that it is divergence-free, $\del^iJ_i=0$. Three-dimensional theories have therefore a global topological symmetry $U(1)_J$ with conserved current $J$.

It is possible to describe the vector superfield equivalently with a linear superfield $\Sigma$:
\begin{equation}
  \begin{aligned}
    \Sigma &= -\tfrac{i}{2} \epsilon^{\alpha\beta} \overline D_\alpha D_\beta V \\
    &= \sigma + \theta\bar\lambda+\bar\theta\lambda + \tfrac{1}{2}
    \theta\gamma^i\bar\theta J_i+i\theta\bar\theta
    D+\tfrac{i}{2}\bar\theta^2\theta \gamma^i\del_i\lambda
    -\tfrac{i}{2}\theta^2\bar\theta \gamma^i\del_i\bar\lambda
    +\tfrac{1}{4}\theta^2\bar\theta^2\del^2\sigma.
  \end{aligned}
\end{equation}
It is furthermore possible to dualize this linear multiplet into a chiral multiplet $\Phi$ with lowest component
\begin{equation}
\phi=\frac{2\pi}{e_3^2} \sigma + i\gamma.
\end{equation}

\medskip
Having collected all the necessary ingredients, we can now write down invariant Lagrangians in three dimensions.

\paragraph{Gauge fields.}
For the gauge part, we can again form the gauge invariant field strength given in Eq.~\eqref{eq:gfs}, resulting in
\begin{equation}
  \begin{aligned}
    \mathcal{L}_{\mathrm{gauge}} &= \frac{1}{e_3^2} \int \dd^2 \theta\, W_\alpha W^\alpha + \mathrm{c.c}\\
    &= \frac{1}{e_3^2} \Tr\left( \tfrac{1}{4}F^{ij}F_{ij}+D_i\sigma
      D^i\sigma+D^2+\lambda^*\gamma^i\D_i\lambda \right).
  \end{aligned}
\end{equation}
Equivalently, we can use the description in terms of the linear superfield $\Sigma$:
\begin{equation}
\mathcal{L}_{\mathrm{gauge}} = \frac{1}{e_3^2} \int \dd^4\theta \, \Sigma^2.
\end{equation}
Unlike in four dimensions, the above are not the only gauge-invariant combinations of the gauge fields. It is possible to add a Chern--Simons term to the Lagrangian:
\begin{equation}
\mathcal{L}_{\mathrm{CS}} = \frac{k}{4\pi}\Tr\left( \epsilon^{ijk}(A_i\del_jA_k+\tfrac{2i}{3}A_iA_jA_k)+2D\sigma-\lambda^*\lambda\right),
\end{equation}
where $k\in \mathbb{Z}$ is the Chern--Simons level.
In the Abelian case, the above can be rewritten very compactly as
\begin{equation}
\mathcal{L}_{\mathrm{CS}} =  \frac{k}{4\pi}\int \dd^4 \theta \,\Sigma V.
\end{equation}

\medskip
As in four dimensions, we can add a Fayet--Iliopoulos term which has the same form as Eq.~\eqref{eq:FI} whenever the gauge group contains a $U(1)$ factor. 

\paragraph{Matter fields.} The Lagrangian for the chiral superfields $Q$ is given by
\begin{equation}
\mathcal{L}_{\mathrm{matter}} = \int \dd^4 \theta K(Q,Q^\dagger) + \int \dd^2 \theta \left( W(Q) + \mathrm{c.c.}\right). 
\end{equation}
For supersymmetric theories, the Kähler potential takes the usual form
\begin{equation}
K=Q\,e^VQ^\dagger,
\end{equation}
so in terms of component fields, we have
\begin{multline}\label{eq:3dmatter}
  \mathcal{L}_{\mathrm{matter}} = \abs{D_i\phi_Q}^2 + \phi^*_Q\sigma^2 \phi_Q + i \phi^*_Q D\phi_Q + i \psi^*\gamma^i D_i\psi + \\ - i \psi^* \sigma \psi + i \phi^*_Q \lambda^* \psi - i \psi^* \lambda \phi_Q+  \abs{F}^2.
\end{multline}
We see that for $\langle\sigma\rangle\neq0$, a mass term is induced for $Q$.

\paragraph{Real and complex mass terms.}
In three dimensions, different types of mass terms can be written for a chiral superfield $Q$. In the vector-like theory, we can add the term
\begin{equation}
W_{mc}=m_{\mathbb{C}}Q \overline Q
\end{equation}
to the superpotential. Since $m_{\mathbb{C}}$ is complex, this is called a \emph{complex mass term}. It is the analog of the usual mass term in four dimensions.

Alternatively, we can induce a mass term via $\langle\sigma\rangle\neq0$. This is called a \emph{real mass term} and can be understood as a modification of the Kähler potential:
\begin{equation}
\int \dd^4 \theta \, Q\,e^{m_{\mathbb{R}}\theta^2} Q^\dagger \sim \left( \frac{m_{\mathbb{R}}}{2} |\phi_Q|^2+i m_{\mathbb{R}} \epsilon^{\alpha\beta}\bar\psi_\alpha\psi_\beta \right).
\end{equation}
Note that the complex mass term preserves parity, while the real real term breaks parity. The physical mass of the chiral superfield is given by the combination
\begin{equation}
m=\sqrt{m_{\mathbb{R}}^2+m_{\mathbb{C}}^2}.
\end{equation}
A real mass term can be induced by turning on a vector superfield $\tilde V$ as a susy-preserving background field,
\begin{equation}
\tilde\sigma=\frac{m_{\mathbb{R}}}{e_3},\ \tilde A_i=\tilde\lambda=\tilde{\bar\lambda}=\tilde D=0.
\end{equation}
Note, that since the real masses come from a background vector multiplet, they cannot appear in holomorphic objects such as the superpotential.

\paragraph{Moduli space.} The moduli space of vacua of three-dimensional $\mathcal{N}=2$ supersymmetric gauge theories contains in general a Coulomb branch and a Higgs branch. By $\mathcal{N}=2$ supersymmetry, both moduli spaces are Kähler manifolds, but neither is protected against loop or perturbative corrections. The scalars of the quark multiplets $\phi_Q$ parameterize the Higgs branch, while \ac{vev}s of $\sigma$ parameterize the Coulomb branch. Due to the term $\phi^*_Q\sigma^2 \phi_Q$ appearing in the scalar potential (see the matter action Eq.~\eqref{eq:3dmatter}), $\langle \phi_Q \rangle\neq 0$ in general requires $\langle \sigma \rangle=0$ and vice versa. From this, we see that the moduli space splits up into distinct branches.

The low-energy effective action has the form
\begin{equation}
S_{\mathrm{eff}} = \int \dd^3x\int \dd^4 \theta\, K(\Phi_i,\bar\Phi_i) + \int \dd^2 \theta \left( W(\Phi_i) + \mathrm{c.c.}\right). 
\end{equation}
On the Higgs branch, the $\Phi_i$ are suitable gauge-invariant combinations of the matter fields $Q,\,\widetilde Q$.
At a generic point of the Coulomb branch, the gauge group is broken to its maximal torus, as the scalar $\sigma$ is in the adjoint representation of $G$ and a generic \ac{vev} breaks $G \to U(1)^{\mathrm{rk}(G)}$.

The action contains the terms 
\begin{equation}
\Tr[A_i,A_j] \quad \text{and} \quad \Tr[A_i,\sigma].
\end{equation}
For a supersymmetric vacuum, they must be zero, which is the case if we choose both $A_i$ and $\sigma$ to lie in the Cartan subalgebra of $G$. 

We cannot dualize $A_i$ to $\gamma$ for non-Abelian gauge theories, but at a generic point of the Coulomb branch, this is nonetheless possible. We can dualize the $\mathrm{rk}(G)$ massless vector multiplets into chiral multiplets $Y_n$, $n=1,\dots,\mathrm{rk}(G)$. The low-energy theory at a generic point of the Coulomb branch thus includes these $\mathrm{rk}(G)$ chiral multiplets $Y_n$. The classical Coulomb branch can be parameterized by the \ac{vev}s of $\sigma_n,\gamma_n$, $n=1,\dots,\mathrm{rk}(G)$, or by the chiral operators
\begin{equation}
Y_n\sim \exp\left( \frac{\sigma_n}{\tilde e_3^2}+i\gamma_n\right), 
\end{equation}
where we choose $\sigma_1\geq\dots\geq\sigma_{\mathrm{rk}(G)}$ in order to fix the Weyl freedom.

The metrics of the classical moduli spaces receive both loop and non-perturbative corrections. A superpotential can only be generated non-perturbatively via instanton effects. These instantons, which correspond to monopoles in four dimensions, can for non-Abelian cases lift most or all of the Coulomb branch, and sometimes also the Higgs branch.

It has been shown that for a pure $SU(2)$ \ac{sym}, the \ac{vev} of $Y_1$ classically breaks $SU(2)$ to $U(1)$. In four dimensions, this breaking is associated to a static monopole solution, while in three dimensions, it corresponds to an instanton. This instanton generates the \ac{ahw} superpotential 
\begin{equation}
W_{\mathrm{AHW}}=\frac{1}{Y_1}.
\end{equation}
In pure $U(N)$ or $SU(N)$ \ac{sym}, the same effect happens whenever two eigenvalues of $\sigma$ approach each other, resulting in an unbroken $SU(2)$. In these cases, the superpotential generated by the instantons takes the form
\begin{equation}
W_{\mathrm{AHW}}=\sum_{n=1}^{N-1}\frac{1}{Y_n}.
\end{equation}
This superpotential lifts the entire Coulomb branch. When matter multiplets are present, this behavior is modified as extra fermion zero modes can prevent instantons from generating a superpotential. The zero-mode counting analysis  can be performed using the Callias index theorem~\cite{Callias:1977kg} (see also \cite{deBoer:1997kr} for a 
more physical approach).
The global charges of the monopole operators can be computed following~\cite{Benna:2009xd}.

\paragraph{Parity anomaly.} Note that the Chern--Simons terms breaks parity. This is referred to as a \emph{parity anomaly} and is related to an induced term at one loop with charged fermions running in the loop \cite{Niemi:1983rq,Redlich:1983dv}. 

Let us consider a $U(1)^r$ gauge theory. The Chern--Simons term takes the form
\begin{equation}
\sum_{m,n=1}^r k_{mn}\int \dd^4\theta\ \Sigma_m V_n.
\end{equation}
If we integrate out the charged fermions, an additional contribution to the Chern--Simons term is induced:
\begin{equation}
k_{mn} + \tfrac{1}{2} \sum_Q (q_Q)_m(q_Q)_n\,\mathrm{sign}(m_Q) \quad \in \mathbb{Z},
\end{equation}
where the sum runs over all fermions in the chiral matter multiplets $Q$, $(q_Q)_n$ is the integer charge of the fermion $\phi_Q$ under $U(1)_n$, and $m_Q$ is the real mass of $\phi_Q$, \emph{i.e.}
\begin{equation}
m_Q = m_{\mathbb{R}}+\sum_{n=1}^r(q_Q)_n\sigma_n.
\end{equation}
If $\sum_Q (q_Q)_m(q_Q)_n$ is odd, $k_{ij} \neq 0$ and the parity is broken.
An induced Chern--Simons term can be avoided by choosing the matter content such that the parity anomaly does not arise.

A similar parity anomaly arises for non-Abelian theories.

\subsection{SQCD with $G=U(N_c)$}\label{sec:3dsqcd}

Here we will study three-dimensional \ac{sqcd} with gauge group $U(N_c)$. The matter consists of $N_f$ chiral multiplets $Q_i$ in the fundamental representation $\mathbf{N}_c$ and  $N_f$ chiral multiplets $\widetilde Q^{\tilde i}$ in the anti-fundamental representation $\overline{\mathbf{N}}_c$. As the matter multiplets are paired, this choice of matter fields does not lead to a parity anomaly and thus no Chern--Simons term is present.

A topological global symmetry similar to the $U(1)_J$ in Abelian gauge theories can be formed from the $U(1)$ gauge group contained in $U(N_c)$. It acts by shifting the dual photon, such that the operators $Y_i$ are charged under it.

\medskip
For $G=U(N_c)$, the adjoint scalar $\sigma$ can be diagonalized to 
\begin{equation}
\sigma=\mathrm{diag}(\sigma_1,\dots,\sigma_{N_c}).
\end{equation}
On generic points of the Coulomb branch, we can as discussed above, dualize the massless gauge fields into $N_c$ dual photons $\gamma_i$. 

After taking into account the instanton effects, only the part of the Coulomb branch remains on which
\begin{equation}
\sigma_1>0=\sigma_2=\dots=\sigma_{N_c-1}>\sigma_{N_c}.
\end{equation}
The Coulomb branch that remains after instanton corrections is thus a two-di\-men\-sional manifold and is parameterized by
\begin{align}
T &=\exp\left( \frac{\sigma_1}{\tilde e_3^2}+i\gamma_1\right), &\widetilde T &=\exp\left( -\frac{\sigma_{N_c}}{\tilde e_3^2}+i\gamma_{N_c}\right).
\end{align}
The monopole operators $T (\widetilde T)$ carry charge $1(-1)$ under the $U(1)_J$ symmetry.

The Higgs branch of $\mathcal{M}_{q}$ is parameterized by the gauge-invariant superfields $M_i^{\tilde i}=Q_i\widetilde Q^{\tilde i}$.

Additional non-perturbative effects lifting the Coulomb branch come into play for $N_f < N_c-1$. At the end of the day, the quantum moduli space $\mathcal{M}_{q}$ is very similar to its four-dimensional counterpart:
\begin{itemize}
\item $N_f < N_c-1$: the classical moduli space is lifted completely, there is no supersymmetric vacuum.
\item $N_f = N_c-1$: a smooth quantum moduli space exists.
\item $N_f = N_c$: a quantum moduli space exists with a dual description at the origin using only chiral multiplets.
\item $N_f \geq N_c+1$: a quantum moduli space exists, but it has a singularity at the origin.
\end{itemize}
Under the (quantum corrected) global symmetries, the fields transform as given in Table~\ref{tab:global3d}\footnote{Note, that often, the choice $\Delta=0$ is used in the literature.}
\begin{table}[]
  \centering
  \begin{tabular}{cccccc}
    \toprule
                                & $SU(N_f)_L$    & $SU(N_f)_R$               & $U(1)_J$ & $U(1)_A$ & $U(1)_R$               \\ \midrule
    $Q^i_a$                     & $\mathbf{N}_f$ & $\mathbf{1}$              & $0$      & $1$      & $ \Delta $             \\
    $\widetilde Q_{\tilde i}^a$ & $\mathbf{1}$   & $\overline{\mathbf{N}}_f$ & $0$      & $1$      & $ \Delta $             \\ 
    $M^i_{\tilde i}$            & $\mathbf{N}_f$ & $\overline{\mathbf{N}}_f$ & $0$      & $2$      & $ 2\Delta $            \\
   $T$                                 & $\mathbf{1}$   & $\,\mathbf{1}$              & $1$   & $-N_f$   & $N_f(1-\Delta)-N_c+1 $ \\ 
  $\widetilde T$                   & $\mathbf{1}$   & $\mathbf{1}$              & $-1$   & $-N_f$   & $N_f(1-\Delta)-N_c+1 $ \\
    \bottomrule
  \end{tabular}
  \caption{Transformation properties of the fields under the global symmetries.}
  \label{tab:global3d}
\end{table}

\section{Dualities in three dimensions}\label{sec:3ddualities}

In this section we discuss some general aspects of dualities between
three-di\-men\-sion\-al $\mathcal{N}=2$
theories. We distinguish three classes of dualities:
\begin{itemize}
\item Aharony-like dualities
\item Giveon--Kutasov-like dualities
\item Mirror symmetry.
\end{itemize} 
The name of the first two classes originates from the prototypical examples of those dualities,
\emph{i.e.} the duality found by Aharony in~\cite{Aharony:1997gp} and the one found by Giveon and Kutasov in~\cite{Giveon:2008zn}.
The main difference between these dualities consists in the presence of a \ac{cs} term 
in the action of the latter type. %
As we will see later, the two dualities can be connected via an \ac{rg} flow
triggered by a real mass term.
The third class of dualities is a generalization of $\mathcal{N}=4$ mirror symmetry to $\mathcal{N}=2$. This last duality will play an interesting role in our discussion. 

In the first part of this section, we will review the main aspects of the three dualities just introduced,
while in the second part we will enumerate the generalizations of Aharony and Giveon--Kutasov 
duality that have been found in the literature.
We will not review the dualities among theories with chiral matter content
described in~\cite{Benini:2011mf}. 

To conclude, observe that in three dimensions, there is no notion of chirality. When we 
call a three-dimensional theory chiral, we are referring to its four-dimensional parent theory.
In the following, we will refer to a four-dimensional theory as a parent theory 
of a three-dimensional theory if they share the same field content,
\emph{i.e.} the same amount of $\mathcal{N}=1$ vector and matter multiplets, and the same interactions among the
matter fields. In this identification, we do not refer to the gauge interactions and to the Coulomb branch of the three-dimensional
theory. 
We call for example four-dimensional $\mathcal{N}=1$ $U(N_c)$ \ac{sqcd} with $N_f$ pairs of 
fundamental and antifundamental quarks the parent theory of both $U(N_c)_k$ and $U(N_c)_0$
three-dimensional $\mathcal{N}=2$ \ac{sqcd} with $N_f$ pairs of fundamental and antifundamental quarks
and $W=0$.

\subsection{Aharony duality}

This duality was originally discussed in~\cite{Aharony:1997gp}. The electric theory
consists of a $U(N_c)$ gauge theory with $N_f$ pairs 
of fundamental and antifundamental quarks.

Observe that, differently from the four-dimensional Seiberg duality, in this case the 
$U(1)$ factor of the gauge group does not decouple in the \ac{ir}.
In other words, it is possible to consider a gauging of the baryonic symmetry.\footnote{There is also 
a version of this duality with an $SU(N)_c)$ gauge group
that we will not discuss here~\cite{Park:2013wta}.}
The dual theory has gauge group $U(N_f-N_c)$, $N_f$ pairs
of dual fundamentals $q$ and antifundamentals $\tilde q$
and a singlet $M$ corresponding to the meson of the electric theory.
The superpotential term $W=M q \tilde q$ is present also in this case.

\medskip
So far we have presented a version of the duality identical to four-dimensional
Seiberg duality in presence of a gauging of the baryonic symmetry.
Nevertheless, in the three-dimensional case this is not the end of the story.
One can see that something is missing \emph{e.g.} by looking at the
global symmetries. In the three-dimensional case, there are two global currents that do not arise in four dimensions.
The first one is the axial symmetry, $U(1)_A$: while this symmetry is anomalous in four dimensions, the absence of 
global anomalies in three dimensions forces us to consider it here.
We will see that this symmetry plays a crucial role in the reduction of Seiberg duality to three dimensions and also 
its interpretation at the~brane level will look quite interesting.

The second symmetry that one has to consider in three dimensions is a topological symmetry,
\emph{i.e.} a symmetry conserved because of the Bianchi identity.
This topological $U(1)_J$ symmetry does not charge the elementary fields but it shifts the dual photon
associated to the monopole operators.
Let us denote the monopole operators with flux $(\pm1,0,\dots,0)$ of the electric theory as
$T$ and $\widetilde T$ and the one of the magnetic theory as $t$ and $\tilde t$.
It has been argued in~\cite{Aharony:1997gp} that the monopoles $t$ and $\tilde t$ in the dual theory have to be set to zero 
in the chiral ring for the duality to hold. This can be achieved by adding a superpotential 
\begin{equation}
\Delta W = t T + \tilde t \widetilde T.
\end{equation}
where the electric monopoles $T$ and $\widetilde T$ are treated as singlets in the dual theory, imposing the
constraints on $t$ and $\tilde t$ from their F-terms.
Observe that the monopole operators $t$ and $\tilde t$ of the dual theory are not elementary degrees of freedom, \emph{i.e.}
the above term does not have to be interpreted as a mass term but has the role of enforcing the constraints on the chiral ring as discussed above.

We conclude this review of the duality with a table of the
charges of the various fields in the magnetic
phase (Tab.~\ref{tab:field-charges}).
\begin{table}[]
  \centering
  \begin{tabular}{cccccc}
    \toprule
                            & $SU(N_f)_L$               & $SU(N_f)_R$               & $U(1)_J$ & $U(1)_A$ & $U(1)_R$               \\ \midrule
    $q^i_a$                 & $\overline{\mathbf{N}}_f$ & $\mathbf{1}$              & $0$      & $-1$     & $1- \Delta $           \\
    $\tilde q_{\tilde i}^a$ & $\mathbf{1}$              & ${\mathbf{N}}_f$          & $0$      & $-1$     & $1- \Delta $           \\ 
    $M^i_{\tilde i}$        & $\mathbf{N}_f$            & $\overline{\mathbf{N}}_f$ & $0$      & $2$      & $ 2\Delta $            \\
    $t$                      & $\mathbf{1}$              & $\mathbf{1}$              & $-1$   & $N_f$    & $N_f(\Delta-1)+N_c+1 $ \\
    $\tilde t$               & $\mathbf{1}$              & $\mathbf{1}$              & $ 1$   & $N_f$    & $N_f(\Delta-1)+N_c+1 $ \\
    \bottomrule
  \end{tabular}
  \caption{Transformation properties of the magnetic fields under the global symmetries.}
  \label{tab:field-charges}
\end{table}

\subsection{Giveon--Kutasov duality}
\label{GKdualaity}
This duality was originally worked out in~\cite{Giveon:2008zn} using a~brane construction which we will discuss in Sec.~\ref{subsec:GK}.
The electric and the magnetic sides of this duality are as follows.
The electric theory has gauge group $U(N_c)_k$, where $k$ represents the integer \ac{cs} level.
There are again $N_f$ pairs of fundamentals $Q$ and antifundamentals $\widetilde Q$ in the spectrum.
The superpotential is
\begin{equation}
\label{eleGK}
W=0.
\end{equation} 
The dual theory in this case has gauge group $U(N_f-N_c+|k|)_{-k}$, $N_f$ 
pairs of fundamentals $q$ and antifundamentals $\tilde q$ in the action
and mesons $M_i^{\tilde j}=Q_i \widetilde Q^{\tilde j}$, interacting through a superpotential 
\begin{equation}
\label{magGK}
W = M q \tilde q.
\end{equation}

\subsection{There and back again}\label{sec:thereback}

Aharony duality and Giveon--Kutasov duality are related and 
it has been shown that they can be connected via a real mass flow in both directions.

The first flow (from Aharony to Giveon--Kutasov) is quite easy to understand, it has been first discussed in~\cite{Willett:2011gp}.
Here one turns on a background gauge field for the global symmetries
and assigns a (large) \ac{vev} to the background scalars $\sigma$.
This turns into a large real mass for some of the fundamental fields that are integrated
out. Integrating out these fields generates \ac{cs} levels and contact terms (or \textsc{bf} couplings).
More in detail, we start from $U(N_c)$ \ac{sqcd}  with $N_f+k$ pairs of fundamentals and antifundamentals
and assign a large real mass to $k$ fundamentals and $k$ antifundamentals.
These large real masses have to be chosen with the same sign. By integrating out the 
massive fermions the theory becomes three-dimensional $\mathcal{N}=2$ $U(N_c)_{\pm k}$ \ac{sqcd} with 
$N_f$ pairs of fundamentals and antifundamentals, with vanishing superpotential,
\emph{i.e.} the electric theory of the Giveon--Kutasov duality.
The sign of the \ac{cs} level $\pm k$ depends on the sign of the real masses.
The same flow has to be realized on the dual side. In this case, the dual theory
before turning on the background gauge fields is 
$U(N_f+k-N_c)$ with $N_f+k$  pairs of fundamentals and antifundamentals, 
$q$ and $\tilde q$ interacting with a meson with $(N_f+k)^2$ component.
One has to turn on the background gauge fields for the global symmetries
consistently with the electric theory. This implies that  $k$ pairs of
fundamentals and antifundamentals acquire a large real mass, as
opposed to the single one of the
electric fields. There are $N_f^2$ components left in the mesonic field and 
the electric monopoles are massive.
By integrating the massive fermions out a \ac{cs} level $\mp k$ is generated.
The dual gauge group is $U(N_f+k - N_c)_{\mp k}$.  
At the end of the day, the electric gauge group is $U(N_c)_k$
and the magnetic one is $U(N_f+|k| - N_c)_{-k}$,
with $k \neq 0$. This corresponds precisely to the duality of Giveon and Kutasov.

The flow from  Giveon--Kutasov duality to Aharony's is more
involved and it requires the notion of topological vacua, discussed
in~\cite{Intriligator:2013lca}. Let us review briefly this
construction. In this case one starts with $U(N_c)_k$ with $N_f+k$
pairs of fundamentals and antifundamentals\footnote{The case with
  $k=1$ was studied in~\cite{Intriligator:2013lca}, and it was
  generalized to higher $k$ in~\cite{Khan:2013bba,Amariti:2013qea}.}.
One can integrate the $k$ pairs of fundamentals and antifundamentals
by turning on some background gauge fields as before. The theory
becomes $U(N_c)_0$ if the sign of these masses is chosen such that the
\ac{cs} term is canceled. The novelty is in the dual flow: in this case the duality
is obtained not only by turning on the masses in the dual phases
correctly, but we also need a Higgsing of the gauge group. This is
due to the presence of a non-trivial \ac{vev}
for the real scalars in the $U(N_f+|k|-N_c)_0$ dual theory. After the
flow, the dual gauge group becomes $U(N_f-N_c) \times U(k)^2$. By
integrating out the massive fermions and keeping the massless matter
fields, the first sector has vanishing \ac{cs} level with $N_f$ pairs of
fundamentals and antifundamentals, while the two $U(k)$ sector have \ac{cs}
level $k/2$, plus $k$ massless fundamenentals and no
antifundamentals. In a slight abuse of language, borrowed from the
four-dimensional case, we can can think of them as
chiral theories. The $U(k)_{k/2}$ theories can be dualized to
singlets, and these singlets interact with the monopoles of the
$U(N_f-N_c)_0$ theory through an \ac{ahw} superpotential. This
completely reconstructs the dual phase of the Aharony duality.

\subsection{Mirror symmetry}

Three-dimensional mirror symmetry, in its original formulation, is a duality 
of theories with $\mathcal{N}=4$ supersymmetry~\cite{Intriligator:1996ex}.
It essentially exchanges the Higgs and the Coulomb branches of moduli space.  At the level of the 
algebra, it exchangers the two $SU(2)$ R--symmetry groups inside 
$SO(4)_R = SU(2)_L \times SU(2)_R$.
The origin of this duality can be easily visualized in the~brane description: it corresponds
to exchanging the NS and the R sector, \emph{i.e.} \NS5 and \D5~branes. In this sense, mirror symmetry is a consequence of S-duality.

Here, we are interested in a less supersymmetric version of mirror symmetry, \emph{i.e.} in theories with 
$\mathcal{N}=2$ supersymmetry
\cite{deBoer:1996mp,deBoer:1996ck,deBoer:1997ka,Aharony:1997bx,Kapustin:1999ha}. In this case, the action of mirror symmetry is less clear than
above, as the R-symmetry group is Abelian and there are mixed branches in the moduli space. 
However, we can still think of the action of this symmetry as exchanging the R and the NS sectors.
The prototypical example of $\mathcal{N}=2$ mirror symmetry  corresponds to the duality between 
$\mathcal{N}=2$ \ac{sqed} with one flavor and a model with three chiral fields, named $X$, $Y$ and $Z$ 
with the superpotential $W=XYZ$. This last model is named XYZ model in the literature.

Observe that this example can be thought of as a limiting case of Aharony duality
 if the electric gauge group is $U(1)$ and $N_f=1$, the dual theory consists of one monopole, 
 one anti-monopole and the meson $M$, interacting through the superpotential. In the following we will
 make a lot of use of this relation, by conjecturing that similar relations hold also for the limiting cases
 of generalized Aharony dualities with tensor matter content.
 In this case, the effective superpotential can be written as
 \begin{equation}
W = t \tilde t \det M,
\end{equation}
corresponding to the one of the XYZ model, by identifying $X$ with $t$, $Y$ with $\tilde t$
and $Z$ with $\det M$.
Note that the topological symmetry of the electric theory becomes non-topological in the
mirror description, \emph{i.e.} a symmetry of the Coulomb branch on one side is a symmetry of the Higgs branch on the other side.
Interestingly, mirror symmetry maps the \ac{fi} term of \ac{sqed} to a real mass in the dual model.

\subsection{Generalizations of the dualities: real groups and tensor matter}

As discussed in the case of Seiberg duality, also in the three-dimensional case
the dualities have been extended to cases with tensor matter and real gauge group.
In this section, we will review the extensions that have been worked out in the literature for 
the case of Aharony and Giveon--Kutasov duality.
This is a necessary step in order to study the reduction of the four-dimensional case to the three-dimensional ones.

\subsubsection{Tensor matter}

\begin{description}[style=nextline]
\item[{$U(N_c)_0$ with adjoint matter}]
The simplest generalization of Aharony duality with tensor matter is the case of 
$U(N_c)_0$ \ac{sqcd} with an adjoint field $X$ with superpotential 
(\ref{KSSe}).
For $N_c <n N_f$, this theory  has a dual description in terms of an $U(n N_f -N_c)_0$
gauge theory, with $N_f$ dual quarks, an adjoint $Y$ and a set of singlets.
There are two types of singlets, the meson 
\begin{equation}
M_j = Q X^j \widetilde Q,\quad j=0,\dots,n-1,
\end{equation}
and a set of fields with the same quantum charges of the electric monopoles
with flux $(\pm 1, \dots, 0)$.
From the quantum charges on can identify these fields with the electric monopoles, denoted 
here as $T_{0}$ and $\widetilde T_0$, to stress that they are the bare monopoles discussed above,
dressed by the adjoint field $X$.
These singlets are defined as
\begin{equation}
T_j = T_0 X^{j}\quad \text{and}\quad  \widetilde T_j = \widetilde T_0 X^j,\quad j=0,\dots,n-1,
\end{equation}
consistently with the constraints on the chiral ring imposed by the superpotential of the 
electric theory.
These singlets play the role of enforcing the correct quantum constraints on the Coulomb branch of the
dual theory. These relations are enforced by a superpotential relation, as in the 
Aharony duality.
The superpotential of the dual theory is given by
\begin{equation}
\label{magN}
W= Y^{n+1} + \sum_{j=1}^{n-1} M_j q Y^{n-1-j} \tilde q + \sum_{j=1}^{n-1} (t_j  T_{n-1-j} 
+ \tilde t_j \widetilde T_{n-1-j}) ,
\end{equation}
where $t_j$ and $\tilde t_j$ are the monopoles of the magnetic theory, with flux $(\pm1, \dots, 0)$.
The first two terms of the superpotential correspond to the generalization of \ac{kss} duality 
in three dimensions, and enforce the constraints on the chiral ring for the
matter fields.
The last two terms generalize the constraints on the chiral ring to impose on the monopole
sector, taking into account the presence of the adjoint field
in the counting of the zero modes. 
This duality was first derived by \ac{kp} %
(see also~\cite{Nii:2014jsa,Amariti:2014iza} for related discussions).

\item[{$U(N_c)_0$ with symmetric flavor}]
This duality has never been 
discussed in the literature so far, but it can be derived from the case of 
$U(N_c)_k$ with symmetric flavor discussed in~\cite{Kapustin:2011vz}.
Another derivation of this duality comes from the dimensional reduction of the four-dimensional case
as  will be discussed in Section~\ref{sec:redux}. 
For completeness, we present here the final result.

This duality has  gauge group $U(N_c)_0$, $N_f$ pairs of fundamentals and antifundamentals and two
tensor matter fields, the field $S$ in the symmetric representation of the gauge group,
and its conjugate $\tilde S$,
with an interaction 
\begin{equation}
W = \Tr (S \tilde S)^{n+1}.
\end{equation}
This induces a truncation on the chiral ring, such that the mesons that we have to consider are
\begin{equation}
M = Q (\widetilde S S)^{j+1} \widetilde Q ,\quad  P = Q (\widetilde S S )^{j} \widetilde S Q, \quad
\tilde P =   \widetilde Q S (\widetilde S S )^{j}  \widetilde Q, \quad j=0,\dots,n-1,
\end{equation}where $M$ is a bifundamental
operator, while $P$ and $\tilde P$  are symmetric in the flavor indices.

The dual theory has $U((2n+1) N_f+2n -N_c)$ gauge group, there are dual fundamentals
and antifundamentals,\ $q$ and $\tilde q$, dual symmetric and conjugate fields $s$ and $\tilde s$,
and the singlet fields are the mesons $M$, $P$ and $\tilde P$.

Also in this case one can work out the details on the monopole sector and study how the
electric monopoles  modify the structure of the dual superpotential. We leave this derivation 
to the interested reader.

\item[{$U(N_c)_0$ with antisymmetric flavor}]

Also this duality has not been discussed in the literature but 
it can be derived from the case of 
$U(N_c)_k$ with symmetric flavor discussed in~\cite{Kapustin:2011vz}.
Another derivation of this duality comes from the dimensional reduction of the four-dimensional case
we will discuss in Section~\ref{sec:redux}.
For completeness, we present here the final result.

This duality has gauge group $U(N_c)_0$, $N_f$ pairs of fundamentals and antifundamentals and two
tensor matter fields, in the antisymmetric and in the conjugate antisymmetric representations.
There is also a superpotential term
\begin{equation}
\Tr (A \tilde A)^{n+1}.
\end{equation}
This induces a truncation on the chiral ring, such that the mesons that we have to consider are
\begin{equation}
M = Q (\widetilde A A)^{j+1} \tilde Q, \quad P = Q (\widetilde A A)^{j} \widetilde A Q ,\quad \tilde P =   \widetilde Q A (\widetilde A A)^{j}  \widetilde Q,\quad j=0,\dots,n-1,
\end{equation}
where $M$ is a bifundamental
operator, while $P$ and $\tilde P$  are antisymmetric in the flavor indices.

The dual theory has $U((2n+1) N_f-2n -N_c)$ gauge group, there are dual fundamentals
and antifundamentals $q$ and $\tilde q$, dual antisymmetric and conjugate fields $a$ and $\tilde a$,
and the singlet fields are the mesons $M$, $P$ and $\tilde P$.
The details on the monopole sector are left to the reader also here.
\end{description}

\subsubsection{Orthogonal case}

In this section we discuss three-dimensional dualities for gauge group $SO(N)$.
Before starting the analysis we will make a small digression about the global properties of
theories with $so(n)$ gauge algebra.
In presence of $N_f$ vectors, there are four possible dual pairs of Aharony type (at zero \ac{cs} level):
\begin{eqnarray}
O(N_c)_+   &\leftrightarrow&    O(N_f-N_c+2)_+,\nonumber\\
O(N_c)_-    &\leftrightarrow&   Spin(N_f-N_c+2), \nonumber\\
Spin(N_c)   &\leftrightarrow&    O(N_f-N_c+2)_-,\nonumber\\
Pin(N_c)     &\leftrightarrow&    Pin(N_f-N_c+2), \nonumber\\
\end{eqnarray}
where the $Pin(N_c)$ group is obtained by gauging the 
charge conjugation symmetries in the $Spin(N_c)$ group.
In the discussion of the D~brane constructions, we will ignore the global properties and refer to the 
duality between an $SO(N_c)$ theory and an $SO(N_f-N_c+2)$ theory.

\begin{description}[style=nextline]
\item[{$SO(N_c)_0$ with $N_f$ vectors}]
This duality was originally proposed by~\cite{Aharony:2011ci} (see also~\cite{Hwang:2011ht} for further discussions).
The electric theory has $SO(N_c)_0$ gauge group and $N_f$ vectors $Q$, with vanishing superpotential.
It is dual to an $SO(N_f-N_c+1)$ theory, with $N_f$ dual vectors $q$, the meson $M = q^2$ in the symmetric representation of the $SU(N)_f$ flavor symmetry group and the extra singlets $T_0$.
The latter have the same quantum numbers as the monopole operators that parameterize the Coulomb branch of the electric theory and they appear as singlets of the dual theory enforcing the constraint of
$t_0=0$ on the magnetic monopole. The superpotential of this dual theory is 
\begin{equation}
W = M qq + t_0 T_0.
\end{equation}
\item [{$SO(N_c)_0$  with $N_f$ fundamentals and an adjoint}]
This duality has been studied in~\cite{Kim:2013cma}.
In this case one has an $SO(N_c)_0$ theory with $N_f$ vectors and one adjoint
$X$, with superpotential 
\begin{equation}
W = \Tr X^{2(n+1)}.
\end{equation}
The dual theory is an $SO((2n+1)N_f-N_c +2)$ gauge theory with $N_f$ vectors and singlets.
The mesonic singlets are $M_j = q X^j \tilde q$ with $j\leq 2n$, (anti)-symmetric in the flavor indices for
(odd) even $n$. There are also singlets associated to the dressed monopole operators of the electric theory.
They are of the form $T_j = X^j T_0$, for $j \leq 2n$, where $T_0$ is the bare monopole, and impose the necessary constraint on the Coulomb branch of the dual theory.
This theory has superpotential 
\begin{equation}
W = Y^{2(n+1)} + \sum_{j=0}^{2n} M_j q Y^{2n-j}  q + \sum_{j=0}^{2n} T_j  t_{2n-j} .
\end{equation}
\item [{$SO(N_c)_0$  with $N_f$ fundamentals and a symmetric tensor}]
This duality has not been discussed in the literature. It may be derived by a 
real mass flow from the case discussed in~\cite{Kapustin:2011vz}, in which the electric theory has a \ac{cs} term.
Another possibility consists in deriving it from dimensional reduction, see
Section~\ref{sec:redux}.
In this case the electric theory has $SO(N_c)$ gauge symmetry, with $N_f$ vectors
$Q$ and a traceless  symmetric tensor $S$ with superpotential
\begin{equation}
W = \Tr S^{n+1}.
\end{equation}
The mesons in the chiral ring are identified as $M_j = Q S^j \widetilde Q$, with $j=0,\dots,n-1$.
The dual theory has $SO(n (N_f+2) - N_c)$ 
gauge group, $N_f$ vectors $q$, a traceless symmetric tensor
$s$ and the mesons $M_j$.
There are also dressed monopole operators acting as singlets in the dual phase, of the form 
$T_j = T_0 S^j $, with $j=0,\dots, n-1$. 
The superpotential of this dual theory is  
\begin{equation}
W =s^{n+1} +  \sum_{j=0}^{n-1}  M_{n-j} q s^j q +\sum_{j=0}^{n-1} t_j T_{n-1-j},
\end{equation}
where $t_j$ are the monopoles that parameterize the Coulomb branch of the dual theory.
\end{description}

\subsubsection{Symplectic case}

\begin{description}[style=nextline]
\item [{$Sp(2N_c)_0$ with $2N_f$ fundamentals}]
This case was first discussed in Aharony's original paper~\cite{Aharony:1997gp}.
The electric theory has $Sp(2N_c)_0$ gauge group with $2N_f$ 
fundamentals $Q$ and $W=0$. The magnetic theory has 
$Sp(2(N_f-N_c-1))_0$ gauge group,
$2N_f$ dual fundamentals and a meson $M = Q^2$ in the
antisymmetric representation of the $SU(2N_f)$
flavor symmetry group.
There is also an extra singlet, corresponding to the 
electric monopole $T_0$, necessary to impose the constraint $t_0=0$
on the monopole that parameterizes  the Coulomb branch of the dual theory.
The superpotential of the dual theory is
\begin{equation}
  W = M q q + t_0 T_0.
\end{equation}
\item [{$Sp(2N_c)_0$  with $2N_f$ fundamentals and an adjoint}] 
This duality has been studied in~\cite{Kim:2013cma}.
In this case one has an $Sp(2N_c)_0$ theory with $2N_f$ fundamentals and one adjoint
$X$, with superpotential 
\begin{equation}
W = \Tr X^{2(n+1)}.
\end{equation}
The dual theory is an $Sp(2((2n+1)N_f-N_c -1))$ gauge theory with $2N_f$ fundamentals and singlets.
The mesonic singlets are $M_j = q X^j \tilde q$ with $j\leq 2n$, (anti)-symmetric in the flavor indices for
(even) odd $n$. There are also singlets associated to the dressed monopole operators of the electric theory.
They are of the form $T_j = X^j T_0$, for $j \leq 2n$, where $T_0$ is the bare monopole, and impose the necessary constraint on the Coulomb branch of the dual theory.
This theory has superpotential 
\begin{equation}
W = Y^{2(n+1)} + \sum_{j=0}^{2n} M_j q Y^{2n-j}  q + \sum_{j=0}^{2n} T_j  t_{2n-j} .
\end{equation}
\item [{$Sp(2N_c)_0$  with $2N_f$ fundamentals and an antisymmetric tensor}]
Here we can consider the case of the traceless antisymmetric tensor.\footnote{At the end of this section, 
we also discuss the case with non-vanishing trace, which is just a singlet that modifies the duality slightly.}
This duality has not been described in the literature, but it can be derived from the reduction of the four-dimensional case
as we will discuss later.
In this case the electric theory has gauge group $Sp(2N_c)$ and there are $2N_f$ fundamental
quarks $Q$ and a traceless antisymmetric tensor $A$. The superpotential is 
\begin{equation}
W = \Tr A^{n+1},
\end{equation}
and there are mesons $M_j = Q A^j Q $ with $j=0,\dots n-1$.
The dual theory has $Sp(2(n (N_f-1)-N_c))$ gauge group with $2N_f$ fundamentals $q$,
a dual antisymmetric field $a$, and the mesons $M_j$.
There are also dressed monopole operators acting as singlets, $T_j = Tr  \, T_0 A^{j}$
for $j=0,\dots, k-1$.
The superpotential of this dual theory is
\begin{equation}
W =\Tr  a^{n+1} +  \sum_{j=0}^{n-1}  \Tr  M_{n-j} q a^j q +\sum_{j=0}^{n-1} t_j T_{n-1-j},
\end{equation}
where $t_j$ are the dressed monopole operators of the dual theory.
\end{description}

\subsection{Dualities with Chern--Simons terms}

Another generalization of the dualities discussed above consists in adding a level $k$
\ac{cs} term for the electric gauge group, \emph{i.e.} dealing with  $U(N_c)_k$, $Sp(2N_c)_{2k}$
and $O(N_c)_{k}$.
Many of these dualities can been derived from a~brane construction  and 
we will review these constructions at the of this chapter.
Another possible way to derive such dualities consists in assigning a large real mass (the same, or at least same sign) to 
$k$ pairs of quarks and antiquarks. This induces a real mass flow, reducing the number of flavors and generating at one loop  
a level $k$ \ac{cs} level.
In the dual theory one can assign the real masses consistently to the dual fields and trigger the dual of
this flow.
The prototypical example of such a flow is the derivation of Giveon--Kutasov duality from Aharony duality~\cite{Willett:2011gp}, as discussed in Sec.~\ref{sec:thereback}.
 Many dualities involving also chiral theories have been constructed in a similar fashion in~\cite{Benini:2011mf}.
Here we will not review these constructions but present a different example: we show how to obtain the duality of 
Niarchos from the duality of unitary theories and adjoint matter with superpotential $X^{n+1}$.
We will leave the derivation of the other dualities to the reader, and limit ourselves to a listing of the various three-dimensional dualities found in the literature. 

\subsubsection{$U(N_c)_k$ with the adjoint}

Let us consider an electric theory with $U(N_c)_0$ gauge group, $N_f+k$
fundamental flavors and an adjoint field $X$ with superpotential
(\ref{KSSe}).
We can assign a  large mass $m$ to $k$ fundamental flavor pairs.
This can be done by combining the effect of the $U(N_f+k)_r \times U(N_f+k)_l$
symmetries.
We can parameterize them as follows:
\begin{align}
  m(Q_a) &= \mu_a + m_A + m_B &
                                m(\widetilde Q_a) &= \nu_a + m_A - m_B,
\end{align}
where $\mu_a$ and $\nu_a$ refer to the $SU(N_f+k)_r$ and the $SU(N_f+k)_r$
part of the symmetry group, and are subjected to the constraint 
$\sum_a \mu_a = \sum_a \nu_a =0$, while $m_A$ and $m_B$ refer to the
real mass of the axial and of the baryonic symmetry respectively.
We assign the masses as
\begin{align}
\mu_{a_1} &= \tilde \mu_{a_1} -  \frac{k m}{k-N_f} 
& \text{with} &&  \sum_{a_1=1}^{N_f} \tilde \mu_{a1} &=0 & \text{and} &&
\mu_{a_2} &=  -  \frac{k m}{k-N_f} \\
\mu_{a_1} &= \tilde \nu_{a_2} -  \frac{N_f m}{k-N_f} 
&  \text{with} &&  \sum_{a_2=1}^{N_f} \tilde \mu_{a1} &=0 & \text{and} &&
\mu_{a_2} &= - \frac{N_f m}{k-N_f}   
\end{align}
and $m_A =  \frac{k m}{k-N_f}+m_A$ where $a_1= 1,\dots,N_f$ and $a_2 = 1,\dots,k$.
By integrating these fields at one loop
we are left with a model with $U(N_c)_k$ gauge group, $N_f$ flavors
and an adjoint with the superpotential (\ref{KSSe}). 
The dual theory, before considering the large-mass limit, had gauge group $U(n (N_f+k)  - N_c)_0$. 
The charged matter fields where 
$N_f+k$ dual fundamental flavors and adjoint $Y$.
There were also  singlets, the mesons $M_j = Q X^j \widetilde Q$ and 
the dressed monopoles $T_j = T_0 X^j$.
The dual real mass flow can be understood from the spectrum of the global symmetries.
By performing the analysis, one obtains $k$ fundamental flavors with a large mass, $-m$.
By integrating them out one obtains a \ac{cs} theory at level $-k$.
Each meson $M_j$ has $2k N_f$ components with real mass $m$ and $k^2$ components
with mass $2m$. The monopoles are charged under $U(1)_A$ and they are all massive.
The final theory has $U(n (N_f+k)  - N_c)_{-k}$ gauge group, with $N_f$ fundamental
pairs and one adjoint $Y$. 
Observe that so far we have been agnostic on the sign of $m$. Both
signs are possible, and by specifying the sign, the analysis one ends up with
the magnetic gauge group in the form 
 $U(n (N_f+|k|)  - N_c)_{-k}$.
There are also mesons $M_j = Q X^j \widetilde Q$,
each with $N_f^2$ components, and the superpotential is
(\ref{KSSm}).
This corresponds to the dual phase originally found in~\cite{Niarchos:2008jb}.

\subsubsection{Other dualities}

We can perform an analogous real mass flow on the other dualities without the \ac{cs}
terms discussed above.
The flows generate \ac{cs} terms in the action and one can end up with
various generalizations of \ac{gk} dualities.

We will not discuss these cases in detail but just review the final
results in Table~\ref{tab:three-d-dualities}, by specifying the
electric and the magnetic gauge groups and the charged fields.
We also specify the paper in which the duality of interest has been
first introduced.
\begin{sidewaystable}
  \centering
  \begin{tabular}{LLLLLl}
    \toprule
    G             & \text{Charged Fields}                                                            & \tilde G                                        &W_e & W_m        &       \text{Ref.}                 \\
    \midrule
    U(N_c)_k      & N_f(F \oplus \tilde F)                                         & U(N_f-N_c+|k|)_{-k}                 &(\ref{eleSei})&(\ref{magSei})  & \cite{Giveon:2008zn}   \\
    U(N_c)_k      & N_f(F \oplus \tilde F) \oplus X                           & U(n(N_f+|k|)-N_c)_{-k}              &(\ref{KSSe})&(\ref{KSSm})   & \cite{Niarchos:2008jb} \\
    U(N_c)_k      & N_f(F \oplus \tilde F) \oplus (S \oplus \tilde S)  & U((2n+1) (N_f+|k|)+2n -N_c)_{-k}      &(\ref{eleS})&(\ref{magS})   & \cite{Kapustin:2011vz} \\
    U(N_c)_k      & N_f(F \oplus \tilde F) \oplus (A \oplus \tilde A)  & U((2n+1) (N_f+|k|)-2n -N_c)_{-k}       &(\ref{eleA})&(\ref{magA})  & \cite{Kapustin:2011vz} \\
    Sp(2N_c)_{2k} & 2 N_f F                                                          & Sp(2(N_f+|k|-N_c-1))_{-2k}         &(\ref{eleSp})&(\ref{magSp})    & \cite{Willett:2011gp}  \\
    Sp(2N_c)_{2k} & 2 N_f F \oplus X                                            & Sp(2((2n+1)(N_f+|k|)-N_c -1))_{-2k}  &(\ref{eleSpAdj})&(\ref{magSpAdj})   & \cite{Kapustin:2011vz} \\
    Sp(2N_c)_{2k} & 2N_f F \oplus A                                             & Sp(2(n (N_f+|k|-1)-N_c))_{-2k}   &(\ref{eleSpA})&(\ref{magSpA})       & \cite{Dolan:2011rp}    \\
    O(N_c)_{k}    & N_f V                                                               & O(N_f+|k|-N_c+2)_{-k}               &(\ref{eleSO})&(\ref{magSO})    & \cite{Kapustin:2011gh} \\
    O(N_c)_{k}    & N_f V \oplus X                                                 & O((2n+1)(N_f+|k|)-N_c +2)_{-k}    &(\ref{eleSOAdj})&(\ref{magSOAdj})      & \cite{Kapustin:2011vz} \\
    O(N_c)_{k}    & N_f V \oplus S                                               & O(n (N_f+|k|+2) - N_c)_{-k}          &(\ref{eleSOS})&(\ref{magSOS})   & \cite{Kapustin:2011vz} \\
    \bottomrule
  \end{tabular}  
  \caption{Summary of the three dimensional dualities with \ac{cs} action discussed here. $F$ is the  fundamental representation,
$\tilde F$ the antifundamental, $S$ the symmetric, and $\tilde S$ to its conjugate, $A$ to the antisymmetric,
$\tilde A$ to its conjugate, $X$ to the adjoint and $V$ to the vector.
There are also singlets in the spectrum of the magnetic phases, corresponding to the generalized mesons
that can be read off from the case without the \ac{cs} term or from the four-dimensional parent theory.
We do not report these singlets here.
Electric and Magnetic superpotentials coincide with those of the
four-dimensional parent theory.
}
  \label{tab:three-d-dualities}
\end{sidewaystable}

\subsection{Checks}

We conclude this overview of $\mathcal{N}=2$ three-dimensional dualities with a discussion of possible checks.
In three dimensions, many of the tools used for the four-dimensional case are not available due to
the absence of anomalies for the continuous symmetries. 
Also at the level of the moduli space, the checks are more involved because of the
presence of a Coulomb branch, parameterized by the monopole operators.
We refer the reader to some relevant papers for the analysis of these checks~\cite{Aharony:1997bx,Intriligator:2013lca}.
In the recent years, new checks have come from the computation of the Witten index~\cite{Witten:1999ds,Intriligator:2013lca}, matching the number of vacua of
the dual theories.
Another class of checks consists in matching the parity anomalies.
Moreover, in presence of \ac{cs} terms, one can further check the contact terms 
for the global symmetries~\cite{Closset:2012vg,Closset:2012vp}.
Many new techniques have been possible thanks to the development of localization techniques.
The partition function on the three-sphere~\cite{Kapustin:2009kz,Jafferis:2010un,Hama:2010av}  and the superconformal index~\cite{Kim:2009wb}
can be used to match the dualities  discussed above (see for example~\cite{Willett:2011gp,Hwang:2011qt,Benini:2011mf}).
Exact mathematical results play a crucial role in these matchings.
Matching the partition function on the three-sphere for example corresponds
to some identities among hyperbolic hypergeometric integrals~\cite{VanDeBult}.
More recently, the topologically twisted index of~\cite{Benini:2015noa} has been used to test some of the dualities 
presented above.

\section{The~brane picture in three dimensions}\label{sec:3dbranes}

In this section, we discuss the realization of the three-dimensional dualities in presence of \ac{cs} terms
 in terms of \D{} and \NS{}~branes.

We begin our analysis in Sec.~\ref{subsec:fivebranes} by reviewing the
generation of \ac{cs} terms in the~brane picture in terms of bound
states of \(1\) \NS{} and \(p\) \D5~branes, \emph{i.e.}
\((1,p)\)--fivebranes. In Section~\ref{subsec:GK}, we review the
simplest example of a three-dimensional $\mathcal{N}=2$ duality,
namely \ac{gk} duality. The generalization to the case of an adjoint
with a polynomial potential is discussed in Sec.~\ref{sec:NiarchosBr}. We
discuss how these results are extended to configurations with
orientifolds in Section~\ref{subsec:orientifold}. We conclude this
section by discussing the problems in extending these results to the
cases with vanishing \ac{cs} level $k_{CS}=0$ in
Section~\ref{subsec:zerolevel}.

\subsection{Chern--Simons action from fivebranes}
\label{subsec:fivebranes}

Three-dimensional field theories are engineered in ten-dimensional
\tIIB string theory. The \NS{}~branes are the same as
in the \tIIA case (Section~\ref{sec:4dbranes}), while the \D{p}~branes have $p=-1,1,3,5,7,9$.
We will mostly consider \NS{}~branes extended along $x_{012345}$, 
\NS{}' along $x_{012389}$,
\D3~branes extended along $x_{0126}$,
\D5~branes along $x_{012789}$,  and \D5' along $x_{012457}$ (see Table~\ref{tab:3d-brane-engineering})
\begin{table}
  \centering
  \begin{tabular}{lCCCCCCCCCC}
    \toprule
           & 0      & 1      & 2      & 3      & 4      & 5      & 6      & 7      & 8      & 9      \\
    \midrule
    \NS{}  & \times & \times & \times & \times & \times & \times &        &        &                 \\
    \NS{}' & \times & \times & \times & \times &        &        &        &        & \times & \times \\
    \D3    & \times & \times & \times &        &        &        & \times &        &                 \\
    \D5    & \times & \times & \times &        &        &        &        & \times & \times & \times \\
    \D5'   & \times & \times & \times &        & \times & \times &        & \times &        &        \\ \bottomrule  \end{tabular}
  \caption{Brane configuration for three-dimensional~brane engineering.}
  \label{tab:3d-brane-engineering}
\end{table}

\begin{figure}[h]
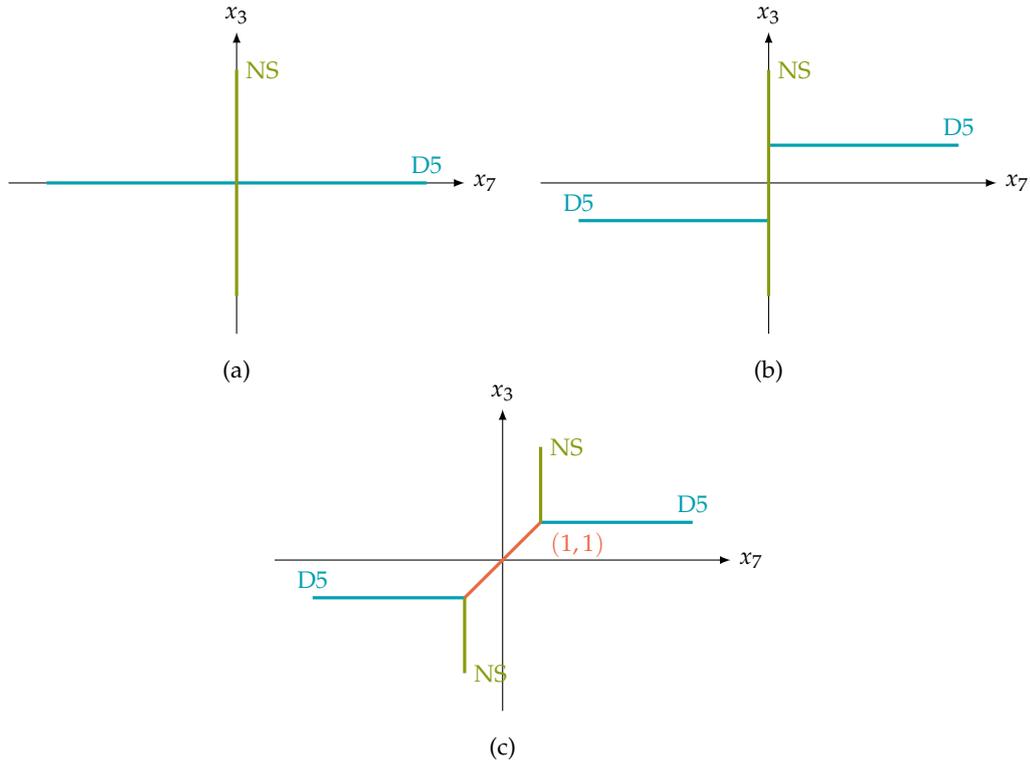

  \centering
  \begin{footnotesize}
    \FigureCS
  \end{footnotesize}
  \caption{Brane engineering of the \ac{cs} term. Start with a
    \D5--\NS{} sytem~(a). The \D5~brane is
    broken at the intersection with the \NS{} into two semi-infinite
   ~branes~(b). Supersymmetry is preserved if the \NS{} segment is
    replaced by a \((1,1)\) bound state~(c).}
  \label{fig:CS-engineering}
\end{figure}

A new ingredient, not present in four dimensions, is the notion of
the \ac{cs} term, which can also be realized in the~brane engineering.
Let us consider first a \D5~brane and an \NS{}~brane. 
Let us suppose that the two~branes intersect in the directions $x_{3,7}$. 
Then one can break the \D5~brane on the \NS{}~brane
into two semi-infinite \D5~branes on opposite sides of the \NS{}.
The two halves can slide along the same or opposite directions 
along $x_3$.   
The first case corresponds to generating opposite real masses for the 
quarks and the antiquarks, such that there is no \ac{cs} generated
\footnote{As anticipated in footnote~\ref{foot:flavor-in-branes-4d} on the bottom of page~\pageref{foot:flavor-in-branes-4d},
this configuration shows that the global symmetry is $U(N_f)^2$, such that the two $U(N_f)$
act on each stack of semi-infinite \D5~branes separately.}.
In the other case, a \ac{cs} term at level one is generated for each 
pair of \D5~branes sliding along opposite directions.
In this case, supersymmetry is preserved by replacing the \NS{}
segment between the two configurations with a bound state of
one \D5 and one \NS{}'~brane, \emph{i.e.} a  $(1,1)$-fivebrane.
This fivebrane is rotated by an angle $\tan \phi = g_s$
with respect to the original configuration (see Figure~\ref{fig:CS-engineering}).
In general, starting with $k$ \D5~branes we end up with a \((1,k)\) fivebrane 
and the angle is $\tan \phi = k g_s$.

Let us now consider a configuration with an \NS{}~brane and a
\((1,k)\) fivebrane and stretch along $x_6$ a stack of $N_c$ \D3
branes. The resulting theory corresponds to a level $k$,
$\mathcal{N}=2$ \(U(N_c)_k\) \ac{sym}-\ac{cs} theory. To see this,
start by looking at the original configuration with $k$ \D5~branes,
before the~brane recombination. In this case, ignoring the Abelian
factors, there should be an $SU(k)^2$ flavor symmetry on the field
theory side\footnote{Remember that there are no anomalies in three
  dimensions.}, corresponding to the independent rotations of the $Q$
and the $\widetilde Q$ flavors. This symmetry is visible at the~brane
level by the fact that we can break the \D5 on the \NS{}' and treat
the two semi-infinite~branes separately. Moving the semi-infinite \D5
branes corresponds to integrating out the corresponding superfields
with a real mass. The \ac{cs} is generated precisely by the real
massive fermions in the one loop diagrams. In this case the fivebrane
is constructed by moving the two semi-infinite stacks on opposite
directions. They have opposite charge under the gauge symmetry
implying that a level $k$ is generated.

In general, by considering a configuration with two fivebranes of the
form $(p_1,q_1)$ and $(p_2,q_2)$ we can realize a \ac{cs} theory with
level $k = q_1 p_2- p_1 q_2$. Observe that the sign of $k$ in our case
depends from having sent to infinity the fundamentals or the
antifundamental flavors; the opposite sign would reflect into an
different choice of the angle, $\tan (\pi - \phi )= k g_s$.

\subsection{Giveon--Kutasov duality}
\label{subsec:GK}

\begin{figure}[t]
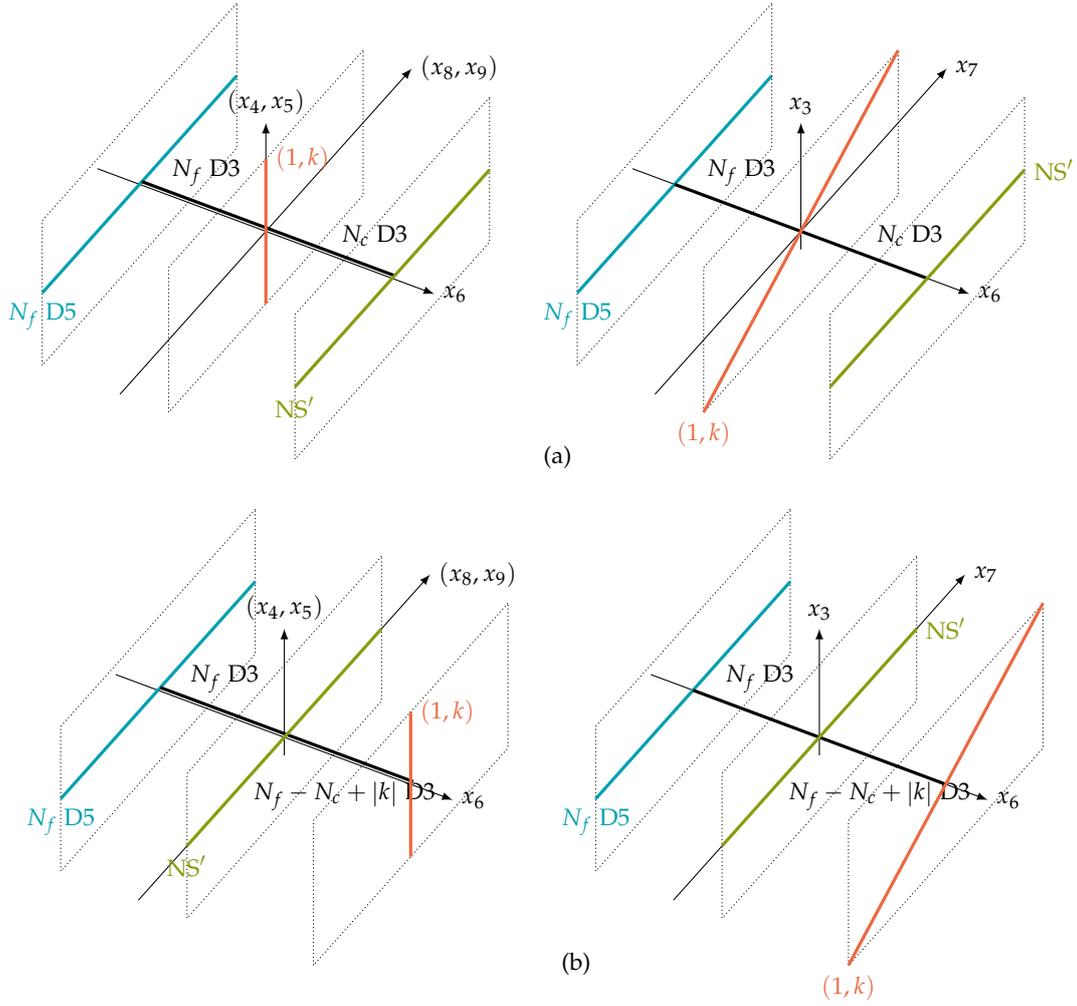

  \centering
  \begin{footnotesize}
    \FigureGKele
    \FigureGKmag
  \end{footnotesize}
  \caption{Giveon--Kutasov duality as Hanany--Witten transition.
    Electric (a) and magnetic (b) phases.}
  \label{fig:GK}
\end{figure}

The duality of Giveon and Kutasov can be now formulated as in the original presentation,
in terms of its~brane engineering.
The electric theory is obtained by considering the $U(N_c)_k$ theory 
discussed above and by adding $N_f$ \D3~branes on the right side of the \NS{}~brane
ending on a stack of $N_f$ \D5~branes.
This configuration corresponds to considering $N_f$ pairs of fundamentals
and antifundamentals $Q$ and $\widetilde Q$.
The dual configuration is obtained by considering the \ac{hw} transition. Note, however that 
the presence of $k$ \D5~branes in the $(1,k)$ bound state modifies the rules of the duality:
when the $(1,k)$ fivebrane crosses the \NS{}~brane, one has to add $|k|$ \D3~branes, independently
 of the sign of $k$, because there are always \D5~branes in the bound state.
The s-rule is not violated as long as $N_f+|k| > N_c$ in this case.
The dual configuration is shown in Figure~\ref{fig:GK}.
In this case the \ac{cs} level is $-k$
and the gauge group becomes $U(N_f-N_c +|k|)_{-k}$.
There are $N_f$ dual pairs of fundamentals $q$ and antifundamentals $\tilde q$
and a meson $M$ as in the four-dimensional case, with a superpotential $W = M q \tilde q$.
Note the following subtlety: in this case the $U(1)$ factor of the gauge group
does not decouple in the \ac{ir}, and we are allowed to consider the baryonic symmetry as
gauged.
Since from type \tIIB, there is no uplift to M-theory we do not expect any bending effect
for the \NS{}~branes. This reflects the fact that the axial symmetry is not anomalous in 
three dimensions, and the two rotations $x_{45}$ and $x_{89}$ are both realized in the quantum theory.

\subsection{Niarchos duality}\label{sec:NiarchosBr}

\begin{figure}[t]
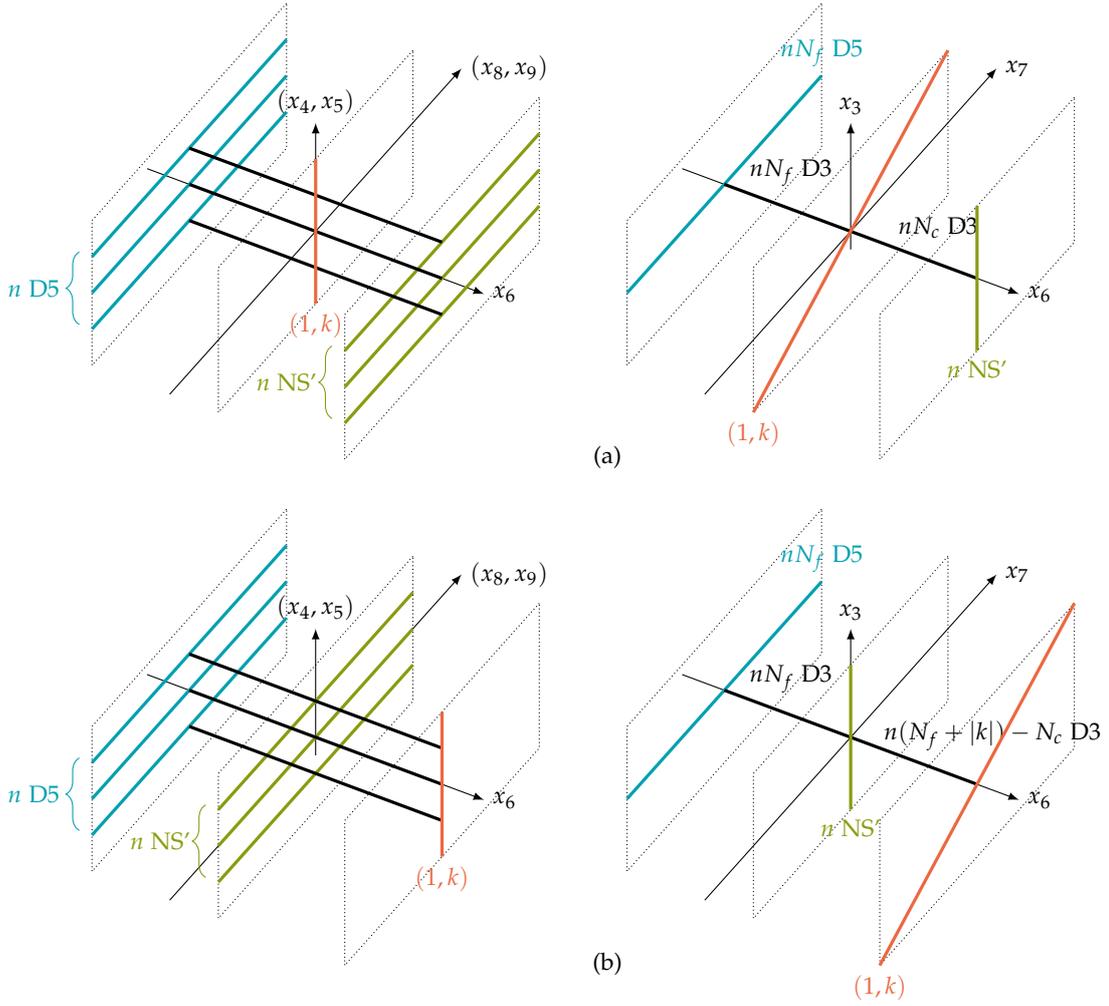

  \centering
  \begin{footnotesize}
    \FigureNiarchosele
    \FigureNiarchosmag
  \end{footnotesize}
  \caption{Niarchos duality as Hanany--Witten transition.
    Electric (a) and magnetic (b) phases.}
  \label{fig:Niarchos}
\end{figure}

A natural extension of \ac{gk} duality has been studied in~\cite{Niarchos:2008jb}, by adding a power law
superpotential for the adjoint field $X$.
This generalizes the Giveon--Kutasov duality in the same way as \ac{kss} duality generalizes Seiberg duality.
In this case, one considers a configuration with one \NS{}' and $n$ \NS{}~branes and $N_c$ \D3s stretched along the segment connecting them.
One can also add flavor by considering $N_f$ \D3~branes ending on $N_f$ \D5s on the right of the $n$ \NS{}~branes.
By replacing the \NS{}'~brane with a $(1,k)$ fivebrane one realizes the electric theory, \emph{i.e.} $SU(N_c)_k$ \ac{cs} matter theory with 
$N_f$ pairs of fundamentals and antifundamentals and one adjoint with superpotential $W = \Tr X^{n+1}$.
The duality is obtained as in the four-dimensional case, by the \ac{hw} transition. In this case we can still proceed as above, first we separate the 
\NS{}~branes along the $x_{89}$ direction, breaking the gauge group into $SU(r_1)_{k}\times SU(r_n)_{k}$ with $r_1+\dots + r_n = N_c$.
Then we dualize each single sector by considering the~brane creation effect as above. We have a gauge group 
$SU(N_f-r_1+|k|)_{-k}\times SU(N_f-r_n+|k|)_{-k}$ as we can see from Figure~\ref{fig:Niarchos}.
By reconnecting the \NS{}~branes we end up with an $SU(n(N_f+|k|)-N_c)_{-k}$ gauge group as expected.

\subsection{Adding orientifolds}
\label{subsec:orientifold}

One can also study three-dimensional dualities with \ac{cs} terms in presence of orientifold planes.
We can construct the electric theories by considering the four-dimensional parent theories
and substituting \D{p}~branes and \O{p} planes with \D{(p-1)} and \O{(p-1)}~branes. The direction on which these~branes are
not extended anymore can be fixed to be $x_3$.
On the other hand the \NS{}~branes are still as above, \emph{i.e.} extended along the non-compact direction $x_3$.
Observe that this is just a mnemonic at this point and it is not related to any compactification
or dimensional reduction of a Seiberg duality.

The \ac{cs} levels are obtained by substituting \NS{}~branes with $(1,k)$ fivebranes, consistently with the constraints from the parity anomalies and the s-rule.
The dual theories are then constructed by the \ac{hw} transition.
The authors are not aware of any of such construction in the literature so far.
For this reason, we will present some simple example in the following to clarify the construction.
The reader can work out further possibilities.

\begin{figure}[]
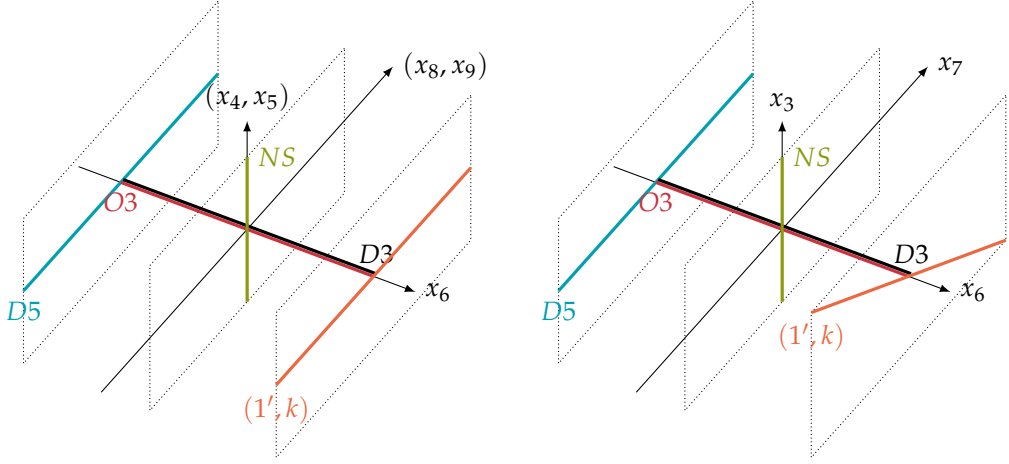

  \centering
  \begin{small}
  \FigureDualityFundamentals
  \end{small}
  \caption{Brane engineering with fundamental matter.}
  \label{fig:duality-fundamentals}
\end{figure}

\begin{figure}[]
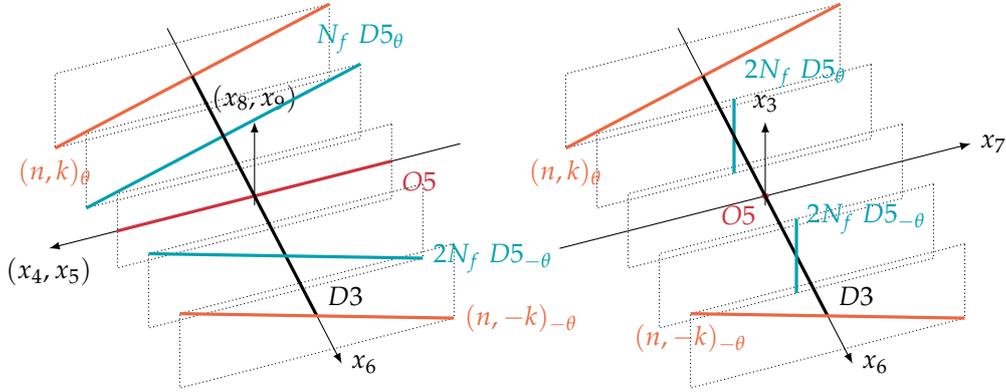

  \centering
  \begin{small}
  \FigureSOSpOfive
  \end{small}
  \caption{Brane engineering for \(SO\) and \(Sp\) theories with
    fundamental matter.}
  \label{fig:SO-Sp-O5}
\end{figure}

\begin{figure}[]
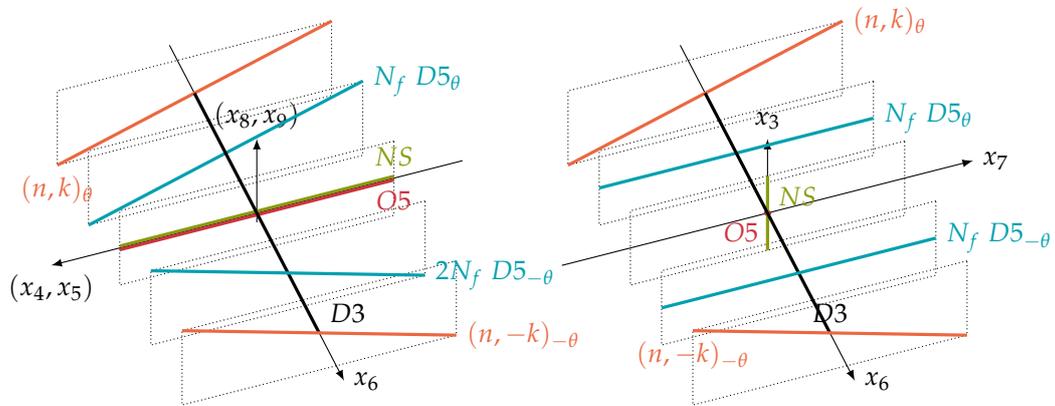

  \centering
  \begin{small}
  \FigureUnitarySymmAntisymm
  \end{small}
  \caption{Brane engineering for \(U(N)\) theories with symmetric and antisymmetric flavors.}
  \label{fig:unitary-symm-antisymm}
\end{figure}

\begin{figure}[]
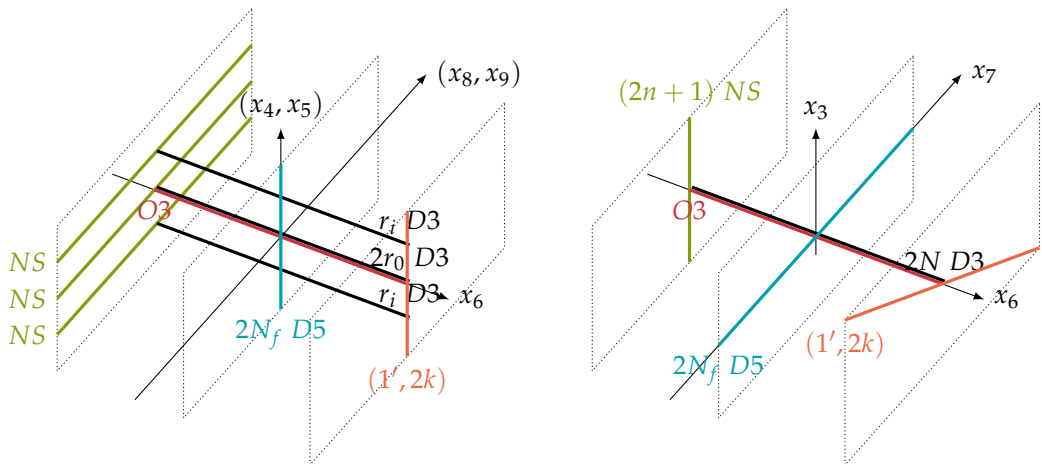

  \centering
  \begin{small}
  \FigureSOSpAdjoint
  \end{small}
  \caption{Brane engineering for \(SO\) and \(Sp\) theories with
    adjoint matter.}
  \label{fig:SO-Sp-Adjoing}
\end{figure}

\subsection{Discussion of the case $k=0$}
\label{subsec:zerolevel}

We conclude this review section with presenting a problem in describing the Aharony 
duality and its generalization in terms of the~brane engineering.
One can imagine to engineer an $U(N_c)_0$ electric theory in terms of $N_c$ \D3~branes
stretched between an \NS{} and an \NS{}'~brane and adding the usual $N_f$ \D3 flavor~branes ending on 
$N_f$ heavy \D5~branes.
The naive guess would result in a dual theory with $U(N_f-N_c)_0$
gauge group and superpotential $W = M q \tilde q$, but in this description there is no trace of any extra superpotential involving 
the monopole operators as singlets in the dual description.
In order to understand the problem we remind the reader that the monopole
operators can be associated to \D1~branes extended along the directions 
$x_3$ and $x_6$ between pairs of \D3~branes ending on the \NS{} and the \NS{}'~brane.
It is unclear how to enforce such operators as singlets of the dual description.
There is however a physical description of the role of the dual superpotential:
it should constrain the monopole operators $t$ and $\tilde t$.
This constrains the freedom of moving the first and the last gauge \D4~brane
along the infinite direction $x_3$.
In principle, this is a quantum effect, but in this case we cannot use the help of 
M-theory to visualize it on the~brane picture.
We will see that there is another way to understand this effect via a T-duality to \tIIA.

\chapter{Reduction of four-dimensional dualities to three dimensions}\label{sec:redux}

In this section we review the reduction of four-dimensional Seiberg duality to three dimensions.

In Section~\ref{sec:redfields}, we review the field theory aspects of the reduction, following largely~\cite{Aharony:2013dha}. We then take the~brane perspective in Section~\ref{sec:redbranes}, where in the first step, we explain how to obtain the $\eta$ superpotential from T--duality (Sec.~\ref{sec:superpotential-from-branes}).
In Section~\ref{sec:Braneology}, we discuss in detail the reduction pf \ac{sqcd} from four to three dimensions. The reduction of four-dimensional 
Seiberg duality with adjoint matter to three dimensions is discussed in Section~\ref{sec:adjSQCD}. Finally, we discuss the reduction of four-dimensional dualities 
in the presence orientifold planes, leading to real gauge groups and tensor matter.

\section{Field theory aspects}\label{sec:redfields}

Here we review the work of~\cite{Aharony:2013dha}, where Seiberg duality was reduced to three dimensions.\footnote{See also to~\cite{Niarchos:2012ah} for an early attempt.}
A four-dimensional theory can be naively reduced to three dimensions by compactifying one dimension on a circle
and shrinking its radius to zero size.
This procedure is well defined and leads to a consistent three-dimensional theory.
It turns out however, that Seiberg duality is not preserved under this naive dimensional reduction.
Roughly speaking, the duality does not commute with the zero-size limit of the circle.
We have seen that Seiberg duality is an \ac{ir} property, valid for energies much below the
holomorphic scales of the electric and of the magnetic theory, 
\begin{equation}
E \ll \Lambda, \widetilde \Lambda.
\end{equation}
These scales are related to the four-dimensional coupling through the relation (\ref{eq:lambdaolo}).
The gauge coupling of the dimensionally reduced theory is related to the one of its parent theory via
\begin{equation}
g_4^2 = r\, g_3^2.
\end{equation} 
We see that in the limit $r \to 0$, the strong coupling scales
of the electric and of the magnetic theory are vanishing. For this reason the limit 
$E \ll \Lambda, \widetilde \Lambda$ cannot be defined. Postulating a three-dimensional duality inherited from four-dimensional Seiberg duality becomes 
an empty statement.

That the duality cannot be preserved under dimensional reduction is also obvious when we consider the global symmetries. In the case of four-dimensional \ac{sqcd}, there is an anomalous axial symmetry $U(1)_A$. When reducing the theory on the circle, this symmetry is however preserved also at the quantum 
level in three dimensions. In order to preserve the four-dimensional duality in three dimensions, we need a mechanism that prevents the generation of this symmetry.
 
\medskip
The above problems can be avoided by reducing the dual phases on the circle, but at finite size $r$, and considering the theories at an
energy $E \ll 1/r$, where the physics is effectively three dimensional.
We have seen that when reducing a vector multiplet to three dimensions, the fourth component $A_3$ becomes a real 
scalar $\sigma$. Gauge invariance of $\int_{S^1} A_3$ requires the identification of
$\sigma$ with $\sigma + r^{-1}$, \emph{i.e.} a compact scalar.
The presence of this compact scalar is reflected in the presence of an additional interaction
for the unlifted coordinates on the Coulomb branch. This interaction is called the 
\ac{ahw} superpotential.
If we consider for example a $U(N_c)$ theory, this term becomes
\begin{equation}
\label{AHW}
W_{AHW}  = \eta T_0 \widetilde T_0,
\end{equation}
where $\eta = e^{-(r g_3)^{-2}}$.
Observe that the superpotential (\ref{AHW}) breaks the axial symmetry explicitly.

One can perform the same reduction on the Seiberg-dual theory, and generate the superpotential 
 (\ref{AHW}) for the Coulomb branch coordinates on the magnetic side.
 This leads to an effective three-dimensional duality between the two theories considered 
 at energies much below the finite radius of $S^1$.
 
 \medskip
It is  possible to obtain Aharony duality from this new effective duality in three dimensions by sending the 
superpotential (\ref{AHW}) to zero consistently in the dual phases.
This limit requires a real mass flow. This is however not enough because, for consistency,  
often a non-trivial map between the vacua of the electric and the magnetic theory has to be considered.
If the vacuum of the electric theory is at the origin of the Coulomb branch, 
\emph{i.e.} $\langle \sigma \rangle =0$, one typically has 
$\langle \widetilde \sigma \rangle \neq 0$ for the dual theory. This corresponds to
the topological vacuum discussed in~\cite{Intriligator:2013lca}.

Consider for example $U(N_c)$ \ac{sqcd} with $N_f+2$ flavors.
We assign real masses to two flavors on the electric side using
opposite masses for the fundamental and the antifundamental, such that no
\ac{cs} term is generated in the action.
In the dual theory, one can assign the masses to the dual quarks and to the mesons consistently with the symmetries.
One is also forced to consider a non-zero \ac{vev} for the last two diagonal entries of $\widetilde \sigma$,
breaking the gauge group to $SU(N_f -N_c) \times U(1)^2$. 
The dual superpotential becomes
\begin{equation}
W = M q \tilde q + M_1 q_1 \tilde q_1 + M_2 q_2 \tilde q_2,
\end{equation}
where the first term is associated to the $U(N_f-N_c)$ sector, while the extra two belong
to the two $U(1)$ sectors. 
The two $U(1)$ sectors correspond to \ac{sqed} 
with one flavor and can be dualized to a set of singlets. 
The superpotential after this duality is given by
\begin{equation}
W = M q \tilde q + M_1 N_1 + M_2 N_2 + N_1 t_1 \tilde t_1  + N_2 t_2 \tilde t_2,
\end{equation}
where we identified the fields $N_i$, $t_i$ and $\tilde t_i$ in each sector
with$X$, $Y$ and $Z$.
By carefully studying the interaction, we see that there are \ac{ahw} interactions between the monopoles
of the $U(N_f-N_c)$ sector and the monopoles of the $U(1)^2$ sector.
This interaction is given by
\begin{equation}
W_{AHW} = t \tilde t_1 + t_1 \tilde t_2 + t_2 \tilde t  .
\end{equation}
By integrating out the massive fields and by identifying the fields $t_1 = T$
and $\tilde t_2 = \widetilde T$, \emph{i.e.} with the monopoles of the electric theory acting as singlets
in the magnetic phase, we get
 \begin{equation}
W = M q \tilde q + t T + \tilde t \widetilde T.
\end{equation}
This shows that the three-dimensional Aharony duality can be obtained by a stepwise procedure from four-dimensional
Seiberg duality. First one reduces the duality on the circle, at finite size. Then one has to perform
a real mass flow on the electric side while carefully analyzing the vacuum structure of the dual theory in order
to recover the right constraint on the Coulomb branch.
To conclude, we can also recover the Giveon--Kutasov duality: this is done by assigning some real 
masses to the flavors of the electric theory, with the same sign, such as to generate a \ac{cs} term.
The real masses of the dual are dictated by the duality and in this case, 
no topological vacua arise. See~\cite{Willett:2011gp} for details on this derivation.

\subsection{$U(N_c)$ with adjoint matter}

In this section, we generalize the reduction of Seiberg duality to the case
of \ac{kss} duality discussed in Section~\ref{sec:kss}.
By reducing the theory on the circle in presence of an adjoint field with superpotential $\Tr X^{n+1}$,
 we have $2n$ unlifted directions on the Coulomb branch of the $U(N_c)$ theory.

There is a superpotential contribution coming from the \ac{kk} mono\-poles in this case as well.
As observed in~\cite{Nii:2014jsa}, in this case the \ac{kk} monopoles have too many zero modes
and should not give rise to a superpotential. 
As one can see from the Callias index theorem,
there  are in total four fermionic zero modes, two coming from the gauginos and two from the adjoint matter
fermions.
Nevertheless the extra fermionic zero modes from the adjoint fermion can be contracted because the 
superpotential gives rise to a potential term $X^{n-1} \psi_X^2$. After taking this contraction into account, 
we are left with the two fermionic zero modes coming from the gauginos,
and the \ac{kk} superpotential for the (dressed) monopoles is generated.
We can parameterize these directions in terms of a set of operators $T_{i}$ and $\widetilde T_i$ and the
$\eta$-superpotential is of the form 
\begin{equation}
W_\eta = \eta \sum_{j=0}^{n-1} T_j \widetilde T_{n-j-1}.
\end{equation}
There is a similar contribution in the magnetic theory, necessary to preserve the duality at finite radius of the circle.
We now can proceed as above with a real mass flow, and by identifying the 
topological vacuum of the dual theory one can reduce this effective duality to the duality of~\cite{Kim:2013cma}.
We can also reduce the duality in a slightly different way, along the lines of~\cite{Nii:2014jsa},
by deforming the duality by considering a polynomial superpotential 
for the adjoint and 
considering the four-dimensional theory in a vacuum of the adjoint. In this way one has 
a set of $n$ decoupled \ac{sqcd}-like theories.
One can perform the reduction in each sector separately, in both the electric and the magnetic phase.
The real mass flow to the duality of~\cite{Kim:2013cma} can be done in each sector 
separately. This gives rise to the expected duality in a broken phase.
The agreement between the two procedures has been shown in~\cite{Amariti:2014iza}.
This second approach will be useful in deriving the reduction of the dualities from the~brane dynamics.

\subsection{Generalizations}

The discussion above can be generalized to cases with real gauge groups and with tensor matter.

Let us first summarize the case with real gauge groups and fundamental matter, discussed 
in~\cite{Aharony:2013dha} for the symplectic case and in~\cite{Aharony:2013kma}
for the orthogonal one.

The second case is more involved because it requires the analysis of the global properties of the
gauge group and we will not elaborate on this subtlety here.
We will restrict the analysis to the gauge algebras $sp(2N_c)$ and $so(N_c)$.

When reducing a theory on the circle, the contribution of the \ac{ahw} superpotential can be formulated algebraically as
\begin{equation}
\label{eq:etagen}
W_\eta = \frac{2 \eta}{\alpha_0^2} e^{\alpha_0^* \Phi},
\end{equation}
where $\alpha_0$ corresponds to the 
affine root of the Lie algebra and  $\alpha_0^*$ corresponds to the associated coroot.
We also identified the Coulomb branch coordinates with 
\begin{equation}
\phi= \sigma/e_3^2+ i \gamma,
\end{equation} where $e_3$ is the gauge coupling and $\gamma$
is the dual photon~\eqref{eq:dualphoton}.

Once the $\eta$-superpotential is added, we can claim an effective duality on the circle as
in the case discussed above. The cases of interest are the $A_N$,  $B_N$,  $C_N$ and  $D_N$ series
corresponding to the $su(N)$, $so(2N+1)$, $sp(2N)$ and $so(2N)$ gauge algebras, respectively.
Note that here, we refer to the $su(N)$ case, while above we discussed the $u(N)$ case.
On the~brane side, we will comment on the difference of the interpretation of these cases.

The conventional Aharony dualities are obtained by a real mass flow. 
In the case of $sp(2N)$, we do not have to consider a topological vacuum on the dual side.
The meson $M$ of the dual description in this case is broken into two massless fields.
One of them corresponds to the meson of the Aharony dual theory, while the other 
corresponds to the monopole $T$ that parameterizes
the Coulomb branch of the electric theory. By a scale matching relation, one can show that
the $\eta$ superpotential  becomes the extra contribution of Aharony duality.
The final form of the superpotential is 
\begin{equation}
W = M q q +t T.
\end{equation}
In the orthogonal case, the three-dimensional Aharony duality is obtained without considering any real mass flow.
In this case, one can eliminate the $\eta$ superpotential in the electric theory safely and in the dual 
theory one has to consider a topological vacuum.
We refer the reader to~\cite{Aharony:2013kma} for details.

\bigskip
Further generalizations are possible. The picture that has emerged so far serves as a guideline for studying
the reduction of all the cases discussed in Section~\ref{sec:4d} in presence of tensor matter and unitary or real gauge group.
In presence of tensor matter, one can break the gauge group into \ac{sqcd} sectors by deforming the theory with a polynomial
superpotential for the tensors. These sectors can have a unitary or
real gauge group depending on the details of the model.
Then one has to reduce each sector separately, according to the rules explained above.
This results in new effective three-dimensional dualities on the circle. One can also flow in each broken sector to an Aharony-like duality
by assigning real masses and choosing the topological vacua when necessary.

This procedure has been used in~\cite{Amariti:2015vwa} to study the reduction of the four-dimensional duality for the $Sp(2N_c)$ 
gauge theory with $2N_f$ fundamentals and one adjoint discussed in Section~\ref{sec:4d-extensions}.
The polynomial superpotentials in the electric and in the magnetic theory are given by
\begin{equation}
\label{lambda}
W_e = \sum_{i=1}^{n+1} \lambda_i \Tr S^{2i}
\quad \rightarrow \quad
W_m = \sum_{i=1}^{n+1} \widetilde \lambda_i \Tr S^{2i}.
\end{equation}
In the electric theory we can break  $Sp(2N_c)$ to $Sp(2 r_0) \times \prod U(r_i)$, with
$\sum r_i = N_c$.
In each sector, there is an $\eta$-superpotential generated on the circle:
for $Sp(2r_0)$  $W=\eta_0 T$ is generated, while for each $U(r_i)$ sector,
a superpotential $W= \eta_i T_i \widetilde T_i$ is generated. The coupling constant  $\eta_i$ corresponds to  $\Lambda_i^{b_i} $. 
In the magnetic theory, the gauge group $Sp(2 \widetilde N_c)$ is broken to 
$Sp(2 \widetilde  r_0)  \times  \prod U(\widetilde   r_i)$, where
$\sum \widetilde  r_i = \widetilde  N_c$.
Also in this case, a \ac{ahw} superpotential is generated in each broken sector.
The reduction to three dimensions is performed via a real mass flow. In the $Sp(2 \widetilde r_0)$ 
sectors, there is no Higgsing, while in the $U(\widetilde r_i)$ case, we have to consider the topological vacua.
By considering the interactions of the monopoles in each Higgsed phase and by using local mirror symmetry when necessary,
we obtain the dual Aharony-like theory. 
More concretely, the massive singlets are integrated out  and the massless ones interact with the monopoles of the $U(\widetilde r_i)$ theories. 
These singlets become the electric monopoles expected in the dual superpotential.

\section{The~brane picture}
\label{sec:redbranes}

In this section, we study the reduction of four-dimensional dualities from the perspective 
of~brane engineering.
We will translate the field theory constructions into the~brane language, which 
offers a unified version of the results that have emerged on the field theory side.
The reduction is performed via a T-duality along a circle
that we will keep finite also after the duality.
In this way, we are able to capture the quantum effects leading to the
$\eta$-superpotential in the~brane picture.
It also offers an explanation of the breaking of the axial symmetry in a geometric 
language.
The real mass flow leading to Aharony-like dualities has a clear explanation in the~brane picture
and the relation to the dualities in the new sectors obtained from the topological vacua 
can be explained in terms of a local mirror symmetry.
In this way, we are able to give a~brane realization of Aharony duality explaining the role
of the extra monopoles in the dual phase.

\medskip

We start our analysis with a general review of the reduction 
by discussing the generation of the non-perturbative superpotential for \ac{sqcd} at finite radius from the~brane perspective. 
We start by discussing the case of pure \ac{sym} ($N_f=0$). This analysis introduces two of the main characters, namely T--duality and \D1 monopoles,
that will be very useful in the following.
After this, we extend the discussion to the case with fundamental
matter and describe  
the reduction of the four-dimensional duality for
$U(N_c)$ \ac{sqcd}, essentially following the discussion in~\cite{Amariti:2015yea}.
Then we explain the generalization to the case with an adjoint field.
In the following, we add orientifold planes in order to connect to the description of the real gauge groups and tensor matter, following the discussion in~\cite{Amariti:2015mva}.

Our goal is to explain the different three-dimensional limits discussed on the field theory side. From the~brane description,
we will see that we naturally require a double-scaling limit of the radius of the circle and the real mass
in order to recover the standard three-dimensional dualities.

\subsection{The \(\eta\) superpotential from T--duality}
\label{sec:superpotential-from-branes}

\subsubsection{Gaugino condensate}
To introduce the relation between D~branes and monopole operators in three dimensions,
we start with a discussion of gaugino condensation in four dimensional $\mathcal{N}=1$ \ac{sym} 
in terms of~brane constructions. This derivation has been presented in~\cite{Davies:1999uw} and we refer the reader there for details.

Consider four-dimensional $\mathcal{N}=1$ $SU(N_c)$ \ac{sym} on
$\setR^3 \times S^1$. The supersymmetric vacuum can be expressed in
terms of a gaugino condensate. This gaugino bilinear has been
obtained in~\cite{Novikov:1985ic,Amati:1988ft} from an instantonic
contribution in four dimensions. Equivalently, on
$\setR^3 \times S^1$, one obtains the same result by splitting 
the instantons in terms of $N_c$ monopoles \cite{Brodie:1998bv}. In the
three-dimensional effective description, $N_c-1$ of them are \ac{bps}
monopoles while the $N_c$-th one is usually called the \ac{kk}
monopole. We will not discuss this construction and refer the
reader to the original reference for details.

Interestingly, the same effect can be reproduced via the~brane engineering of the 
theory in \tIIA. 
As discussed in Sec.~\ref{sec:4dbranes}, the $\mathcal{N}=1$ \ac{sym} theory  is described by an \NS5~brane, an \NS5'~brane and $N_c$ \D4~branes extended as shown in Table~\ref{tab:IIA-sqcd-no-flavors}. The \D4s are suspended between the \NS5 and the \NS5'. The four-dimensional gauge coupling is \(e_4^2 = g_4^2/\ell_6 =  (2\pi)^2 \sqrt{\alpha'} / \ell_6 \), where \(\ell_6 \) is the distance between the \NS5~branes and \(g_4^2 = (2\pi)^2 \sqrt{\alpha'}\) is the \D{}~brane coupling constant.

\begin{table}
  \centering
  \begin{tabular}{LCCCCCCCCCC}
    \toprule
         & 0      & 1      & 2      & 3      & 4      & 5      & 6      & 7 & 8      & 9      \\
    \midrule
    \NS  & \times & \times & \times & \times & \times & \times &        &   &                 \\
    \NS' & \times & \times & \times & \times &        &        &        &   & \times & \times \\
    \D4  & \times & \times & \times & \times &        &        & \times &   &                 \\
    \bottomrule
  \end{tabular}
  \caption{Brane configuration for \(\mathcal{N}=1, d =4 \) \ac{sym}. The \D4~branes are suspended between the two \NS5s.}
  \label{tab:IIA-sqcd-no-flavors}
\end{table}

\subsubsection{\(\eta\) superpotential}
If \(x_3\) is compact, \(x_3 \sim x_3 + 2 \pi R_3\), we can perform a T--duality along that direction.
In the resulting \tIIB frame, the \D4s have turned into \D3~branes
while the \NS{} and \NS'~branes are unchanged.
This configuration describes $U(N_c)$ \ac{sym} on $\setR^3 \times S^1$.
As shown in~\cite{Witten:1982df}, the entire Coulomb branch is lifted and there are $N_c$ isolated vacua.
To see that in the string theory picture, observe that the vacua correspond to
stable supersymmetric configurations of the~brane system.
Again, in absence of \D5~branes there is a
repulsive force between the \D3s so that the unique  stable
configuration is obtained when the \D3~branes are distributed along $x_3$ at equal distances
All moduli are lifted as the \D3~branes cannot move freely due to the repulsive force.

\begin{figure}
  \begin{center}
\begin{tikzpicture}
    \begin{scope}[shift={(-3,-2)}]
      \begin{footnotesize}
      \draw[->] (-0.2,0) arc (-140:135:.2 and .4);
      \draw[->] (-0.4,.25) -- (.6,.25);
      \draw (0.6,.45) node[]{\(x_6\)};
      \draw (0.1,.9) node[]{\(x_3\)};
      \end{footnotesize}
    \end{scope}

      \begin{scope}[very thick]
        \begin{footnotesize}
          \draw[fill=black!10,black!10] (0,-.5) rectangle (7,.5); 
          \draw[D3col] (0,-.5) -- (7,-.5);
          \draw[D3col] (0,.5) -- (7,.5);
          \draw[D3col] (.5,1.5) -- (6.7, 1.5);
          \draw[D3col] (.5,-1.5) -- (6.7, -1.5);

          \draw[NScol,fill=white] (0,0) ellipse (1 and 2);
          \draw[NScol,dashed] (6,0) ellipse (1 and 2);

          \draw[D3col] (0,2) -- (6,2);
          \draw[D3col] (0,-2) -- (6,-2);

          \draw[D5col,fill=white] (6.9,.5) circle (5pt);
          \draw[D5col] (6.9,.5) node[cross=4pt]{};
          \draw[D5col,fill=white] (6.9,-.5) circle (5pt);
          \draw[D5col] (6.9,-.5) node[cross=4pt]{};

          \draw[D5col,fill=white] (6,2) circle (5pt);
          \draw[D5col] (6,2) node[cross=4pt]{};
          \draw[D5col,fill=white] (6,-2) circle (5pt);
          \draw[D5col] (6,-2) node[cross=4pt]{};

          \draw[D5col,fill=white] (6.7,1.5) circle (5pt);
          \draw[D5col] (6.7,1.5) node[cross=4pt]{};
          \draw[D5col,fill=white] (6.7,-1.5) circle (5pt);
          \draw[D5col] (6.7,-1.5) node[cross=4pt]{};

          \draw[D5col,fill=white] (5.1,.8) circle (5pt);
          \draw[D5col] (5.1,.8) node[cross=4pt]{};
          \draw[D5col,fill=white] (5.1,-.8) circle (5pt);
          \draw[D5col] (5.1,-.8) node[cross=4pt]{};

          \draw[NScol] (-.8,-2) node[]{\NS5};
          \draw[NScol] (6.7,-2) node[]{\NS5'};
          \draw[D3col] (3,-1) node[]{\D3};         
          \draw (4,0) node[]{\D1};
          \draw[D5col] (7.4,.5) node[]{\D5};

        \end{footnotesize}
      \end{scope}
    \end{tikzpicture}
  \end{center}
  
  \caption{\D1~branes (in gray) stretched between \D3~branes with
    compact \(x_3\). The \D5~branes are represented by \(\otimes\) symbols, the \NS5 by a continuous line and the \NS5' by a dashed line.}
  \label{fig:D1-between-D3}
\end{figure}
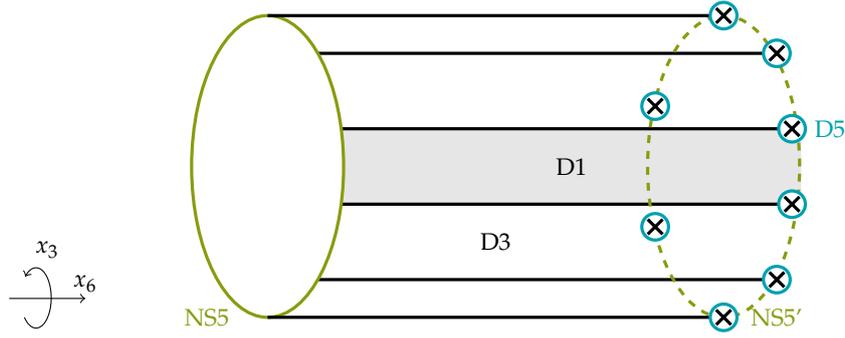

The repulsive force is a non-perturbative quantum effect. 
From the three-dimensional field theory point of view, a
non-perturbative superpotential induced by three-dimensional
instantons is generated.  These instantons are monopole configurations of the four-dimensional theory.
In the~brane picture, these monopoles are represented by Euclidean \D1--strings stretched between 
each pair of \D3~branes and the \NS{} and \NS'~branes, as depicted in Figure~\ref{fig:D1-between-D3} 
as shaded area along $x_6$ and $x_3$, respectively. 
The effect of the monopoles has been computed
in~\cite{deBoer:1997ka,Hanany:1996ie} and is given by the exponential $e^{-S}$, where $S$ is the \D1 world-sheet action.
This action has two pieces, the Nambu--Goto part and a contribution
from the boundary of the \D1s:
\begin{itemize}
\item The Nambu--Goto term is proportional to the area of the \D1 and involves the scalar $\sigma$ that parameterizes the position of the \D3;
\item The boundary term is proportional to the dual photon $\gamma$.
\end{itemize}
Putting them together, we obtain
\begin{equation}
  \label{WD1}
  W = \sum_{i=1}^{N_c-1} e^{\Phi_i-\Phi_{i+1}},
\end{equation}
where $\phi_i = \sigma_i/e_3^2+i \gamma_i $ and $e_3^2 = (\sqrt{\alpha'}/R_3)
(g_3^2 /\ell_6) = 2 \pi \sqrt{\alpha'}/ (R_3 \ell_6)$ is the
three-dimensional gauge coupling\footnote{Here \(g_3^2 = 2 \pi \), and \( \sqrt{\alpha'}/R_3\) is the contribution of the \tIIB dilaton so that \(e_4^2/ e_3^2 = 2 \pi R_3\).}.       
The result can be naturally expressed in terms of Coulomb branch coordinates
and we see explicitly the monopole operators $e^{\Phi_i}$.

Now, if $x^3$ is compact there is another contribution from \D1~branes stretching from the $N_c$th to the $1$st \D3~branes,
which are respectively at $\sigma_{N_c}$ and $\sigma_1 + 2 \pi R_3$. 
By following the calculation of~\cite{Davies:1999uw,Katz:1996th} one finds
\begin{equation}
  \label{etaWW}
  W_\eta = \eta \, e^{\Phi_{N_c}-\Phi_1}  ,
\end{equation}
where $\eta$ incorporates the radius dependence from the position of the first
\D3~brane. This is the $\eta$--superpotential $\eqref{eq:etagen}$
which plays an essential role in the reduction from four to three
dimensions as shown in Section~\ref{sec:redfields}.
From the point of view of field theory, the Euclidean \D1~branes are
the \acs{bps} monopoles and the one coming from the extra \D1 is the \ac{kk} monopole.

\subsubsection{Affine algebra and Toda chain}

\begin{figure}
  \centering
  \begin{tikzpicture}
  \begin{scope}[thick]
    \begin{footnotesize}
      \begin{scope}
        \foreach \position/\Dname in {-2/{\(\sigma_1\)}, -1/{\(\sigma_2\)}, 1/{\(\sigma_{N-1}\)}, 2/{\(\sigma_N\)}}
        {
          \draw (\position, 1) -- (\position, -1);
          \draw (\position, -1.3) node {\Dname};
        }
        \foreach \leftpos/\rightpos/\heightpos in {-2/-1/-.1, -1/-.5/.2, .5/1/-.1 ,1/2/.1 }
        {
          \draw (\leftpos, \heightpos) -- (\rightpos, \heightpos);
        }
        
        \draw (0,0) node {\(\dots\)};
         \draw[->] (-3,-2) -- (-2,-2);
      \draw[->] (-2.8,-2.2) -- (-2.8,-1.2);
      \draw (-1.8,-2) node[]{\(x_3\)};
      \draw (-2.8,-1) node[]{\(x_{0,1,2}\)}; 

      \end{scope}

      \begin{scope}[node distance=4em, ch/.style={circle,draw,on chain,inner sep=2pt},chj/.style={ch,join},every path/.style={shorten >=4pt,shorten <=4pt},line width=1pt,baseline=-1ex,start chain,shift={(4,0)}]
        \dnode{\(\sigma_1 - \sigma_2\)}
        \dnode{\(\sigma_2 - \sigma_3\)}
        \dydots
        \dnode{}
        \dnode{\(\sigma_{N-1} - \sigma_N\)}
      \end{scope}

      \begin{scope}[shift={(0,-4)}]
        \draw (1.5,0) arc (0:240:1.5);

        \draw [dash pattern= on 2pt off 20pt on 2pt off 3pt on 2pt off 3pt on 2pt off 3pt on 2pt off 3pt on 2pt off 3pt on 2pt off 30pt ] (1.5,0) arc (0:-90:1.5);

        \draw [blue] (0,1.5) arc (90:150:1.5);

        \foreach \braneanglepos/\Dname in {30/{\(\sigma_{N-1}\)}, 90/{\(\sigma_N\)}, 150/{\(\sigma_{1}\)}, 210/{\(\sigma_{2}\)}}
        {
          \tikzcrosscircle{({1.5 * cos(\braneanglepos)},{1.5 * sin(\braneanglepos)})}
          \draw ({1.5 * cos(\braneanglepos)},{1.5 * sin(\braneanglepos)}) node [right=1ex] {\Dname};
        }

        \tikzcircleaxis{(-1.7,-1.7)}{\(x_3\)}
      \end{scope}

      \begin{scope}[start chain,node distance=3ex and 4em, ch/.style={circle,draw,on chain,inner sep=2pt},chj/.style={ch,join},every path/.style={shorten >=4pt,shorten <=4pt},line width=1pt,baseline=-1ex, shift={(4,-4)}]
        \dnode{\(\sigma_1 - \sigma_2\)}
        \dnode{\(\sigma_2 - \sigma_3\)}
        \dydots
        \dnode{}
        \dnode{\(\sigma_{N-1} - \sigma_N\)}
        \begin{scope}[start chain=br going above]
          \chainin(chain-3);
          \node[ch,join=with chain-1,join=with chain-5,label={[inner sep=1pt]10:\(\sigma_N - \sigma_1\)}] {};
        \end{scope}
        \draw[blue] (3.14,.83) circle (.1);

      \end{scope}
    \end{footnotesize}
  \end{scope}

\end{tikzpicture}
   \caption{Branes and Dynkin diagrams for \(A_N\) and \(\widetilde
    A_N\).  The left column shows the S-dual configuration of \F1--strings stretched between \D3~branes.
  In the right column we depict the corresponding \(A_N\) and
  \(\widetilde A_n\) Dynkin diagrams.
  After compactification a new string appears between
  \(\sigma_1\) and \(\sigma_N\) and corresponds to the affine node
  (in blue in the~brane cartoon and in the Dynkin diagram).}
  \label{fig:Dynkin-An}
\end{figure}
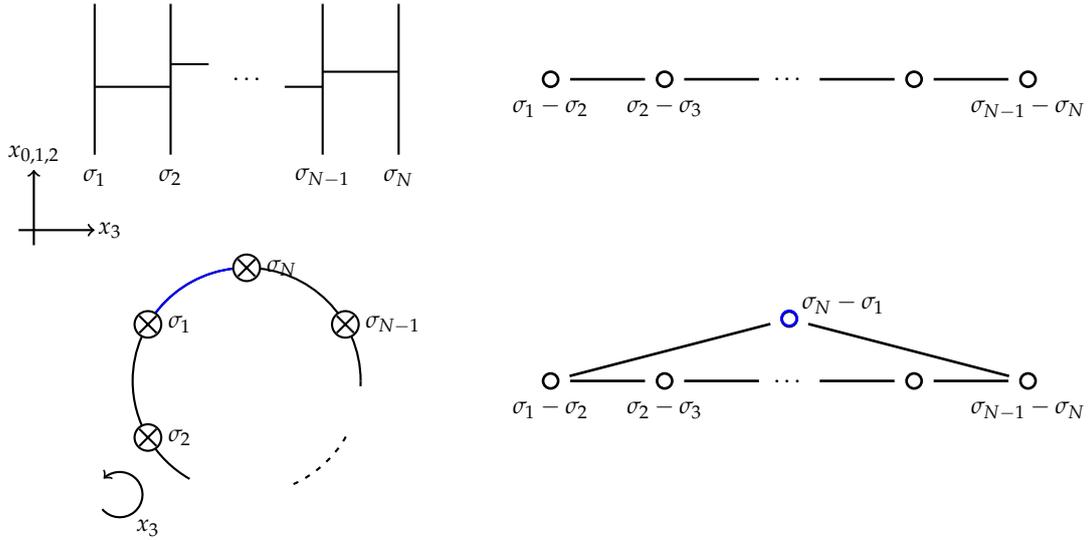

There is an algebraic approach which allows an elegant description of
this construction. The fundamental (in the sense of~\cite{Weinberg:1979zt,Weinberg:1982ev}) \acs{bps} monopoles are labeled by the simple co-roots of the Lie co-algebra.
For unitary gauge groups $G$ this corresponds to placing the \(i\)--th \D1~brane between the \(i\)--th and the \((i+1)\)--th  \D3~brane (for $i=1,\dots,\rank(G)$).
It is useful to study the S--dual configurations where \D1~branes become \F1--strings.
In this picture the \D3~branes are still distributed on the circle and connected by \F1--strings, as depicted in the upper left corner of Figure~\ref{fig:Dynkin-An}.
The spectrum of the allowed \acs{bps} \F1--strings is given by the simple roots of the corresponding Lie algebra~\cite{Garland:1988bv,Hanany:2001iy}.
For a unitary gauge group the simple roots of the $A_N$ series correspond to $\sigma_i-\sigma_{i+1}$,
\emph{i.e.} to the difference between the positions of two consecutive \D3~branes.

The picture incorporates very naturally the \acs{kk} monopoles due to
the compact direction. It turns out that the extra \F1 string, which
winds around the circle connecting, for unitary $G$, the $1$st and the
$n$-th \D3~brane, can be accounted for by extending the Dynkin diagram
to its \emph{affine} version.

Summing the contributions from the \acs{bps} and the \acs{kk}
monopoles we obtain the superpotential on the compact Coulomb branch,
finding an affine Toda potential for the associated affine
algebra~\cite{Davies:2000nw}
\begin{equation}
  \label{eq:W-affine}
  W(\Sigma) = \sum_{i=1}^{\rank(G)} \frac{2}{\alpha_i^2} \exp[\alpha_i^* \cdot \Phi] +  \frac{2 \eta}{\alpha_0^2}
\exp[\alpha_0^* \cdot \Phi] \; ,
\end{equation}
where the scalar component of $\Phi$ is $\phi= \sigma/e_3^2+ i  \gamma$, $\alpha_i$ are the simple roots and $\alpha^*_i$ are the associated co-roots.
The extra simple root $\alpha_0$ corresponds to the \ac{kk} monopole
and the corresponding contribution to $\eqref{eq:W-affine}$ is commonly referred to as $\eta$--superpotential.

In the unitary case, we have the affine algebra  $\widetilde A_N$, where the extra simple root is associated to
the combination $\sigma_{N}-\sigma_1$ (see Figure~\ref{fig:Dynkin-An}).

For $SU(N_c)$ theories
\footnote{In the $U(N_c)$ case the same result holds but the last (affine) root splits in two terms
$T =e^{\phi_1}$ and  $\widetilde T=e^{-\phi_N}$.}
the superpotential associated to this diagram is
\begin{equation}
\label{etaWSU}
W = \sum_{i=1}^{N_c-1} \frac{1}{Y_i}  + \eta Y_{N_c} \; ,
\end{equation}
where $Y_i = e^{(\sigma_i-\sigma_{i+1})/e_3^2+i(\gamma_i - \gamma_{i+1})}$.
The last term in (\ref{etaWSU}) breaks explicitly the $U(1)_A$ symmetry
in the three-dimensional field theory. This symmetry is associated to the rotation
in the $(4,5)$ plane in the~brane picture. The geometric realization
of the breaking of this symmetry for compact $x_3$ has been discussed in~\cite{Amariti:2015yea}.

\subsubsection{Matter fields}

The picture is richer if we introduce matter fields.
Now, two directions of the moduli space remain unlifted, which can be seen from the~brane picture as follows.
In the \tIIA frame, fundamental matter is associated to $N_f$
\D6~branes extended along $0123789$ and sitting on the \NS'~brane
(see Sec.~\ref{sec:sqcd}). In the T--dual
frame they become \D5~branes. Strings stretched between the stack of \D3~branes and the \D5s correspond to 
$N_f$ massless fundamentals $Q$ and anti-fundamentals $\widetilde Q$ (see Sec.~\ref{sec:3dbranes}).

When \D5~branes sitting at $x_3 =0$ intersect the worldsheet of the \D1--strings,
they contribute two additional zero modes to the \D1--instanton and the superpotential in Eq.~(\ref{WD1}) is not generated.
In this sense, the \D5~branes screen the repulsive force between the \D3~branes~\cite{deBoer:1997kr,Elitzur:1997hc}.
Without loss of generality, we can chose the screening to happen
between the first and the $N_c$th \D3~brane, which are now free to
move without being subjected to any force. One modulus is nevertheless
lifted by the interaction mediated by the superpotential Eq.~(\ref{etaWW}). 

This is different from the non-compact case where there is no
superpotential $\eqref{etaWW}$ and the forceless motion of the
$1$st and the $N_c$th \D3~brane leads to a two-dimensional moduli space.

\subsection{Reduction of SQCD}
\label{sec:Braneology}

Having understood the origin of the \(\eta\) superpotential, we can now
consider the~brane realizations of the four-dimensional dualities and
relate them to three dimensions via T--duality.

As a first example, consider the reduction of four-dimensional Seiberg duality for \textsc{sqcd} 
to three dimensions.
It is convenient to consider the four-dimensional gauge symmetry to be $U(N_c)$ rather than $SU(N_c)$. 
In field theory, this enhancement is obtained by gauging the baryonic
symmetry (see Sec.~\ref{sec:4dsqcd}). 

As in Sec.~\ref{sec:sqcd}, we start by considering a stack of $N_c$ \D4~branes, one \NS5~brane, one \NS5'~brane and $N_f$ \D6~branes. In this \tIIA description, the~branes are extended as shown in Table~\ref{tab:IIA-sqcd}.
\begin{table}
  \centering
  \begin{tabular}{lcccccccccc}
    \toprule  & 0 & 1 & 2 & 3 & 4 & 5 & 6 & 7 & 8 & 9\\
    \midrule
    NS & $\times$ & $\times$ & $\times$ & $\times$ & $\times$ & $\times$ &  &  & \\
    NS' & $\times$ & $\times$ & $\times$ & $\times$ &  &  &  &  & $\times$ & $\times$\\
    \D4 & $\times$ & $\times$ & $\times$ & $\times$ &  &  & $\times$ &  & \\
    \D6 & $\times$ & $\times$ & $\times$ & $\times$ &  &  &  & $\times$ & $\times$ & $\times$\\ \bottomrule
      \end{tabular}
  \caption{Brane configuration for \(\mathcal{N}=1, d =4 \) \textsc{sqcd} with \(N_f \) flavors. The \(N_c\) \D4~branes are extended between the \NS5 and the \(N_f\) \D6~branes sit on the \NS5'.}
  \label{tab:IIA-sqcd}
\end{table}
The two possible configurations with the \NS5 either on the left or the
right of the  \NS5' are related by a \ac{hw} transition
(Sec.~\ref{sec:hw}). What we want to do is to compactify the two phases along $x_3$ and study the 
resulting theories at finite radius.

\subsubsection{Duality at finite radius}
\label{sec:duality-finite-radius}

If we consider $x_3$ to be compact, we can perform a T--duality and obtain an 
effective three-dimensional $\mathcal{N}=2$
theory. In this case, the \D4 and the \D6~branes become \D3s and \D5s respectively, while the
\NS{} and \NS'~branes are left unchanged. In Figure~\ref{fig:D6-two-sides},
the electric and magnetic Seiberg--dual theories are shown.
In the \tIIB description the~branes are extended as shown in Table~\ref{tab:IIB-sqcd}.
This is a $U(N_c)$ gauge theory with $N_f$ fundamentals and anti-fundamentals as discussed in Section~\ref{sec:superpotential-from-branes}, with global symmetry group $SU(N_f)_L \times SU(N_f)_R \times U(1)_R \times U(1)_J$.
The axial symmetry $U(1)_A$ under which $Q$ and $\widetilde Q$ have the same charge is broken by the superpotential in Eq.~(\ref{etaWW}).
\begin{table}
  \centering 
  \begin{tabular}{lcccccccccc}
    \toprule  & 0 & 1 & 2 & 3 & 4 & 5 & 6 & 7 & 8 & 9\\
    \midrule
    NS & $\times$ & $\times$ & $\times$ & $\times$ & $\times$ & $\times$ &  &  & \\
    NS' & $\times$ & $\times$ & $\times$ & $\times$ &  &  &  &  & $\times$ & $\times$\\
    \D3 & $\times$ & $\times$ & $\times$ &  &  &  & $\times$ &  & \\
    \D5 & $\times$ & $\times$ & $\times$ &  &  &  &  & $\times$ & $\times$ & $\times$\\ \bottomrule
  \end{tabular}
  \caption{Brane configuration for \(\mathcal{N}=2, d =3 \) \textsc{sqcd}. The \(N_c\) \D3~branes are extended between the \NS5 and the \(N_f\) \D5~branes sit on the \NS5'.}
  \label{tab:IIB-sqcd}
\end{table}

Let us see how the global symmetries are realized in the~brane picture
and how having a compact \(x_3\) direction is the geometric
interpretation of the fact that
the three-dimensional theory does not have an axial symmetry, as
discussed in Sec.~\ref{sec:redfields}.
For ease of exposition, we concentrate on the electric frame.

\begin{itemize}
\item The rotations in the $(4, 5)$ and in the $(8, 9)$~plane discussed in section \ref{sec:sqcd} still correspond to
the  \(R\)--symmetry  and to the axial symmetry.
While admissible in three dimensions, on $\mathbb{R}^3 \times S^1$ the latter is broken as will be discussed in the next paragraph.
\item The non-Abelian flavor symmetry comes from the stack of $N_f$ \D5 branes.
The branes can be broken at the intersection with the \NS5', leading
to two semi-infinite stacks, $\D5_L$ and $\D5_R$, one extended in
$x_7>0$ and the other in $x_7<0$: this corresponds to the fact that
the symmetry has a notion of chirality.
The flavor group is $U(N_f) \times U(N_f)$. One combination of the two $U(1)$ factors appears as baryonic and one as axial symmetry. 
The $U(1) \subset U(N_f)$ real mass, \emph{i.e.} the \ac{vev} of the scalar in the corresponding $U(1)$ vector multiplet, 
is a collective shift of the stack of semi-infinite \D5s. 
The \emph{axial} $U(1)$ subgroup in $U(N_f) \times U(N_f)$ comes from
separating the two stacks and moving them in opposite directions; 
the \emph{diagonal} subgroup corresponds to shifting both stacks of \D5 together.
This is the same as a shift of the stack of the \D3~branes, 
which is a $U(1) \subset U(N_c)$ gauge transformation (the gauged baryonic symmetry).
The separation of the stacks acts on the fundamentals and the
anti-fundamentals (this is the axial symmetry) and corresponds to turning on a real mass for $U(1)_A$.\\
On a circle (when $x_3$ is compact) the $U(1)_A$ is broken. In the~brane picture 
this breaking can be visualized as follows. 
When moving the stack of $\D5_L$s in $x_3 < 0 $ and the stack of $\D5_R$s in $x_3 > 0$,
charge conservation requires the generation of a $(1,N_f)$ fivebrane along the directions $x_3$ and $x_7$~\cite{Aharony:1997ju,Kitao:1998mf} (see Figure~\ref{U(1)A}).
The \NS{}~brane now cannot close anymore on the circle without breaking supersymmetry.
This obstruction precludes the realization of the $U(1)_A$ symmetry in
the~brane setup. As we have seen before, a compact \(x_3\) leads to an
\(\eta\)-superpotential, here we find a
geometrical origin  for the breaking the $U(1)_A$ symmetry.
\begin{figure}
  \centering
    \begin{tikzpicture}
    \begin{scope}[shift={(-2,-2)}]
      \begin{footnotesize}
      \draw[->] (-0.2,0) arc (-140:135:.2 and .4);
      \draw[->] (-0.4,.25) -- (.6,.25);
      \draw (0.6,.45) node[]{\(x_7\)};
      \draw (0.1,.9) node[]{\(x_3\)};
      \end{footnotesize}
    \end{scope}
    \begin{scope}[thick]
      \begin{footnotesize}
        \draw[thin] (0,0) ellipse (1 and 2);
        \draw[thin] (4,2) arc (90:-90:1 and 2);
        \draw[thin] (0,2) -- (4,2);
        \draw[thin] (0,-2) -- (4,-2);

        \draw (2.5,2) arc (90:-90:1 and 2);
        \draw[dashed] (2.5,-2) arc (-90:-270:1 and 2);

        \draw (1,0) -- (5,0);

        \draw (2.4,.4) node[]{\(N_f\) \(\D5_L\)};
        \draw (4.2,-.4) node[]{\(N_f\) \(\D5_R\)};

        \draw (3.8,1.2) node[]{\NS5'};
      \end{footnotesize}
    \end{scope}

    \begin{scope}[shift={(6.5,0)},thick]
      \begin{footnotesize}
        \draw[thin] (0,0) ellipse (1 and 2);
        \draw[thin] (4,2) arc (90:-90:1 and 2);
        \draw[thin] (0,2) -- (4,2);
        \draw[thin] (0,-2) -- (4,-2);

        \draw (3,2) arc (90:30:1 and 2);
        \draw (1.5,-2) arc (-90:-30:1 and 2);

        \draw (0.86,-1) -- (2.36,-1) -- (3.86,1) -- (4.86,1);

        \draw (1.6,-.7) node[]{\(N_f\) \(\D5_L\)};
        \draw (4.3,.6) node[]{\(N_f\) \(\D5_R\)};

        \draw (3.2,1.4) node[]{\NS5'};
        \draw (1.8,-1.4) node[]{\NS5'};

        \draw (3.7,-.2) node[]{\((1,N_f)\)};
 
      \end{footnotesize}
    \end{scope}
\end{tikzpicture}
  \caption{Geometric realization of the breaking of \(U(1)_A\)
    for compact \(x_3\). If the stacks of \D5s are moved apart to give a
    real mass, the \NS5'~brane cannot be closed in the direction \(x_3\) without a displacement in \(x_7\).}
  \label{U(1)A}
\end{figure}
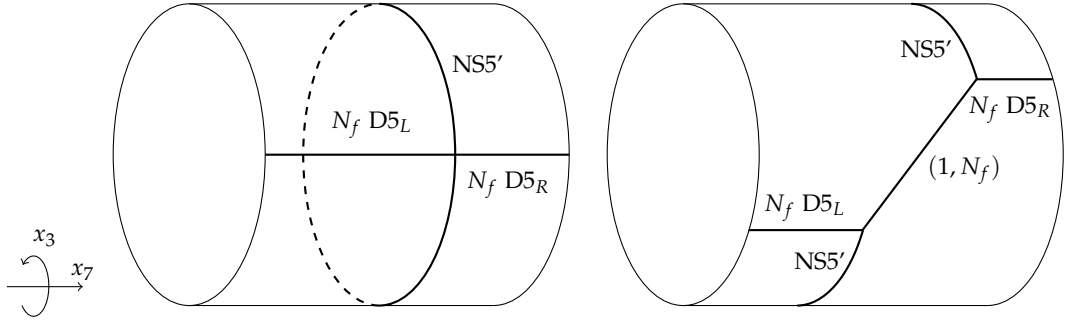

The situation is different for $U(1)_B$.
This symmetry is realized by the motion of the entire stack of \D5~branes with respect to the stack of \D3~branes
along \(x_3\). The \D5~branes 
can slide together on the \NS{}~brane along $x_3$, without generating any $(1,N_f)$ fivebrane, and the symmetry is 
still realized in the~brane picture. 
Observe that in the three-dimensional picture this symmetry is gauged
since we can understand it as coming from the motion of the
\D3~branes as opposed to the motion of the \D5~branes.
\item The last symmetry is the topological $U(1)_J$ that shifts the
  dual photon. This is an effect that is not visible in the \tIIB description
and a geometric interpretation would require the lift of the configuration
to M--theory where the dual photon is associated to the  $x_{10}$ direction.
This is not surprising, since already in gauge theory this symmetry is
not manifest in the Lagrangian but comes from the Bianchi identity. In
the \tIIB description, however
we still have control of the real mass parameter associated to this
symmetry. This is the \ac{fi} term coming from to the displacement of the \NS{} and \NS'~branes along $x_7$.
\end{itemize}

The same reduction can be performed in the magnetic picture where the
discussion is essentially equivalent.
The superpotential $M\,q \tilde q$ is combined with the $\eta'$ term. At finite radius, this theory can be treated as an effective three-dimensional theory if the
radius of the T--dual circle is large enough, and in this sense we
have a new three-dimensional duality.

\subsubsection{Flowing to Aharony duality}
\label{sec:aharony-duality}

In Section~\ref{sec:redfields} we have seen how to flow to Aharony
duality by turning on 
real mass terms for some of the quarks. Now we would like to reproduce
this flow in terms of the~brane engineering, where the real masses
correspond to the motion of the \D5~branes in the $x_3$ direction, as
discussed in the previous section.

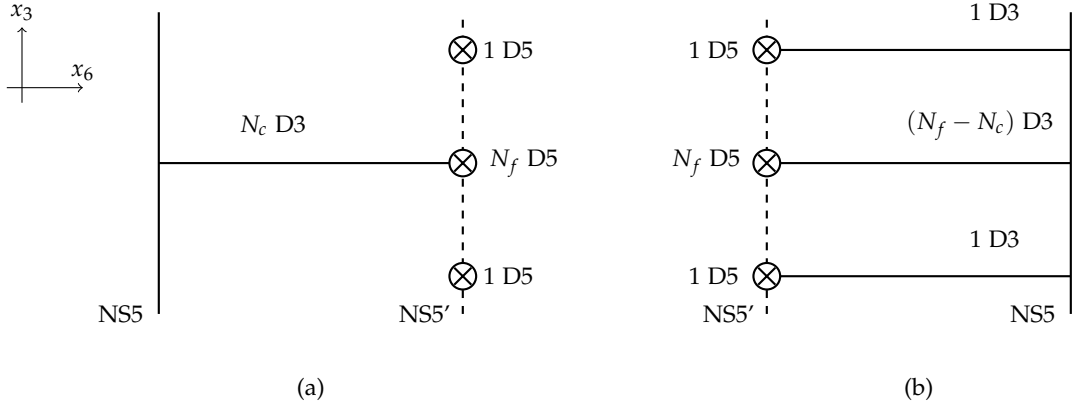
\begin{figure}
  \begin{center}
    \begin{tikzpicture}
      \begin{scope}[shift={(-2,1)}]
        \begin{footnotesize}
          \draw[->] (0,3) -- (1,3);
          \draw[->] (.2,2.8) -- (.2,3.8);
          \draw (1,3.2) node[]{\(x_6\)};
          \draw (.2,4) node[]{\(x_3\)};
        \end{footnotesize}
      \end{scope}
      \begin{scope}[thick]
        \begin{footnotesize}

          \draw (0,1) -- (0,5); 
          \draw (-0.5,1) node[]{\NS5};
          \draw (0,3) -- (4,3); 
          \draw (1.5,3.5) node[]{\(N_c\) \D3}; 
          \draw[dashed] (4,1) -- (4,5);
          \draw (3.5,1) node[]{\NS5'};
          \draw[fill=white] (4,3) circle (5pt);
          \draw (4,3) node[cross=4pt]{};
          \draw (4.8,3) node[]{\(N_f\) \D5};
          \draw[fill=white] (4,1.5) circle (5pt);
          \draw (4,1.5) node[cross=4pt]{};
          \draw (4.6,1.5) node[]{\(1\) \D5};
          \draw[fill=white] (4,4.5) circle (5pt);
          \draw (4,4.5) node[cross=4pt]{};
          \draw (4.6,4.5) node[]{\(1\) \D5};
          \node at (2,0) {(a)};
        \end{footnotesize}
        
      \end{scope}
      \begin{scope}[shift={(8,0)},thick]
        \begin{footnotesize}
          \draw (4,1) -- (4,5);
          \draw (3.5, 1) node[]{\NS5};
          \draw (0,3) -- (4,3);
          \draw (2.8, 3.5) node[]{\((N_f - N_c)\) \D3};
          \draw (0, 1.5) -- (4,1.5);
          \draw (3, 2) node[]{\(1\) \D3};
          \draw (0, 4.5) -- (4,4.5);
          \draw (3, 5) node[]{\(1\) \D3};
          \draw[dashed] (0,1) -- (0,5);
          \draw (-0.5,1) node[]{\NS5'};
          \draw[fill=white] (0,3) circle (5pt);
          \draw (0,3) node[cross=4pt]{};
          \draw (-0.8,3) node[]{\(N_f\) \D5};
          \draw[fill=white] (0,1.5) circle (5pt);
          \draw (0,1.5) node[cross=4pt]{};
          \draw (-0.7,1.5) node[]{\(1\) \D5};
          \draw[fill=white] (0,4.5) circle (5pt);
          \draw (0,4.5) node[cross=4pt]{};
          \draw (-0.7,4.5) node[]{\(1\) \D5};
          \node at (2,0) {(b)};
        \end{footnotesize}

      \end{scope}
    \end{tikzpicture}
  \end{center}
\caption{Electric and magnetic~brane system for \textsc{sqcd} on $\setR^3\times S^1$}
\label{fig:electric-magnetic-SQCD}
\end{figure}

Start with $N_f + 2$ flavors. The flow is generated by breaking the  
$SU(N_f+2)^2$ flavor symmetry down to $SU(N_f)^2 \times U(1)_A$. 
At energy scale $E < 1/ \widetilde R_3$ (where $\widetilde R_3 = \alpha'/R_3$ is the T--dual radius),
$x_3$ is effectively non-compact. This is the large-mass limit in which the $\eta$--superpotential disappears and the axial $U(1)_A$ is restored. 
In the~brane description we obtain this breaking by moving 
one \D5~brane in the $x_3>0$ direction and one \D5 in the $x_3<0$
direction (we put the stack of $N_c$ \D3~branes at $x_3=0$).
This is the configuration in Figure~\ref{fig:electric-magnetic-SQCD}(a). 
The result is a $U(N_c)$ gauge theory with 
$N_f$ fundamentals and anti-fundamentals and vanishing superpotential,
namely the electric theory studied by
Aharony in~\cite{Aharony:2013dha}.

The dual picture is shown in Figure~\ref{fig:electric-magnetic-SQCD}(b).
Now each \D5~brane drags one \D3 along the $x_3$ direction and we
have three separate gauge sectors that form $U(N_f-N_c) \times U(1)^2$, as expected from field theory.
The chiral multiplets connecting the two sectors acquire a large mass whenever the
two \D5~branes are separated along $x_3$. Now we have one fundamental and one anti-fundamental massless flavor and a meson with a superpotential interaction in each $U(1)$ sector.
The three sectors interact on the Coulomb branch.
As we discussed above, the \D3s separate along $x_3$ at equal distance.
In the large-mass limit, when the two \D5s are far away in the $x_3$ direction, two \D3s of the stack of
$N_f-N_c$ \D3~branes are free to move, and this parameterizes the Coulomb branch.
While in the electric phase, these~branes can be pushed to infinity, in
the magnetic phase there are the two extra \D3-branes representing the
\(U(1)\) sectors.
This explains the interaction between the $U(N_f-N_c)$ sector and the
$U(1)$ sectors, mediated by an \ac{ahw} 
superpotential, coming from the broken $U(N_f-N_c+2)$ theory.
In the~brane picture this can be understood as coming from \D1--strings extended
between the \NS{}~branes, the two \D3s that are pushed far away in the $x_3$ direction by 
the \D5s, and the two \D3s that parameterize the Coulomb branch of the $U(N_f-N_c)$ sector.

This can be made precise in analogy with the construction discussed in
Section~\ref{sec:superpotential-from-branes}. The non-perturbative
\ac{ahw} superpotential due to the \D1~branes is 
\begin{equation}
\label{eq:monmox}
W_{\text{AHW}} = e^{\Phi^{(1)}-\Phi_{\widetilde N_c}}+ e^{\Phi_1- \Phi^{(2)}}+ \eta' e^{\Phi^{(2)}-\Phi^{(1)}}
=
\tilde t_1 t + \tilde t t_2 + \eta'  \tilde t_2 t_1 ,
\end{equation}
where $\Phi^1$ ($\Phi^2$) is the Coulomb branch coordinate of the $1$st ($2$nd) $U(1)-$sector.
Similarly, $\Phi_1$ and $\Phi_{\widetilde N_c}$ are the Coulomb branch
coordinates that remain unlifted by the potential $\eqref{WD1}$.

In the limit of large $R_3$, the two $U(1)$ sectors at $x_3 > 0 $ and
$x_3 < 0$ can be dualized separately.
It is known that the \ac{hw} exchange of the \NS{} and the \D5~branes
in a $U(1)$ gauge theory with one pair of fundamental and anti
fundamental leads to the \textsc{xyz} model.
Our $U(1)$ sectors are similar to the gauge theory dual of the \textsc{xyz}
models, but for the additional superpotential interaction $M^{(i)} q^{(i)} \tilde q^{(i)}$ 
in each sector $i = 1,2$.
After the duality we obtain a deformation of the \textsc{xyz} model, where the
singlets $X, Y$ and $Z$ are identified with the meson $N^{(i)}=
q^{(i)} \tilde q^{(i)}$ and the Coulomb branch coordinates $t_i,\tilde t_i$.
At the end of the day, the superpotential in each sector is
\begin{equation}
  W_{\text{m}}^{(i)} = M^{(i)} N^{(i)} + N^{(i)} t_i \tilde t_i .
\end{equation}
Putting it all together, after integrating out the massive fields, 
the superpotential $W_{\text{m}}^{(1)} +  W_{\text{m}}^{(2)} + W_{\text{AHW}} $
becomes the interaction $t T + \tilde t \widetilde T$ expected for the Aharony duality.
The fields 
$\tilde t_1$ and $t_2$  have become the singlets $T$ and $\widetilde
T$ describing the monopole operators of the electric phase. 

\bigskip

If we do not take \(R_3\) to be infinite, the two~branes reconnect on the circle, at the \emph{mirror} 
point $x_3=x_3^\circ$. The magnetic dual is then obtained by a \ac{hw} transition,
generating a confining gauge theory at $x^\circ_3$. 
This is a $U(2)$  gauge theory with two flavors. 
The three-dimensional limit is recovered with a double scaling on the relative positions of these
 \D{}~branes at $x_3^\circ$ and the radius of the circle.
The new three-dimensional confining sector in the magnetic theories
is locally dualized to its confined counterpart.
This is a local mirror symmetry\footnote{See~\cite{Collinucci:2016hpz} for a related discussion of local mirror symmetry.}.
The singlets of this confining sector are coupled to the monopole of
the gauge symmetry living at $x_3=0$,
reproducing the gauge theory duality in the pure three-dimensional
limit.

In the reduction of \ac{sqcd}, the double-scaling limit is just an
alternative description. It will become more important because it
provides a unifying description when discussing more general dualities.

\subsubsection{Giveon--Kutasov duality}
\label{sec:GK-duality}

We can now move on to discuss the \textsc{rg} flow of four-dimensional Seiberg duality 
to \ac{gk} duality for three-dimensional $U(N_c)_k$ \textsc{sqcd}.
Start with $N_f+k$ \D5~branes and separate $k$ \D5s on the
\NS'~brane to obtain two semi-infinite stacks.
If we move half of the stack to $x_3>0$ and the other half to $x_3<0$,
a  $(1,k)$ five-brane is created and in the limit when the $(1,k)$ fivebrane becomes infinite, a \ac{cs} term 
is generated~\cite{Kitao:1998mf,Aharony:1997ju} (we had already
encountered this effect when discussing the breaking of axial symmetry
in Sec.~\ref{sec:duality-finite-radius}). This reproduces the electric side of the \ac{gk} duality.
In the dual phase, the construction is more involved. The number of \D3s created 
by this process is still $N_f+k+2$, the gauge theory is now $U(1)^2 \times U(N_f-N_c+k)$.
The motion of the \D5s along $x_3$ generates a non-trivial \ac{fi} term for the $U(1)$ sectors. This term is proportional 
to the axial real mass (since in the flow $U(1)_J$ and $U(1)_A$ mix).
This gives a mass to the monopoles and finally one is left with the dual \ac{gk} configuration.

\subsection{The case of Adjoint SQCD}\label{sec:adjSQCD}

Consider now the reduction of four-dimensional 
Seiberg duality with adjoint matter to three dimensions.

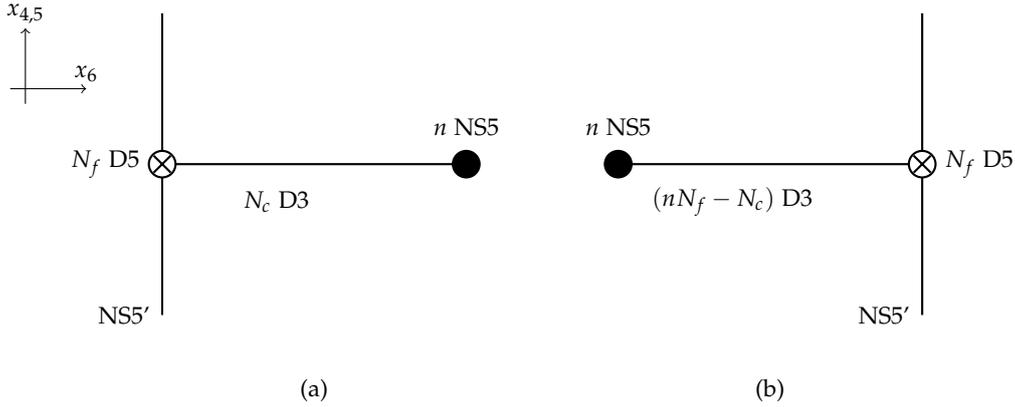
\begin{figure}
  \begin{center}
    \begin{tikzpicture}
      \begin{scope}[shift={(-2,1)}]
        \begin{footnotesize}
          \draw[->] (0,3) -- (1,3);
          \draw[->] (.2,2.8) -- (.2,3.8);
          \draw (1,3.2) node[]{\(x_6\)};
          \draw (.2,4) node[]{\(x_{4,5}\)};
        \end{footnotesize}
      \end{scope}

      \begin{scope}[thick]
        \begin{footnotesize}

          \draw (0,1) -- (0,5);
          \draw (-0.5,1) node[]{\NS5'};
          \draw (0,3) -- (4,3);
          \draw (1.5,2.5) node[] {\(N_c\) \D3};
          \draw[fill=black] (4,3) node{} circle (5pt);
          \draw (4,3.5) node{\(n\) \NS5};
          \draw[fill=white] (0,3) circle (5pt);
          \draw (0,3) node[cross=4pt]{};
          \draw (-0.75,3) node[]{\(N_f\) \D5};
          \node at (2,0) {(a)};
        \end{footnotesize}
      \end{scope}
      \begin{scope}[shift={(6,0)},thick]
        \begin{footnotesize}
          \draw (4,1) -- (4,5);
          \draw (3.5,1) node[]{\NS5'};
          \draw (0,3) -- (4,3);
          \draw (1.5,2.5) node[]{\((n N_f - N_c)\) \D3};
          \draw[fill=black] (0,3) node{} circle (5pt);
          \draw (0,3.5) node[]{\(n\) \NS5};
          \draw[fill=white] (4,3) circle (5pt);
          \draw (4,3) node[cross=4pt]{};
          \draw (4.75,3) node[]{\(N_f\) \D5};
          \node at (2,0) {(b)};
        \end{footnotesize}
      \end{scope}
    \end{tikzpicture}
  \end{center}
  \caption{Electric and magnetic version of \ac{kss} from the~brane picture.
}
  \label{figKSS}
\end{figure}

In the~brane description we introduce a 
set of $n$ \NS5~branes instead of only one. As discussed above, this induces the superpotential $W = \Tr X^{n+1}$.
The electric theory is given in Figure~\ref{figKSS}(a).
First we separate the  $n$ \NS{}~branes to induce a superpotential in
the adjoint and break the gauge symmetry to $\prod_{i=1}^{n} U(r_i)$.
This corresponds to separating the $N_c$ \D3~branes along $(8,9)$.
Finally, we move the \NS{}s and the \D3~branes in $(8,9)$ and come back to 
the original stack (Figure~\ref{figKSS}(b)).

\begin{figure}
  \begin{center}
    \begin{tikzpicture}
      \begin{scope}[shift={(-2,-3.5)}]
        \begin{footnotesize}
          \draw[->] (0,3) -- (1,3);
          \draw[->] (.2,2.8) -- (.2,3.8);
          \draw (1,3.2) node[]{\(x_6\)};
          \draw (.2,4) node[]{\(x_{4,5}\)};
        \end{footnotesize}
      \end{scope}

      \begin{scope}[thick]
        \begin{footnotesize}
          \draw (0,0) -- (0,8);
          \draw (-0.5,0) node[]{\NS5'};

          \draw (0,7) -- (2.5,7);
          \draw (1.5,6.5) node[]{\(r_1\) \D3};
          \draw[fill=black] (2.5,7) node{} circle (5pt);
          \draw (2.5,7.5) node[]{\NS5};
          \draw[fill=white] (0,7) circle (5pt);
          \draw (0,7) node[cross=4pt]{};
          \draw (-0.75,7) node[]{\(N_f\) \D5};

          \draw (0,5) -- (2.5,5);
          \draw (1.5,4.5) node[]{\(r_2\) \D3};
          \draw[fill=black] (2.5,5) node{} circle (5pt);
          \draw (2.5,5.5) node[]{\NS5};
          \draw[fill=white] (0,5) circle (5pt);
          \draw (0,5) node[cross=4pt]{};
          \draw (-0.75,5) node[]{\(N_f\) \D5};

          \draw (0,3) -- (2.5,3);
          \draw (1.5,2.5) node[]{\dots};
          \draw[fill=black] (2.5,3) node{} circle (5pt);
          \draw (2.5,3.5) node[]{\NS5};
          \draw[fill=white] (0,3) circle (5pt);
          \draw (0,3) node[cross=4pt]{};
          \draw (-0.75,3) node[]{\(N_f\) \D5};

          \draw (0,1) -- (2.5,1);
          \draw (1.5,0.5) node[]{\(r_n\) \D3};
          \draw[fill=black] (2.5,1) node{} circle (5pt);
          \draw (2.5,1.5) node[]{\NS5};
          \draw[fill=white] (0,1) circle (5pt);
          \draw (0,1) node[cross=4pt]{};
          \draw (-0.75,1) node[]{\(N_f\) \D5};

          \draw (1,-.5) node[]{(a)};

        \end{footnotesize}
      \end{scope}

      \begin{scope}[shift={(5,0)},thick]
        \begin{footnotesize}
          \draw (1.5,0) -- (1.5,8);
          \draw (1,0) node[]{\NS5'};

          \draw (0,7.05) -- (2.5,7.05);
          \draw (0, 6.95) -- (1.5,6.95);
          \draw (.5,6.5) node[]{\((N_f - r_1)\) \D3};
          \draw[fill=black] (2.5,7) node{} circle (5pt);
          \draw (2.5,7.5) node[]{\NS5};
          \draw[fill=white] (0,7) circle (5pt);
          \draw (0,7) node[cross=4pt]{};
          \draw (-0.75,7) node[]{\(N_f\) \D5};

          \draw (0,5.05) -- (2.5,5.05);
          \draw (0,4.95) -- (1.5,4.95);
          \draw (.5,4.5) node[]{\((N_f - r_2)\) \D3};
          \draw[fill=black] (2.5,5) node{} circle (5pt);
          \draw (2.5,5.5) node[]{\NS5};
          \draw[fill=white] (0,5) circle (5pt);
          \draw (0,5) node[cross=4pt]{};
          \draw (-0.75,5) node[]{\(N_f\) \D5};

          \draw (0,3.05) -- (2.5,3.05);
          \draw (0,2.95) -- (1.5,2.95);
          \draw (.5,2.5) node[]{\dots};
          \draw[fill=black] (2.5,3) node{} circle (5pt);
          \draw (2.5,3.5) node[]{\NS5};
          \draw[fill=white] (0,3) circle (5pt);
          \draw (0,3) node[cross=4pt]{};
          \draw (-0.75,3) node[]{\(N_f\) \D5};

          \draw (0,1.05) -- (2.5,1.05);
          \draw (0,0.95) -- (1.5,0.95);
          \draw (.5,0.5) node[]{\((N_f - r_n)\) \D3};
          \draw[fill=black] (2.5,1) node{} circle (5pt);
          \draw (2.5,1.5) node[]{\NS5};
          \draw[fill=white] (0,1) circle (5pt);
          \draw (0,1) node[cross=4pt]{};
          \draw (-0.75,1) node[]{\(N_f\) \D5};

          \draw (1,-.5) node[]{(b)};

        \end{footnotesize}
      \end{scope}

      \begin{scope}[shift={(9,0)},thick]
        \begin{footnotesize}
          \draw (2.5,0) -- (2.5,8);
          \draw (2,0) node[]{\NS5'};

          \draw (0,7) -- (2.5,7);
          \draw (1.2,6.5) node[]{\((N_f - r_1)\) \D3};
          \draw[fill=black] (0,7) node{} circle (5pt);
          \draw (1.9,7.5) node[]{\(N_f\) \D5};
          \draw[fill=white] (2.5,7) circle (5pt);
          \draw (2.5,7) node[cross=4pt]{};
          \draw (0,7.5) node[]{\NS5};

          \draw (0,5) -- (2.5,5);
          \draw (1.2,4.5) node[]{\((N_f - r_2)\) \D3};
          \draw[fill=black] (0,5) node{} circle (5pt);
          \draw (1.9,5.5) node[]{\(N_f\) \D5};
          \draw[fill=white] (2.5,5) circle (5pt);
          \draw (2.5,5) node[cross=4pt]{};
          \draw (0,5.5) node[]{\NS5};

          \draw (0,3) -- (2.5,3);
          \draw (1.2,2.5) node[]{\dots};
          \draw[fill=black] (0,3) node{} circle (5pt);
          \draw (1.9,3.5) node[]{\(N_f\) \D5};
          \draw[fill=white] (2.5,3) circle (5pt);
          \draw (2.5,3) node[cross=4pt]{};
          \draw (0,3.5) node[]{\NS5};

          \draw (0,1) -- (2.5,1);
          \draw (1.2,0.5) node[]{\((N_f - r_n)\) \D3};
          \draw[fill=black] (0,1) node{} circle (5pt);
          \draw (1.9,1.5) node[]{\(N_f\) \D5};
          \draw[fill=white] (2.5,1) circle (5pt);
          \draw (2.5,1) node[cross=4pt]{};
          \draw (0,1.5) node[]{\NS5};

          \draw (1,-.5) node[]{(c)};
        \end{footnotesize}
      \end{scope}

    \end{tikzpicture}
\end{center}
\caption{Duality in the \ac{kss} model
}
\label{figKSSduality}
\end{figure}
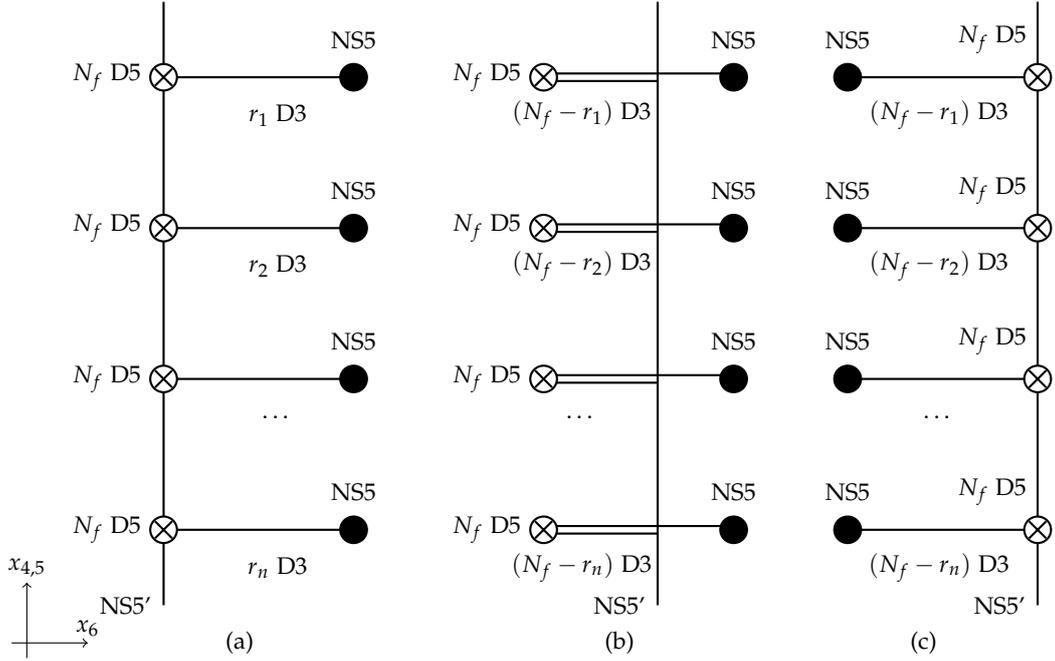

Repeating the procedure outlined above, we T--dualize along $x_3$
so that the \D4 and the \D6~branes turn into \D3s and \D5s, while 
the \NS{}~branes are unchanged (Figure~\ref{figKSSduality}).
The electric configuration (Figure~\ref{figKSSduality} (a))
 consists of a set of \(n\) decoupled \textsc{sqcd} models, each with
 $N_f$ flavors. The corresponding gauge groups are $U(r_i)$, with $\sum_{i=1}^{n} r_i=N_c$.
There is no adjoint matter anymore.
Just like in the previous construction, the superpotential
(\ref{eq:etagen}) is generated since \(x_3\) is compact.
The same procedure can be implemented in the magnetic case.
This reproduces the duality with the $\eta$--superpotential that we
have discussed above.
The flow to the \ac{kp} duality is obtained by separating the
\D5~branes along $x_3$, just like in the case of \textsc{sqcd}.
The final duality is obtained by reconstructing the stack of $n$
\NS{}~branes, which sends the $\eta$-superpotential to zero.

\subsubsection{S-confining theories}

We wish to conclude this section with a comment on the case
$N_f=N_c+1$. In this case, there is no Seiberg duality in four dimensions, but at low energy there is an effective 
description of this confining theory in terms of mesonic and baryonic operators.
This case can also be visualized as a limiting case of Seiberg duality. At the~brane level, one can understand the effective theory by 
studying the role of stringy \D0 instantons on the \D4 gauge~branes after performing an \ac{hw} transition.
The reduction of confining gauge theories in three dimensions has been discussed in~\cite{Csaki:2014cwa,Amariti:2015kha},
while the interpretation in terms of \D-branes has been given in~\cite{Amariti:2015xna}, by elaborating on the 
role of the stringy instanton after T-duality.

\subsection{Orientifolds}\label{sec:redOrienti}

In this final section we discuss the reduction of four-dimensional dualities 
in presence of real gauge groups and tensor matter.
In the geometric picture this implies adding orientifold planes to the
brane systems.

First we will discuss how the orientifold modifies the 
Coulomb branch and then we will see how this affects the reduction of
the dualities, comparing to the unitary cases discussed above.

\subsubsection{Monopole superpotentials from the~branes}

As we have seen in Section~\ref{sec:superpotential-from-branes}, the
repulsive interaction between the \D3~branes can be understood in
terms of Euclidean \D1~branes stretched between the \NS~and the
\D3~branes and the \(\eta\)-superpotential is reproduced by the
extra \D1~brane between the first and the last \D3.
Moreover, an algebraic analysis of the superpotential leads to the
notion of the Toda chain for the affine \(\widetilde A_N\) algebra (Eq.~\eqref{eq:W-affine}).

A similar analysis is possible for real gauge groups
realized with \O3 or \O5 planes.
Let us first discuss the configurations with \O3 planes on $\mathbb{R}^3 \times S^1$.
As reviewed in Sec.\ref{sec:orientifolds}, the four possible orientifolds
$\O3^+$, $\O3^-$, $\widetilde{\O3}^+$ and $\widetilde{\O3}^-$
project a unitary group to $Sp(2N_c), SO(2N_c), Sp(2N_c)$ and $SO(2N_c+1)$.
If one of the directions is compact, orientifolds must come in pairs and
in our case,\footnote{For simplicity we do not distinguish between $\O3^+$ and $\widetilde{\O3}^+$ planes.
} there are six possible pairs~\cite{Hanany:2001iy}.
Brane configurations with $(\O3^-,\O3^-)$,  $(\widetilde{\O3}^-,\O3^-)$ and $(\O3^+,\O3^+)$
are associated to affine Dynkin diagrams.
The other pairs $(\O3^{+},\O3^{-})$, $(\O3^{+},\widetilde{\O3}^{-})$ and $(\widetilde{\O3}^{-}, \widetilde{\O3}^{-})$ correspond to twisted affine Dynkin diagrams.
Here we are interested only in the ``affine'' pairs,
since they can be obtained via T--duality from \tIIA configurations with $O4$~planes.
The same holds for $O6$~planes. Now a
$\O5^+$ is associated to an $SO(N_c)$  and $\O5^-$ to an $Sp(2N_c)$ gauge group.
The pairs $(\O5^\pm,\O5^\pm)$ are obtained by T--duality from a \tIIA configuration with $\O6^{\pm}$ planes.

In the presence of matter fields, we have seen how the two stacks of
\D5~branes reconnect at the mirror point $x^\circ_3$, and in the limit
$R_3 \rightarrow 0$ they correspond to a set of massive singlets.
On the magnetic side,
a \D3 is created whenever a \D5 crosses an \NS{} in the \acs{hw} transition,
generating an extra gauge theory at $x^\circ_3$, which reproduces the topological vacua of the magnetic field theory.

In the following we consider the effects of orientifold planes in this picture.
Their \D-brane charge modifies the standard~brane creation effect in the \ac{hw} transition.
The extra sector at $x^{\circ}_3$ in the magnetic theory remains confining,
and we can use the same picture to describe the reduction of dualities with orientifolds from four to three dimensions.

\subsubsection{$Sp(2N_c)$ with fundamentals}

\label{sec:symplectic}

Let us move now to the duality for $Sp(2N_c)$
with $2N_f$ fundamentals.

There are two ways to engineer this theory, by either considering an $\O4^{+}$~plane,
or an $\O6^{-}$~plane. The two resulting theories differ in the representation
of the matter fields under the global symmetries.
When adding a larger number of \NS{}~branes, the two constructions give rise to different
two-index matter fields, adjoint or antisymmetric, which we will study separately.
\begin{itemize}
\item In the $\O4^{+} $~plane case we consider a stack of \(2N_c\) \D4~branes and an  $\O4^{+} $~plane
stretched between an \NS{} and an \NS'~brane. We consider also \(2N_f\) \D6~branes
on the \NS'~brane. The naive construction has an $SO(2N_f)$ global
symmetry that is enhanced to $SU(2N_f)$ by the same mechanism that
doubles the global symmetry for unitary gauge groups (see Sec.~\ref{subsec:fivebranes}). 
The dual theory is obtained by a \ac{hw} transition that exchanges the \NS{} and the \NS'~branes.
Every time a \D6~brane crosses a non-parallel fivebrane, a
\D4~brane is generated.  We need to pay attention to the linking
number: an $\O4^{+}$ has to be treated as a stack of $-4$
\D6~branes~\cite{Elitzur:1997hc}.
After the transition we obtain the dual picture, in which the net number of \D4~branes is
$2(N_f-N_c-2)$.
\begin{figure}
  \centering
  \begin{tikzpicture}
  \begin{scope}[thick]
    \begin{footnotesize}
      \draw[NScol] (3,-2) -- (3,2);
      \draw[NScol] (3,-2) node[right=1ex] {\NS'};
      
      \draw[NScol] (-3, 2) -- (-1, -2);
      \draw[NScol] (-1,-2) node[right=1ex] {\NS{}};

      \draw [O3col, dashed] (-4,0) -- (4, 0); 
      \draw[O3col] (-4,.3) node[] {\(\O4^+\)};
      \draw[O3col] (0,.3) node[] {\(\O4^-\)};
      \draw[O3col] (4,.3) node[] {\(\O4^+\)};

      \draw[D4col] (-2.5,1) -- (3, 1);
      \draw[D4col] (-1.5,-1) -- (3, -1);
      
      \draw[D4col] (2, 1.3) node[] {\(N\) \D4};
      \draw[D4col] (2, -.7) node[] {\(N\) \D4};

      \tikzcrosscircle{(-2.25,.5)}
      \tikzcrosscircle{(-1.75,-.5)}
      \draw[D6col] (-2.25,.5) node[right=1ex] {\(F\) \D6};
      \draw[D6col] (-1.75,-.5) node[right=1ex] {\(F\) \D6};

    \end{footnotesize}
  \end{scope}
\end{tikzpicture}
  \caption{Brane cartoon for the realization of an \(Sp(2N_c)\) theory in the electric phase with an \O4~plane.}
  \label{fig:O4+_branes}
\end{figure}
\item A similar theory can be constructed by using an $\O6^{-}$~plane.
Consider two \NS5~branes, $2N_c$ \D4s, $2N_f$ \D6's  and an
$\O6^{-}$~plane as in Figure~\ref{fig:O6-_plane_duality_4D}(a). If
all the \NS5~branes are parallel, the system has $\mathcal{N}=2$ supersymmetry.
The orientifold projects the $SU(2N_c)$ gauge group to $Sp(2N_c)$
(where, as usual, $Sp(2)\simeq SU(2)$).
The theory has an $SU(N_f)$
global symmetry with $F$ flavors. We expect this symmetry to be enhanced to $SU(N_f)^2$.
Here we rotate the \NS5~branes and the \D6'~branes by an angle $\theta$ as in
Figure~\ref{fig:O6-_plane_duality_4D}(a), and we have two stacks of $\NS_{\pm \theta}$ and $\D6_{\pm \theta}$.
For generic angles, the $\mathcal{N}=2$
adjoint is massive. If $\theta=\pi/2$ the orientifold is parallel
to the $NS_{\pm \theta}$~branes and this field is massless and has to be considered in the low-energy spectrum. 
This model (for $\theta \neq \pi/2$) has a dual description as discussed above. In this case the
$\O6^{-}$ behaves like a stack of $- 4$  \D6~branes in the \ac{hw} transition. The~brane picture becomes the one shown in Figure~\ref{fig:O6-_plane_duality_4D}(b) where the dual gauge group is again $Sp(2(N_f-N_c-2))$.
\begin{figure}
  \centering
  \begin{tikzpicture}[>=latex]
\tikzstyle{line} = [draw, -latex']
  \begin{scope}[thick]
    \begin{footnotesize}
      \draw (-7,0) node {(a)};
      \draw (-7,-4) node {(b)};
      
      \begin{scope}
        \draw (-4, 0) -- (4, 0);
        \draw (.5, .3) node {\D4}; 

        \draw (-4, -1) -- (-4, 1);
        \draw (-5, -1) -- (-3, 1); 
        \draw (-5, -1) node [left=.7ex] {\(\NS{}_\theta\)};

        \draw (4, -1) -- (4, 1);
        \draw (5, -1) -- (3, 1);
        \draw (5, -1) node [right=.7ex] {\(\NS{}_{-\theta}\)};

        \draw [red] (0, -1) -- (0, 1); 
        \draw (0, 1.3) node [] {\O6}; 

        \draw [dashed] (-2, -1) -- (-2, 1); 
        \draw [dashed] (-3, -1) -- (-1, 1); 
        \draw (-3, -1) node [left=.7ex] {\(\D6_\theta\)};

        \draw [dashed] (2, -1) -- (2, 1); 
        \draw [dashed] (3, -1) -- (1, 1);
        \draw (3, -1) node [right=.7ex] {\(\D6_{-\theta}\)};
        
        \draw [->,line, dashed] (-2, 1.2) to [bend left] (-1, 1.2);
        \draw [->,line] (-4, 1.2) to [bend left] (-3, 1.2);

        \draw [->,line, dashed] (2, 1.2) to [bend right] (1, 1.2); 
        \draw [->,line] (4, 1.2) to [bend right] (3, 1.2);
      \end{scope}

      \begin{scope}[shift={(0,-4)}]
        \draw (-4, 0) -- (4, 0);
        \draw (.5, .3) node {\D4}; 

        \draw (-.5, -1) -- (-2.5, 1); 
        \draw (-.5, -1) node [left=.7ex] {\(\NS{}_\theta\)};

        \draw (.5, -1) -- (2.5, 1);
        \draw (.5, -1) node [right=.7ex] {\(\NS{}_{-\theta}\)};

        \draw [red] (0, -1) -- (0, 1); 
        \draw (0, 1.3) node [] {\O6}; 

        \draw [dashed] (-5, -1) -- (-3, 1); 
        \draw (-5, -1) node [left=.7ex] {\(\D6_{-\theta}\)};

        \draw [dashed] (5, -1) -- (3, 1);
        \draw (5, -1) node [right=.7ex] {\(\D6_{\theta}\)};
        
      \end{scope}
      
    \end{footnotesize}
  \end{scope}
\end{tikzpicture}
  \caption{Brane cartoon for the realization of an \(Sp(2N_c)\) theory in the (a) electric and (b) magnetic phase with an \O6~plane.}
  \label{fig:O6-_plane_duality_4D}
\end{figure}
\end{itemize}

\paragraph{O3 planes.}
Once more, the three-dimensional system is obtained by compactifying
the $x_3$--direction and T--dualizing. The novelty is that the  $\O4^{+}$~plane
becomes a pair of ($\O3^+,\O3^+$)~planes. 
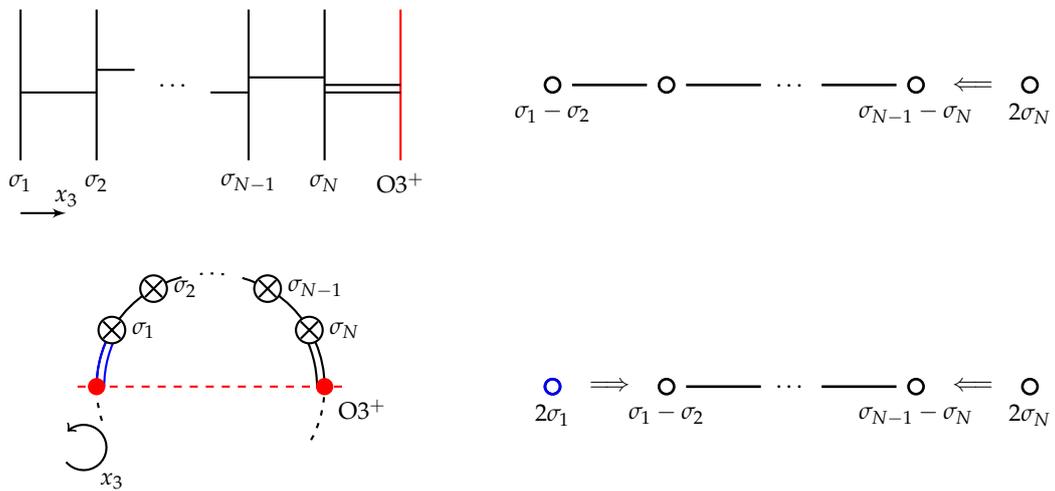
\begin{figure}
  \centering
  \begin{tikzpicture}
  \begin{scope}[thick]
    \begin{footnotesize}
      \begin{scope}
        \foreach \position/\Dname in {-2/{\(\sigma_1\)}, -1/{\(\sigma_2\)}, 1/{\(\sigma_{N-1}\)}, 2/{\(\sigma_N\)}}
        {
          \draw (\position, 1) -- (\position, -1);
          \draw (\position, -1.3) node {\Dname};
        }
        
        \draw[red] (3, 1) -- (3, -1);
        \draw (3, -1.3) node {\(\O3^+\)};

        \foreach \leftpos/\rightpos/\heightpos in {-2/-1/-.1, -1/-.5/.2, .5/1/-.1 ,1/2/.1, 2/3/-.1, 2/3/0 }
        {
          \draw (\leftpos, \heightpos) -- (\rightpos, \heightpos);
        }
        
        \draw (0,0) node {\(\dots\)};
        \tikzhorizaxis{(-1.7,-1.7)}{\(x_3\)}

      \end{scope}

      \begin{scope}[node distance=4em, ch/.style={circle,draw,on chain,inner sep=2pt},chj/.style={ch,join},every path/.style={shorten >=4pt,shorten <=4pt},line width=1pt,baseline=-1ex,start chain,shift={(5,0)}]
        \dnode{\(\sigma_1 - \sigma_2\)}
        \dnode{}
        \dydots
        \dnode{\(\sigma_{N-1} - \sigma_N\)}
        \dnodenj{\(2 \sigma_N\)}
        \path (chain-4) -- node[anchor=mid] {\(\Longleftarrow\)} (chain-5);
      \end{scope}

      \begin{scope}[shift={(.5,-4)}]
        \draw[dash pattern= on 55pt off 7pt on 1pt off 3pt on 1pt off 3pt on 1pt off 7pt ] (1.5,0) arc (0:180:1.5);
        \draw (1.4,0) arc (0:30:1.5);
        \draw [blue] (-1.4,0) arc (180:150:1.5);
        \draw [blue] (-1.5,0) arc (180:150:1.5);

        \draw [dashed, red] (-1.75, 0 ) -- (1.75, 0);
        \draw (2, -.3) node {\(\O3^+\)};

        \draw [dash pattern= on 2pt off 2pt on 2pt off 3pt on 2pt off 3pt on 2pt off 3pt on 2pt off 100pt on 2pt off 3pt on 2pt off 3pt ] (1.5,0) arc (0:-180:1.5);

        \foreach \braneanglepos/\Dname in {30/{\(\sigma_{N}\)}, 60/{\(\sigma_{N-1}\)}, 120/{\(\sigma_{2}\)}, 150/{\(\sigma_{1}\)}}
        {
          \tikzcrosscircle{({1.5 * cos(\braneanglepos)},{1.5 * sin(\braneanglepos)})}
          \draw ({1.5 * cos(\braneanglepos)},{1.5 * sin(\braneanglepos)}) node [right=1ex] {\Dname};
        }

        \draw[red, fill] (1.5,0) ellipse (.1 and .1); 
        \draw[red, fill] (-1.5,0) ellipse (.1 and .1); 

        \tikzcircleaxis{(-1.7,-1)}{\(x_3\)}
      \end{scope}

      \begin{scope}[start chain,node distance=1ex and 4em, ch/.style={circle,draw,on chain,inner sep=2pt},chj/.style={ch,join},every path/.style={shorten >=4pt,shorten <=4pt},line width=1pt,baseline=-1ex, shift={(5,-4)}]
        \dnodenj{\(2 \sigma_1\)}
        \dnodenj{\(\sigma_1 - \sigma_2\)}
        \dydots
        \dnode{\(\sigma_{N-1} - \sigma_N\)}
        \dnodenj{\(2 \sigma_N\)}
        \path (chain-1) -- node{\(\Longrightarrow\)} (chain-2);
        \path (chain-4) -- node{\(\Longleftarrow\)} (chain-5);

        \draw[blue] (0,0) circle (.1);
      \end{scope}
    \end{footnotesize}
  \end{scope}

\end{tikzpicture}
  \caption{Dynkin and affine Dynkin diagrams and spectrum of \acs{bps} \F1--strings
  associated to the fundamental monopoles for $Sp(2N)$ theories
   in the linear case and on the circle. The affine root is represented in blue on the affine 
   Dynkin diagram and in the~brane cartoon.
   }
  \label{fig:F1Cn}
\end{figure}
In this case the superpotential can be read off from the top half of Figure~\ref{fig:F1Cn},
\begin{equation}
  W = \sum_{i=1}^{N_c-1} \frac{2}{Y_i} + \frac{1}{Y_{N_c}} \; ,
\end{equation}
where $Y_i = e^{(\sigma_i-\sigma_{i+1})/e_3^2+i(\gamma_i-\gamma_{i+1})}$
and $Y_{N_c}=e^{2(\sigma_{N_c}/e_3^2+i \gamma_{N_c})}$.
The extra root in the affine case is proportional to the variable $Y_0 = e^{2(\sigma_1/e_3^2+i \gamma_1)}$ (shown in blue on
the bottom half of Figure~\ref{fig:F1Cn}) and it gives rise to the $\eta$-superpotential
\begin{equation}
  W = \eta \eu^{2 \sigma_1/{e_3^2} + 2 i \gamma_1} \; .
\end{equation}
The same result is obtained after the \ac{hw} transition.

\bigskip

Now we want to flow to Aharony duality.
We start by considering $2(N_f+1)$ \D5~branes in the electric theory.
We separate two \D5~branes from the stack and move them in opposite directions along the circle to reconnect them on the other side.
Since the \D5s intersect the \NS{}~brane in this configuration,
there are no massless fields in this extra sector.
If we take the $r \rightarrow 0$ limit for this configuration, we obtain
an $Sp(2N_c)$ theory with $2N_f$ fundamentals.

Next we turn to the dual theory. In this case, if we perform a \ac{hw} transition,
there are $2(N_f-N_c-1)$ \D3s at the origin. On the other side of the circle, the two \D3s created
by the \D5s crossing the \NS{}~brane are destroyed by the extra orientifold plane located there.
This example shows one of the general aspects of our analysis. 
In principle, it is not necessary to reconnect the \D5~branes at $x_3^\circ$.
For example for unitary gauge groups, the double scaling was realized by putting the \D5~branes at $x_3 < x_3^\circ$~\cite{Amariti:2015yea}.
Here, in order to avoid the orientifold creating a negative number of \D3~branes in the \ac{hw} transition, we have to reconnect the \D5s at $x_3^\circ$.

Even if there is no gauge symmetry, an extra meson arising from the \D5~brane remains massless. This suggests that we cannot simply decouple this sector before considering the effect of this massless field in the ordinary dual gauge theory.
In fact in this case, the two \D5~branes attract the~branes labeled by $\sigma_1$ and
$-\sigma_1$ in the dual gauge sector.
This attractive force is reflected in the scale-matching relation between $Y_1$ and
the meson $M_{2N_f+1,2N_f+2}$.
It corresponds to the superpotential interaction
\begin{equation}
W = \widetilde \eta y_{low} M_{2N_f+1,2N_f+2} \; ,
\end{equation}
\emph{i.e.} the low-energy description of the $\eta$--superpotential.
In the large-mass limit
the effect of this interaction has to be considered.

This reproduces the field theory expectation:
the dual theory is an $Sp(2(N_f-N_c-1))$ theory
with $2N_f$ fundamentals, an antisymmetric
meson $M$ and superpotential
\begin{equation}
W = M q q + t T \; ,
\end{equation}
where we identified the broken component of the electric singlet \(M\) that parameterizes
a direction in the dual Higgs branch, with the electric monopole \(T\) that parameterizes the Coulomb branch of the electric phase. This is commonly the case when dealing with mirror symmetry. In fact, the electric singlet
describes the Higgs branch of the dual phase, \emph{i.e.} the Coulomb branch of the
electric theory.

\paragraph{O5 planes.}

In the case of the \O6~plane realization the reduction works
essentially in the same way.,
We reduce the duality to three dimensions by compactifying the $x_3$--direction. After T--duality, the \tIIB system contains a pair of \O5~planes and describes a theory with the same $\eta$--superpotential as above. By considering $N_f+2$ flavors and integrating out two of them, we recover the usual Aharony duality. At the~brane level,
this is obtained by introducing $N_f+2$ $\D5_{\pm \theta}$.
The real masses are realized like the construction with the \O3~plane. The orientifold identification is however different: in this case
we have a unitary symmetry. Moving a pair of $\D5_{\pm \theta}$ along $x_3$ gives a mass to one flavor.

\begin{figure}
  \centering
    \begin{tikzpicture}
    \begin{scope}[shift={(-7,-2)}]
      \begin{footnotesize}
      \draw[->] (-0.2,0) arc (-140:135:.2 and .4);
      \draw[->] (-0.4,.25) -- (.6,.25);
      \draw (0.6,.45) node[]{\(x_7\)};
      \draw (0.1,.9) node[]{\(x_3\)};
      \end{footnotesize}
    \end{scope}
    \begin{scope}[thick]
      \begin{footnotesize}
        \newcommand*\leftplane{-5}
        \newcommand*\rightplane{5}
        
        \draw[thin] (\leftplane,0) ellipse (1 and 2);
        \draw[thin] (\rightplane,2) arc (90:-90:1 and 2);
        \draw[thin,dashed] (\rightplane,-2) arc (-90:-270:1 and 2);

        \draw[thin] (\leftplane,2) -- (\rightplane,2);
        \draw[thin] (\leftplane,-2) -- (\rightplane,-2);

        \draw[D3col] (-4.1,1) -- (5.8,1);
        \node at (-2,.5) {\(2 \pqty{N_f -N_c - 1}\) \D3};
        \node at (2,.5) {\(2 \pqty{N_f -N_c - 1}\) \D3};
        \draw[D5col,fill=D5col] (-4.1,1) node{} circle (5pt) node[above right] {\(N_f\) \D5};
        \draw[D5col,fill=D5col] (5.8,1) node{} circle (5pt) node[above right] {\(N_f\) \D5};

        \draw[O3col,fill=O3col] (0,1) node{} circle (5pt) node[above right] {\O5};

        \draw[D3col,dotted] (-5.9,-1) -- (4,-1);
        \node at (-3.5,-1.5) {\(0\) \D3};
        \node at (1,-1.5) {\(0\) \D3};
        \draw[D5col,fill=D5col] (-5.9,-1) node{} circle (5pt) node[above right] {\(2\) \D5};
        \draw[D5col,fill=D5col] (4.1,-1) node{} circle (5pt) node[above right] {\(2\) \D5};

        \draw[O3col,fill=O3col] (-2,-1) node{} circle (5pt) node[above right] {\O5};

      \end{footnotesize}
    \end{scope}
    \end{tikzpicture}
  \caption{Aharony flow \(Sp\) in \O5 picture}
  \label{fig:Aharony-flow-Sp-O5}
\end{figure}
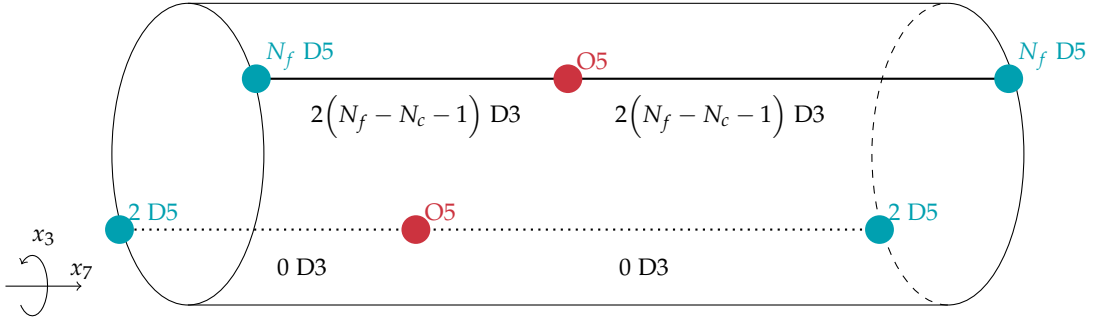

One can flow to Aharony duality by taking the double scaling limit as above (see Figure~\ref{fig:Aharony-flow-Sp-O5}).
First we move the \D5~branes in the $x_3$--direction, assigning the real masses.
We reconnect them on the other side of the circle. They reconnect at
$x_3 = x_3^\circ$, where the second orientifold plane is located.
The extra sector does not have massless degrees of freedom, and we can take the
$r \rightarrow 0$ limit to obtain a three-dimensional $Sp(2N_c)$ theory
with $2N_f$ flavors.
In the dual picture, after the \ac{hw} transition, \D3s are created when the~branes cross each other.
The orientifold cancels two \D3s every time an
\NS5~brane crosses it and 
the net effect  at  $x_3 = x_3^\circ $ is that no
\D3~branes remain.
The final configuration is reproduced in Figure~\ref{fig:Aharony-flow-Sp-O5}.

Like in the case with \O3~planes, here we have an extra sector with massless singlets (coming from the original mesons). The $r \rightarrow 0$ limit has to be taken by considering the effect of this sector on the  $Sp(2(N_f-N_c-1))$ theory. This is the same mechanism introduced above: the $\eta$--superpotential is absorbed in a scale matching, the meson couples with the magnetic monopoles, and in the final three-dimensional dual theory the extra interaction between the electric and magnetic monopoles takes place.

\subsubsection{Other cases}

The prescription explained so far can be generalized to other examples with $Sp(2N_c)$ gauge group
and tensor matter (see \cite{Amariti:2015mva} for details).
Theories with $SO(N_c)$ gauge groups deserve some more comments. 
In this case, one has to consider  $\O{}^+$ and $\widetilde{\O{}}^+$ orientifolds. In the reduction on the circle, the $\eta$ superpotential
is associated to an affine $B_N$ algebra for the $SO(2N+1)$ case and a $D_N$ affine algebra for $SO(2N)$.
The flow to Aharony duality does not require any real masses. In the
electric theory, one can simply take the 
large $r'$ limit. In the dual picture, an extra \D3~brane is generated by the \ac{hw} transition (the orientifold carries D~brane charge and for consistency, 
an extra \D3 has to be considered at the mirror point). This reproduces the Higgsing required in field theory.
The final form of the Aharony duality is eventually recovered with a local mirror symmetry on this sector, 

\bigskip

By now it is clear that we can extract a general recipe 
to  construct three-dimensional $\mathcal{N}=2$ dualities from four-dimensional $\mathcal{N}=1$ Seiberg-like
dualities.
Consider a system with \D4, \D6, \NS~branes (and possibily orientifolds).
Separate the stacks of NS~branes if present, in order to fix a set of \ac{sqcd} sectors. 
Then compactify the direction $x_3$ and T-dualize along it. By following the rules 
of the reduction in each single sector, one obtains an effective three-dimensional duality at finite size.
The flow to Aharony duality is performed in the $U(N)$, $Sp(2N)$ and $SO(N)$ sectors as explained 
above and this reconstructs the dualities discussed in Section~\ref{sec:3ddualities}. 
One can also generate \ac{cs} terms and flow to the generalizations of the Giveon--Kutasov duality that we presented in Section~\ref{GKdualaity}.

\chapter{4d/3d reduction and localization}\label{sec:loc}

A further check for \ac{susy} dualities independent of the brane picture is provided by the one-loop
exact results obtained from localization (see for example~\cite{Pestun:2007rz}).
This procedure corresponds to the computation of the partition
function of a supersymmetric gauge theory on a curved manifold.
The supersymmetric action on the curved manifold can be obtained by a background (super)-gravitational
setup and in general it is possible to turn on background vector multiplets for the global symmetries as well~\cite{Festuccia:2011ws}.
The partition function is exact at one loop and can be obtained by modifying the action by a $Q$-exact deformation ($Q$ being the preserved supercharge). 
This leads to a dramatic simplification of the modes contributing to the partition function
and the final result is expressed in terms of the classical action and the one-loop determinants
for the matter and for the gauge fields. In some cases, depending on the choice of the background, non-perturbative contributions may arise as well.

\section{Reducing the four-dimensional superconformal index to the three-dimensional partition function}\label{sec:3dindex}

The partition function appears in general as a matrix integral over the gauge group (in some cases
also a sum over the monopole sector may be necessary). Its dependence on the global symmetries 
is encoded in the dependence on some fugacities or real masses, depending on the manifold and 
on the dimensionality.
Here we focus on the squashed three-sphere $S_b^3$ partition function for three-dimensional $\mathcal{N}=2$ theories~\cite{Hama:2011ea}.
The sphere is defined by its action on the Euclidean coordinates $x_i$,
 \begin{equation}
 \frac{x_1^2+x_2^2}{b^2} +  \frac{x_3^2+x_4^2}{1/b^2} = 1,
 \end{equation}
where $b$ represents the real squashing parameter.

Before discussing the details of the three-dimensional partition function we discuss its derivation from 
four dimensions, along the lines of~\cite{Aharony:2013dha}. It has been observed that $Z_{S_b^3}$ can be formally obtained by  suitable limit on the fugacities of the superconformal index.
The four-dimensional superconformal index is a Witten-like index, computed by placing a four-dimensional theory
on $S^3 \times \setR$~\cite{Kinney:2005ej,Romelsberger:2005eg}. Alternatively it can be defined as the supersymmetric partition function 
on $S^3 \times S^1$. The two definitions differ by an overall contribution of the supersymmetric Casimir
energy~\cite{Assel:2015nca}, but this discrepancy is irrelevant when checking a duality (provided the anomalies match).

The definition of the Witten index on $S^3 \times \setR$ is given by
\begin{equation}
  I = Tr(-1)^F e^{-\beta H} (pq)^{\frac{\Delta}{2}}
  p^{j_1+j_2-R/2} q^{j_1-j_2-R/2}
  \prod_{a} u_a^{q_a}. 
\end{equation}
In the formula above, $F$ stands for the fermion number, while $H$ corresponds to the Hamiltonian
on the curved manifold.
The fugacities $p$ and $q$ refer to the $SU(2)_l \times SU(2)_r$ isometry of the three-sphere and their third spin components are  $j_1$ and $j_2$. 
The fugacities of $SU(2)_l \times SU(2)_r$ satisfy the constraints
\begin{align}
  Im(pq) &= 0, &   \abs{p/q} &= 1, &   \abs{pq} &< 1  .
\end{align}
The $U(1)$ R-charge is $R$.
The fugacities $u_a$ are in the Cartan of the global (and gauge) symmetry group.
The index is non-vanishing only on states with $H = 0$.
The index can be computed in two steps: first one calculates the single particle index
and then takes the so-called plethystic exponential. 
In localization, this is equivalent to calculating the one-loop determinants 
for the matter and the vector multiplets. They can be formulated in terms of elliptic Gamma functions
by
\begin{equation}
\Gamma_e (y;p,q) \equiv \Gamma_e(y) \equiv
\prod_{j,k=0}^{\infty} \frac{1-p^{i+1} q^{j+1}/ y}{1-p^i q^j y}.
\end{equation}
Here $y$ is a collective fugacity for both global and local symmetries.
The last step consists in integrating the index over the holonomy of the gauge group,
obtaining
\begin{equation}
\label{eq:SCI}
I_{G} = 
\frac{\kappa^{r_G}}{|W|}
\oint_{T^{r_G}}
\frac{d{ z_i}}{2 \pi i z_i}
\prod_{\alpha \in G_+} \Gamma_e^{-1}(z^{\pm \alpha})
\prod_{\rho \in \mathcal{R}_I,\widetilde{\rho} \in \widetilde{\mathcal{R}}_I} \Gamma_e (z^\rho u^{\widetilde \rho} (pq)^{\frac{R_I}{2}}),
\end{equation}
with $\kappa=(p;p)(q;q)$ and $(x;p) = \prod_{k=0}^{\infty} (1-x p^k)$.
We used the notation $\alpha \in G_{+}$ for the positive roots of the gauge group $G$, while
$|W|$ is the dimension of the Weyl group. The weight $\rho_I$ refers to the representation
of each multiplet under the gauge group, while $\widetilde{\rho}_I$ stands for the flavor symmetry. 
The fugacities $z$ and $u$ 
are in the Cartan of the gauge and of the flavor symmetry, respectively. 
The integral in (\ref{eq:SCI})  is over $T^G$, corresponding to the maximal Abelian torus of the gauge group.
The $R$ charges are denoted by $R_I$.

The next step consists in obtaining the partition function on $Z_{S_b^3}$.
One can use a direct approach and derive it from a purely three-dimensional 
perspective, as done in~\cite{Kapustin:2009kz,Jafferis:2010un,Hama:2010av,Hama:2011ea}.
Here, we prefer to review the derivation from the \ac{kk} reduction of
the superconformal index on  $S_\mathbf{b}^3 \times \tilde S^1$, following the discussion of~\cite{Aharony:2013dha}
(see also~\cite{Dolan:2011rp,Gadde:2011ia,Imamura:2011uw,Niarchos:2012ah}).
Let $\tilde r_1$ be the radius of $\tilde S^1$, the fugacities that enter in the definition of
the index become
\begin{align}
\label{eq:fugreal}
  p &= e^{2 \pi i \tilde r_1 \omega_1},& q &= e^{2 \pi i \tilde r_1 \omega_2}, &
u_a &= e^{2 \pi i \tilde r_1 \mu_a}, &  z_i &= e^{2 \pi i \tilde r_1 \sigma_i}. 
\end{align}
The parameters in the RH sides of (\ref{eq:fugreal}) represent the real scalars  
of the local and of the background symmetries of the three-dimensional vector multiplets.
The parameters $\omega_{1,2}$ satisfy the relations $\omega_1 = i b$ and $\omega_2 = ib^{-1}$
and $\omega \equiv \frac{\omega_1+\omega_2}{2}$.

The reduction of the four-dimensional superconformal index to the three-dimensional partition function on $S_\mathbf{b}^3$
corresponds to the $\tilde r_1 \rightarrow 0$ limit,
\begin{equation}
  \label{eq:4D3D}
  \lim_{
    \tilde r_1 \rightarrow 0}
  \Gamma_e(e^{2\pi i \tilde r_1 x}; e^{2\pi i \tilde r_1 \omega_1}, 
  e^{2 \pi i \tilde r_1 \omega_2})  
  =
  e^{\frac{i \pi^2}{
      6\tilde r_1 \omega_1 \omega_2}
    (x-\omega)}
  \Gamma_h(x; \omega_1, \omega_2) .
\end{equation}
This formula corresponds to the \ac{kk} reduction of the states contributing to the 
index to the states contributing to $X_{S_b^3}$. In the language of localization it
gives the one-loop determinants for the matter and for the vector multiplets contributing to the three-dimensional
partition function.
It corresponds to the reduction of the elliptic gamma functions $\Gamma_e$, representing the one-loop determinants for the four-dimensional fields
to the hyperbolic gamma function $\Gamma_h$ (see for example~\cite{VanDeBult}).
This is defined as
\begin{equation}
  \label{eq:Gammahvbd}
  \Gamma_h(x;\omega_1,\omega_2) \equiv
  \Gamma_h(x)\equiv 
  e^{
    \frac{i \pi}{2 \omega_1 \omega_2}
    ((x-\omega)^2 - \frac{\omega_1^2+\omega_2^2}{12})}
  \prod_{j=0}^{\infty} 
  \frac
  {1-e^{\frac{2 \pi i}{\omega_1}(\omega_2-x)} e^{\frac{2 \pi i \omega_2 j}{\omega_1}}}
  {1-e^{-\frac{2 \pi i}{\omega_2} x} e^{-\frac{2 \pi i \omega_1 j}{\omega_2}}}.
\end{equation}
There is a divergent prefactor in (\ref{eq:4D3D}) corresponding to the four-dimensional
gravitational anomalies.

Summarizing, the partition function 
$Z_{S_b^3}$ is given by
\begin{equation}
  \label{eq:SquashedSphere}
  Z_{G;k}(\lambda;\vec{\mu}) = \frac{1}{|W|}\int
  \prod_{i=1}^{G} \frac{d \sigma_i}{\sqrt{-\omega_1 \omega_2}}
  e^{\frac{i k \pi \sigma_i^2}{\omega_1 \omega_2}+\frac{2 \pi i \lambda \sigma_i}{\omega_1 \omega_2}}
  \frac{\prod_I \Gamma_h\left(
  \omega \Delta_I + \rho_I(\sigma)+\widetilde \rho_I(\mu)\right)
  }{\prod_{\alpha \in G_+}\Gamma_h\left(\pm\alpha(\sigma)\right)},
\end{equation}
where $\sigma$ and $\mu$  are real parameters in the Cartan of the gauge and of the flavor symmetries respectively. The \ac{fi} term is identified with $\lambda$ and the $R$ charge by $\Delta_I$.
A last comment is needed regarding the Gaussian factor in the integrand. It refers to the contribution of a possible 
\ac{cs} term at level $k$. It cannot be obtained from the dimensional reduction, and it corresponds, in localization, to the contribution of the classical action. 
Similarly, one can turn on a \ac{cs} term for
the flavor symmetry, associated to the contact terms of the flavor symmetry~\cite{Closset:2012vg,Closset:2012vp}.

Despite the fact that the three-dimensional \ac{cs} term cannot be obtained from the reduction of the index, one can mimic 
its generation by integrating out charged fermions with a large real mass.
This boils down to the following limit on the hyperbolic Gamma functions:
\begin{equation}
\label{eq:limmass}
  \lim_{x\rightarrow \pm \infty} \Gamma_h(x) = e^{\pm \frac{i \pi}{\omega_1 \omega_2} (x-\omega)^2}.
\end{equation}
There is another interesting point regarding the R-charges $\Delta_I$: it is not necessary to 
impose the exact $R$-charge, obtained by minimizing the partition function, in the analysis.
This is because one can always turn on an imaginary part for the
fugacities of the flavor symmetries, in order to parameterize the mixing of a trial $U(1)_R$ 
with the other global symmetries.

\section{Relation to the reduction of dualities}\label{sec:relationloc}

The derivation of the partition function as a \ac{kk} reduction of 
the four-dimensional superconformal index can be used to test the reduction
of four-dimensional Seiberg duality to three dimensions studied above.
One of the main ingredients that has not been discussed yet is the 
generation of the $\eta$-superpotential leading to the new effective dualities on the circle.
Such a superpotential prevents the generation of the axial symmetry and is necessary 
in order to preserve the four-dimensional duality in three dimensions. On the index/partition function side however,
the presence of a superpotential is not captured by localization.Nevertheless,s the constraints imposed on the symmetries can be reformulated as a
balancing condition on the fugacities. 
The condition in four dimensions corresponds to the existence of an anomaly-free R-symmetry.
When reducing the index to the partition function, this condition does not signal an
anomaly-free R-symmetry anymore, due to the absence of anomalies in three dimensions.
It signals instead the presence of the $\eta$-superpotential discussed on the field theory side.
Imposing this condition on the four-dimensional identity between two Seiberg-dual phases then reduces
the identity to the one of an effective duality in three dimensions. Observe that the divergent contributions,
signaling the presence of the four-dimensional gravitational anomalies, cancel among the dual phases, 
leading to an identity for effective three-dimensional theories.
The more conventional relation for Aharony and Giveon--Kutasov like dualities 
are obtained by a real-mass flow, exploiting the formula (\ref{eq:limmass}).
Note that in this case, it is also necessary to match the divergent prefactors to ensure that
the correct leading saddles are picked up.
This strategy has been used in~\cite{Aharony:2013dha} to reduce the identity for Seiberg duality with fundamental matter
for unitary and symplectic gauge group. In~\cite{Aharony:2013kma}, it has been observed that an analogous 
discussion for the orthogonal case fails, due to the fact that the partition function for the effective duality 
is divergent. In this case, a double-scaling limit may be necessary to obtain the expected three-dimensional identity.
These results can be extended to more complicated setups.

\subsection{Application: reduction of the KSS duality}

Here we discuss as an example the reduction for \ac{kss} duality to three dimensions, following the 
derivation of~\cite{Amariti:2014iza}.
By dimensionally reducing the integral identity between the indices of the four-dimensional \ac{kss}
duality, we obtain an identity for the partition functions of the effective three-dimensional duality with the  
$\eta$-superpotential.
The identity for Kim--Park duality is then obtained by a real mass deformation.
The duality of~\cite{Niarchos:2008jb,Niarchos:2009aa} with \ac{cs} terms is obtained by a further real mass flow.

Let us consider four-dimensional \ac{kss} duality.
The electric phase is a $U(N_c)$ gauge theory with one adjoint and $N_f+2$ (anti-)fundamental flavors.
Its superconformal index is
\begin{multline}
  \label{eq:index-el} 
  I_{el} = \frac{(p;p)^{N_c}(q;q)^{N_c}}{N_c!}
  \Gamma_e \big( (pq)^{\frac{1}{k+1}} \big)^{N_c}
  \int \prod_{i=1}^{N_c} \frac{d z_i}{2 \pi i z_i}
  \prod_{i< j} \frac{\Gamma_e \big((pq)^{\frac{1}{k+1}}(z_i/z_j)^{\pm 1}\big)}{ \Gamma_e\big( (z_i/z_j)^{\pm 1} \big)}
  \\
  \times 
  \prod_{a,b=1}^{N_f+2} \prod_{i=1}^{N_c}
  \Gamma_e \big(
  (pq)^{\frac{R_Q}{2}} s_a z_i\big)
  \Gamma_e \big(
  (pq)^{\frac{R_Q}{2}} t_b^{-1} z_i^{-1}
  \big),
\end{multline}
where  $\Gamma_e( (z)^{\pm 1}) \equiv \Gamma_e(z) \Gamma_e (1/z)$.
The fugacities $z_i$ refer to the $U(N_c)$ gauge symmetry 
while $s_a$ and $t_a$ to $SU(N_f)_L$ and $SU(N_f)_R$, respectively.

The magnetic theory has gauge group  $U(k(N_f+2)-N_c)\equiv U(\tilde N_c+2 k)$,
one adjoint,
$N_f+2$ (anti-)fundamental flavors and the $k$ electric mesons.
Its superconformal  index is
\begin{multline}
  \label{eq:index-mag}
  I_{mag}= \frac{(p;p)^{\tilde N_c+2k}(q;q)^{\tilde N_c+2k}}{(\tilde N_c+2k)!}
\Gamma_e\big( (pq)^{\frac{1}{k+1}} \big)^{\tilde N_c +2k}
\prod_{j=0}^{k-1} \prod_{a,b=1}^{N_f+2}
\Gamma_e\big((pq)^{\frac{j}{k+1}+R_Q} s_a t_b^{-1}\big) \\
\times \int \prod_{i=1}^{\tilde N_c+2k} \frac{d z_i}{2 \pi i z_i}
\prod_{i<j} \frac{\Gamma_e\big((pq)^{\frac{1}{k+1}}(z_i/z_j)^{\pm 1}\big)}{
  \Gamma_e\big( (z_i/z_j)^{\pm 1}\big)} \\
\times 
\prod_{a,b=1}^{N_f+2} \prod_{i=1}^{\tilde N_c+2k}
\Gamma_e\big(
(pq)^{\frac{R_q}{2}} t_a^{-1} z_i\big)
\Gamma_e\big(
(pq)^{\frac{R_q}{2}} s_b z_i^{-1}
\big) .
\end{multline}
\ac{kss} duality predicts the integral identity $I_{el} = I_{mag}$, provided that
the chemical potentials
satisfy the balancing condition
\begin{equation}
\label{eq:bal-cond-4d}
\prod_{a=1}^{N_f+2} s_a t_a^{-1} = (pq)^{N_f+2-2N_c/(k+1)},
\end{equation}
equivalent to the requirement of an anomaly free R-current.

The identity (\ref{eq:index-mag}) can be reduced to an identity between a pair
of three-dimensional partition functions with the procedure described above.
First we redefine the chemical potentials as
\begin{align}
  \label{eq:antonello-catthivo}
p &= e^{2 \pi i r \omega_1},
&
q &= e^{2 \pi i r \omega_2},
&
z &= e^{2 \pi i r \sigma},
&
s_a &= e^{2 \pi i r m_a},
&
t_a &= e^{2 \pi i r \tilde m_a},
\end{align}
where $\sigma$, $m_a$ and $\tilde m_a$ parameterize the Cartan of the $U(N_c)$ gauge and the $SU(N_f)_L \times SU(N_f)_R$ flavor group respectively.

By taking the limit $r \rightarrow 0$ and dropping the divergent prefactors, corresponding to the 
contributions of  the gravitational anomaly, that match in the dual phases,
we obtain the partition function for the three-dimensional electric theory with the $\eta$ superpotential
\begin{eqnarray}
\label{eq:Z-eta-el}
Z_{el}=W_{N_c,0}\big(\mu;\nu;\omega \Delta_X;\lambda\big),
\end{eqnarray}
where the function $W_{N_c,k}$ is 
\begin{multline}
\label{eq:Zdef} 
W_{N_c,K} \big( \mu; \nu ; \tau; \lambda \big) = 
\frac{\Gamma_h(\tau)^{N_c}}{N_c !} 
\int \prod_{i=1}^{N_c} d \sigma_i 
e^{\frac{i \pi }{2\omega_1 \omega_2} (2 \lambda tr \sigma - 2 K tr \sigma^2)} \\
\times \prod_{1 \le i < j \le N_c }   
 \frac{\Gamma_h(\tau \pm (\sigma_i - \sigma_j))}{\Gamma_h(\pm(\sigma_i -\sigma_j))}
 \prod_{i=1}^{N_c} \prod_{a,b=1}^{N_f} \Gamma_h(\mu_a + \sigma_i) \Gamma_h(\nu_b -\sigma_i) \;
\end{multline}
and $\Delta_X = 2/(k+1)$ is the $R$ charge of the adjoint field.
From (\ref{eq:antonello-catthivo}) we obtain 
\begin{align} 
   \mu_a &= \omega \, \Delta_Q + m_a &  \nu_a &= \omega \, \Delta_Q - \tilde m_a .
\end{align} 
The balancing condition (\ref{eq:bal-cond-4d}) reduces to 
\begin{equation} \label{eq:bal-cond-3d} 
  \sum_{a=1}^{N_f+2} (\mu_a + \nu_a) = \omega (N_f +2 -N_c \, \Delta_X),
\end{equation} 
reflecting the presence of the $\eta$-superpotential.
In the magnetic phase the index reduces to the partition function 
\begin{equation} 
\label{eq:Z-eta-mag}
Z_{mag} = \prod_{a,b=1}^{N_f+2}\prod_{j=0}^{k-1}
\Gamma_h\big(\mu_a+\nu_b+ j \omega \Delta_X\big) 
W_{\tilde N_c +2k,0}\big(\omega \Delta_X -\nu;\omega \Delta_X -\mu; \omega \Delta_X;-\lambda\big).
\end{equation}
The four-dimensional integral identity $I_{el} = I_{mag}$ reduces to $Z_{el} = Z_{mag}$,
with the constraint (\ref{eq:bal-cond-3d}).

Next we flow to Kim--Park duality by turning on some large real masses.
The parameters $\mu$ and $\nu$ become
\begin{align}
\mu_a &=
\begin{cases}
m_a+m_A+\omega \Delta_Q \\
M-\frac{m_A N_f}{2}+\omega \Delta_{Q_M}\\
-M-\frac{m_A N_f}{2}+\omega \Delta_{Q_M} \\
  \end{cases} &
 \nu_a &=
 \begin{cases}
-\tilde m_a + m_A+\omega \Delta_Q \\
-M-\frac{m_A N_f}{2 }+\omega \Delta_{Q_M}  \\
M-\frac{m_A N_f}{2}+\omega \Delta_{Q_M}     
 \end{cases}
\end{align}
where $a=1,\dots,N_f$.
The global $SU(N_f+2)^2$ symmetry gets broken to $SU(N_f)^2
\times U(1)_A$ in the large-$M$ limit, where 
(\ref{eq:Z-eta-el}) becomes 
\begin{equation}
  Z_{el} = e^{-\frac{i \pi}{2 \omega_1 \omega_2} (4 M N_c (m_A N_f-2 \omega  (\Delta_{Q_M}-1)))}
  W_{N_c,0}\big(\mu;\nu;\omega \Delta_X;\lambda\big) .
\end{equation}
In the magnetic case the situation is more complicated.
While the real mass follows from the one discussed in the electric theory,
through the duality map,
there is a non-trivial Higgsing in the dual gauge symmetry,
\begin{equation}
  \sigma_i=
  \begin{cases}
    0& i=0,\dots,k N_f-N_c \\
M& i=kN_f-N_c+1,\dots,k(N_f+1)-N_c\\
-M& i=k(N_f+1)-N_c+1,\dots,k(N_f+2)-N_c
  \end{cases}
\end{equation}
In the large-$M$ limit (\ref{eq:Z-eta-mag}) becomes 
\begin{multline} 
  Z_{mag} = e^{-\frac{i \pi}{2 \omega_1 \omega_2} \left(4 M N_c (m_A N_f-2 \omega  (\Delta_{Q_M}-1))\right)}
\prod_{j=0}^{k-1} \prod_{a,b=1}^{N_f}
    \Gamma_h\left(\mu_a+\nu_b+ j \, \omega \Delta_X  \right)   \\
 \times 
W_{k N_f  -N_c,0}\left(\omega \Delta_X -\nu_a;\omega \Delta_X-\mu_a; 
\omega \Delta_X;-\lambda\right) \times Z_+   Z_- .
\end{multline}
The additional terms $Z_{\pm}$ are the partition functions of the two $U(k)$ sectors
\begin{equation}
Z_{\pm}=\prod_{j=0}^{k-1} \Gamma_h(m_{j, \pm}) \; W_{k,0}
\left(\mu_{\pm},\nu_{\pm},  \omega \Delta_X, \lambda_{\pm}\right)
\end{equation}
with
\begin{align}
 m_{j,\pm} &= - m_A N_f+ 2 \omega  \Delta_{Q_M} + j \, \omega \Delta_X  \, ,
&
\mu_{\pm}=\nu_{\pm} &= \frac12 m_A N_f+\omega\left(\Delta_X-\Delta_{Q_M}\right)
\end{align}
and effective \ac{fi} terms 
\begin{equation}
  \lambda_\pm = -\lambda \pm  
  \left(N_f \, m_A + \omega  \left(2 N_f \Delta_Q + 2 \Delta_{Q_M} - (N_f +1) (k-1) \Delta_X \right)\right) .
\end{equation}
The $U(k)^2$ sector can be dualized to a set of $6k$ singlets. 
On the partition function it corresponds to the integral identity~\cite{VanDeBult}
\begin{equation}
\label{eq:IdA.5}
W_{N_c,0}(\mu;\nu;\omega \Delta_X;\lambda) \!= \!
\prod_{j=0}^{N-1}
\!\Gamma_h  \Big(\omega -\frac{\mu+\nu}{2} -j \, \omega \Delta_X\pm \frac{\lambda}{2}\Big) \,
\Gamma_h(\mu+\nu + j \,  \omega \Delta_X ) .
\end{equation}
In this case, there are $4k$ massive singlets that are integrated out.
This is reflected in the relation $\Gamma_h (z) \Gamma_h(2\omega-z)=1$ on the partition function.
We are left with the relation
\begin{equation}
Z_+ Z_- = \prod_{j=0}^{k-1} \Gamma_h \Big(
\pm\frac{\lambda }{2}-m_A N_f+\omega  \left(\left(j-N_c+1\right) \Delta_X +N_f \left(1-\Delta _Q\right)\right)
\Big) .
\end{equation}
This corresponds to the $2k$ singlets which remain light.
They have exactly the correct global charges for the superpotential coupling
with the magnetic monopoles, and  we identify them with the electric monopoles,
acting as singlet in the dual phase. 
The final identity for the Kim--Park duality is then 
\begin{multline} 
\label{eq:parkPF}
W_{N_c,0}\!\left(\mu;\nu;\omega \Delta_X;\lambda\right)
=
\prod_{j=0}^{k-1} \prod_{a,b=1}^{N_f} \Gamma_h(\mu_a+\nu_b+ j \, \omega \Delta_X)  \\
\times \prod_{j=0}^{k-1} \Gamma_h \Big(
\pm\frac{\lambda }{2}-m_A N_f+\omega  \left(\left(j-N_c+1\right) \Delta_X +N_f \left(1-\Delta _Q\right)\right)
\Big)
\\
\times W_{k N_f-N_c,0}\left(\omega \Delta_X-\nu;\omega \Delta_X-\mu;\omega \Delta_X;-\lambda\right).
\end{multline}

Eventually, the duality of~\cite{Niarchos:2008jb,Niarchos:2009aa}
is obtained by a further real mass flow. By integrating out the massive fields 
with (\ref{eq:limmass}) we obtain
\begin{multline}
  \label{eq:identity-Niarchos}
W_{N_c,K}\left(\mu;\nu;\omega \Delta_X;\lambda\right) 
=  e^{\frac{i \pi}{2 \omega_1 \omega_2} \, \phi} \zeta^{-k(2+K^2)} \; \,
\prod_{j=0}^{k-1}\prod_{a,b=1}^{N_f} \!  \left(\mu_a+\nu_b+j \, \omega \Delta_X \right)  \\
\times W_{k(N_f+K)-N_c,-K}\left(\omega \Delta_X -\nu;\omega \Delta_X-\mu;\omega \Delta_X;-\lambda\right) ,
\end{multline}
where $\zeta=\exp\big(\frac{\pi i (\omega_1^2 +\omega_2^2)}{24 \omega_1 \omega_2}\big)$ and the extra phase is
\begin{equation}
  \label{eq:phase}
  \begin{aligned}
\phi ={}&
\omega  m_A \big(2 k N_f \left(2(K- N_f) \Delta _Q- 2 N_c \Delta_X  - 2 N_f  +K (k-1)\Delta_X \right)\big)
  \\
&+ 2 k m_A^2 N_f \left(K-N_f\right)-\frac{k \lambda ^2}{2}+
k K \sum _{a=1}^{N_f} \left(\mu _a^2+\nu _a^2\right)
 \\
&-
k \omega^2 \Big( \Big(2 N_c \left(N_c+(k+1) N_f \left(\Delta _Q-1\right)-k K\right)+\frac{k^2 \left(11 K^2+2\right)+K^2-2}{12}\Big) \Delta_X^2
  \\
&-
2 N_f \Big(K \big(\Delta _Q^2+(k-1) \Delta _Q \Delta_X +\frac53 (\Delta_X-1) -1 \big)- N_f \left(\Delta _Q-1\right)^2\Big) \Big) .
\end{aligned}
\end{equation}
Equation (\ref{eq:identity-Niarchos}) is the identity between the electric and the magnetic partition functions of the duality in~\cite{Niarchos:2008jb,Niarchos:2009aa}.

\section*{Acknowledgments}

The authors are grateful to acknowledge the Simons Center for Geometry
and Physics, Stony Brook University where some of the work was performed.

The work of A.A. and S.R. is supported by the Swiss National Science Foundation (\textsc{snf}) under grant number \textsc{pp}00\textsc{p}2\_157571/1.

\printbibliography

\listoffigures

\end{document}